\documentclass[superscriptaddress,aps,pre,preprint]{revtex4-1}
\usepackage{graphicx} 
\usepackage{amsmath}
\usepackage[utf8]{inputenc}
\usepackage{booktabs}
\usepackage{enumitem}
\graphicspath{{./pictures/}}
\begin{document}
%\linenumbers
\title{The role of edge states for early-warning of tipping points}
\author{Johannes Lohmann}
\email{johannes.lohmann@nbi.ku.dk}
\affiliation{Physics of Ice, Climate and Earth, Niels Bohr Institute, University of Copenhagen, Denmark}
\author{Alfred B. Hansen}
\affiliation{Physics of Ice, Climate and Earth, Niels Bohr Institute, University of Copenhagen, Denmark}
\author{Alessandro Lovo}
\affiliation{ENS de Lyon, CNRS, Laboratoire de Physique, F-69342 Lyon, France}
\author{Ruth Chapman}
\affiliation{Physics of Ice, Climate and Earth, Niels Bohr Institute, University of Copenhagen, Denmark}
\author{Freddy Bouchet}
\affiliation{LMD/IPSL, CNRS, ENS, Université PSL, École Polytechnique, Institut Polytechnique de Paris, Sorbonne Université, Paris, France}
\author{Valerio Lucarini}
\affiliation{School of Computing and Mathematical Sciences, University of Leicester, Leicester, UK}

%%%% Abstract text to be placed here %%%%%%%%%%%%
\begin{abstract}
Tipping points (TP) are often described as low-dimensional bifurcations, and are associated with early-warning signals (EWS) due to {\it critical slowing down} (CSD). CSD is an increase in amplitude and correlation of noise-induced fluctuations away from a reference attractor as the TP is approached. But for high-dimensional systems it is not obvious which variables or observables would display the critical dynamics and carry CSD. Many variables may display no CSD, or show changes in variability not related to a TP. It is thus helpful to identify beforehand which observables are relevant for a given TP. Here we propose this may be achieved by knowledge of an unstable {\it edge state} that separates the reference from an alternative attractor that remains after the TP. This is because stochastic fluctuations away from the reference attractor are preferentially directed towards the edge state along a most likely path (the {\it instanton}). As the TP is approached the edge state and reference attractor typically become closer, and the fluctuations can evolve further along the instanton. This can be exploited to find observables with substantial CSD, which we demonstrate using conceptual dynamical systems models and climate model simulations of a collapse of the Atlantic Meridional Overturning Circulation (AMOC). 
\end{abstract}
%%%%%%%%%%%%%%%%%%%%%%%%%%%

%\rsbreak

%%%%%%%%%% Insert the texts which can accomdate on firstpage in the tag "fmtext" %%%%%

\maketitle

\section{Introduction}

Tipping points (TP) in complex systems are often assumed to obey the normal form of a saddle-node bifurcation, which gives rise to universal early-warning signals (EWS) due to critical slowing down (CSD) \cite{NOR81,WIS84,CRO88,HEL04}. In particular, when proceeding towards the bifurcation, saddle and node approach each other and eventually collide and disappear, at which point the largest Jacobian eigenvalue of the linearization around the node crosses the imaginary axis. This leads to an increase in variance and autocorrelation of fluctuations around the node as the bifurcation is approached. 
A general framework based on response theory that details the behaviour of a system and its EWS in the proximity of TPs has been recently presented \cite{GUT22}. 
%%% First: talk about variables. Here it is true that some variables show NO CSD
In practice, these EWS may be not so universal (independent of the system's details), since for high-dimensional systems it is not a priori clear which state variables dominate the dynamics close to the TP, and by extension, which variables would display the expected CSD. 
With the example of TPs in ecological systems it has been pointed out that if the components of the leading eigenvector are dominant in a small sub-set of variables, many other variables will not display any CSD \cite{BOE13}. 
% or could even show the opposite if associated with eigenvectors that become more contracting.
Such TPs without EWS have been termed ``silent catastrophes''. 

%%% Second: Talk about observables. 
For practical applications, one can take the viewpoint that not the actual state variables are measured, but instead observables \cite{MOR23}. If these are defined as scalar functions of all state variables, CSD should in theory be detectable because the divergence in variance of the single critical degree of freedom is imprinted. 
However, because a warning has to be issued before reaching the actual bifurcation with diverging variance, and due to the constraints of finite data and a finite speed of the underlying control parameter change, the EWS in an a priori chosen observable can be easily masked by the variability in non-critical degrees of freedom. Furthermore, in practice, an observable may be independent of a sub-set of state variables, which can lead to a genuine absence of CSD. 
%%% NOTE: is this also a question of the low-noise limit?
%%% NOTE: I guess the line between variable and observable is blurry. There can be 
%%% a set of dynamical equations for observables, which is equivalent to that of the 
%%% ``original'' state variables. 

For TPs in the climate system \cite{LEN08} this leads to a problem of interpretability, because often not all relevant state variables or observables are measured directly and over a long enough time period to establish a baseline of fluctuations prior to anthropogenic global warming. In the case of a potential future collapse of the Atlantic Meridional Overturning Circulation (AMOC), for instance, while past studies argued for CSD in historical fingerprints (or proxies) of the main ``zeroth-order'' observable (the strength of the AMOC streamfunction) \cite{BOE21a,MIC22,DIT23}, a recent study using a coupled climate model showed that the AMOC strength is not associated with any CSD before an AMOC collapse \cite{VWE24}. Observations of CSD in historical climate data may even lead to statistical false positives due to the large number of possible variables available. %cite amazon studies. 

Which observables should be used to detect CSD? Previous work explored data-driven approaches to EWS in multivariate systems, such as by extracting the leading modes of the system using empirical orthogonal functions (EOF), principal oscillation patterns (POP) and Min/Max autocorrelation factors \cite{HEL04,BAT13,KWA18,PRE19,WEIN19}. 
Such dimension reduction methods must capture the degrees of freedom not only of maximum variability, but of maximum {\it increase} in variability (hence CSD), which may not always be the case. Other approaches more directly target the underlying changes in stability, such as by fitting multiple time series to a multivariate autoregressive process and finding a diagonalized Jacobian matrix \cite{WIL15}, or by estimating a multidimensional Langevin equation \cite{MOR24}.
%%% Morr 2024: estimating multidimensional langevin equation.
%check and revise the following: 
Such methods are promising when all variables are measured, or perhaps when as a preliminary step the phase space can be reconstructed by a time-delay embedding of long observational time series \cite{MUS21}.
%%% NOTE prettyman 2019 argue that first EOF is most likely to also be the relevant one for the tipping point.
%%% not sure I buy the argument. 

%NOTE better justification of why we do it differently:
%data-driven methods have potential weaknesses in that not all data is available,
%and that it is hard to find the relevant ``critical'' mode of CSD, which may be masked
%by other dominant variability
%we want to establish beforehand which variables or observables should be measured. 
%``physics-based'' approaches to find the correct degrees of freedom are good,  
%but it can still be that the CSD is hidden in variables that have not been considered
%by investigation of the physical tipping mechanisms in such complex systems
%here we want to instead explore a general guideline, using the edge state, which 
%may or may not be applied in practise. 

%%% NOTE: introduce generalization of the above saddle-node bifurcation. 
%%% saddle = edge state. 
%%% can be an unstable periodic orbit (with stable manifold), or chaotic saddle. 
Here we explore an alternative approach to finding observables that carry EWS, based on the global properties of the system via knowledge of the system's so-called edge state. The edge state is a (possibly chaotic) saddle in phase space, and its stable manifold is the basin boundary separating two co-existing attractors \cite{ITA01, Skufca2006, LUC17}. 
%%% NOTE: originally separating the super-transient of turbulence and the only true node. 
We assume that a) the system exhibits fluctuations away from a base attractor due to an external source of (weak) noise, b) there exists an alternative attractor, and c) the system experiences a slow change in a control parameter such that a TP is approached where the base attractor disappears (saddle-node bifurcation) or loses stability (chaotic attractor) and is replaced by a chaotic transient. 
In this case, there should be a specific direction in phase space along which there are the largest increases in the amplitude and autocorrelation of fluctuations away from the base attractor as the TP is approached. For a given parameter value, the preferred direction of fluctuations away from an attractor is known from large-deviation theory in the low-noise limit. 
Here, the rare fluctuations furthest away from the base attractor are concentrated around a most likely noise-induced transition path (known as instanton), and deviations from this path are exponentially unlikely \cite{FRE84, Graham1987, Graham1991, BOU16, LUC19, MAR21}. The path connects the base attractor with the set of minimum quasipotential on the basin boundary. This set is usually believed to coincide with the dynamical edge state (see discussion) \cite{MAI97,BOU16}. According to Kramer's law and its generalizations \cite{BOU16}, the probability of visiting the vicinity of the edge state decays exponentially with the height of the quasipotential barrier. This barrier is lowered as the TP is approached, and thus for a given, weak noise level, excursions in this vicinity will only begin to occur close to the TP. As a result, fluctuations that evolve a significant distance along the instanton away from the attractor become less rare and the variability is increasingly oriented towards the edge state.
%%% NOTE: say here that there could be boundary crises where the attractor collides
%%% with a region of the boundary far from the edge state 
%%% then, generalization required. 
In general, the initial segment of the instanton pointing away from the base attractor, and a vector that points from the attractor towards the edge state do not need to be aligned. But as the TP is approached this alignment will become stronger.

We propose that this can be exploited to construct observables that show the largest increases in fluctuations indicative of an approaching TP, especially if the knowledge of the edge state allows one to find low-dimensional projections of the dynamics onto observables where the edge state ``stands out'' from the base and alternative attractor. 
In this paper, this principle is illustrated with conceptual models displaying fixed point dynamics, and then further exploited to study the case of a collapse of the Atlantic Meridional Overturning Circulation (AMOC) simulated by a global ocean model with chaotic attractors. 

Depending on the model complexity, different levels of detail may be achievable. As prerequisite the alternative attractor needs to be known, which is relatively simple by forward simulations from different initial conditions. Computing the edge state can be achieved in complex models by an edge tracking algorithm \cite{Skufca2006, Vollmer2009, LUC17, LOH24b}. %other techniques? feedback control?
Even better observables to use for EWS may be obtained by also computing the instanton paths. In principle this can be achieved by long simulations with added noise, but much more efficiently using rare-event algorithms \cite{GIA06, RAG18}. In such algorithms, an ensemble of trajectories is propagated forward for small time intervals and repeatedly 
the trajectories most likely to undergo a transition are selected and replicated, which is based on a score function that needs to be designed in order to quantify the movement towards the alternative attractor. % or the edge state
This allows one to sample also very rare transitions between attractors for low noise strengths. Finally, although it may be unachievable for complex models, even more precise results may be obtained by direct methods to find the path that minimizes the Freidlin-Wentzell action, such as by nonlinear optimization \cite{WEI02, WEI04, GRA17, KIK20} or solving the Hamilton-Jacobi equation \cite{CAM12}. 
%%% NOTE: maybe even mention Onsager-Machlup action. 

% STRUCTURE OF THE PAPER.
The paper is structured as follows. We first demonstrate our approach of constructing observables for EWS using conceptual models (Sec.~\ref{sec:conceptual}), starting with a bi-stable gradient system in two dimensions (Sec.~\ref{sec:gradient}), followed by two non-gradient, conceptual models of the AMOC. The second half of the paper presents an analysis of the fluctuations leading up to a TP of the AMOC in a global ocean model driven by surface temperature and salinity noise (Sec.~\ref{sec:veros}). Here, we first present an overview of the attractors and edge states (Sec.~\ref{sec:stability_landscape}), and then describe the dynamics as noise is added (Sec.~\ref{sec:veros_noise}). Thereafter, we investigate the possibility of finding the instanton in the low-noise limit using a rare-event algorithm, and, since this proves difficult, use higher noise levels to determine an approximate instanton path via the edge state  (Sec.~\ref{sec:veros_transitions}). Finally, for lower noise levels and an approach to the TP, we show that changes in fluctuations and observables displaying CSD can indeed be related to the edge state (Sec.~\ref{sec:veros_ews}). 

\section{Results}

\subsection{Conceptual models}
\label{sec:conceptual}

\subsubsection{Gradient system}
\label{sec:gradient}

\begin{figure}%[floatfix]%!htb
\includegraphics[width=0.99\textwidth]{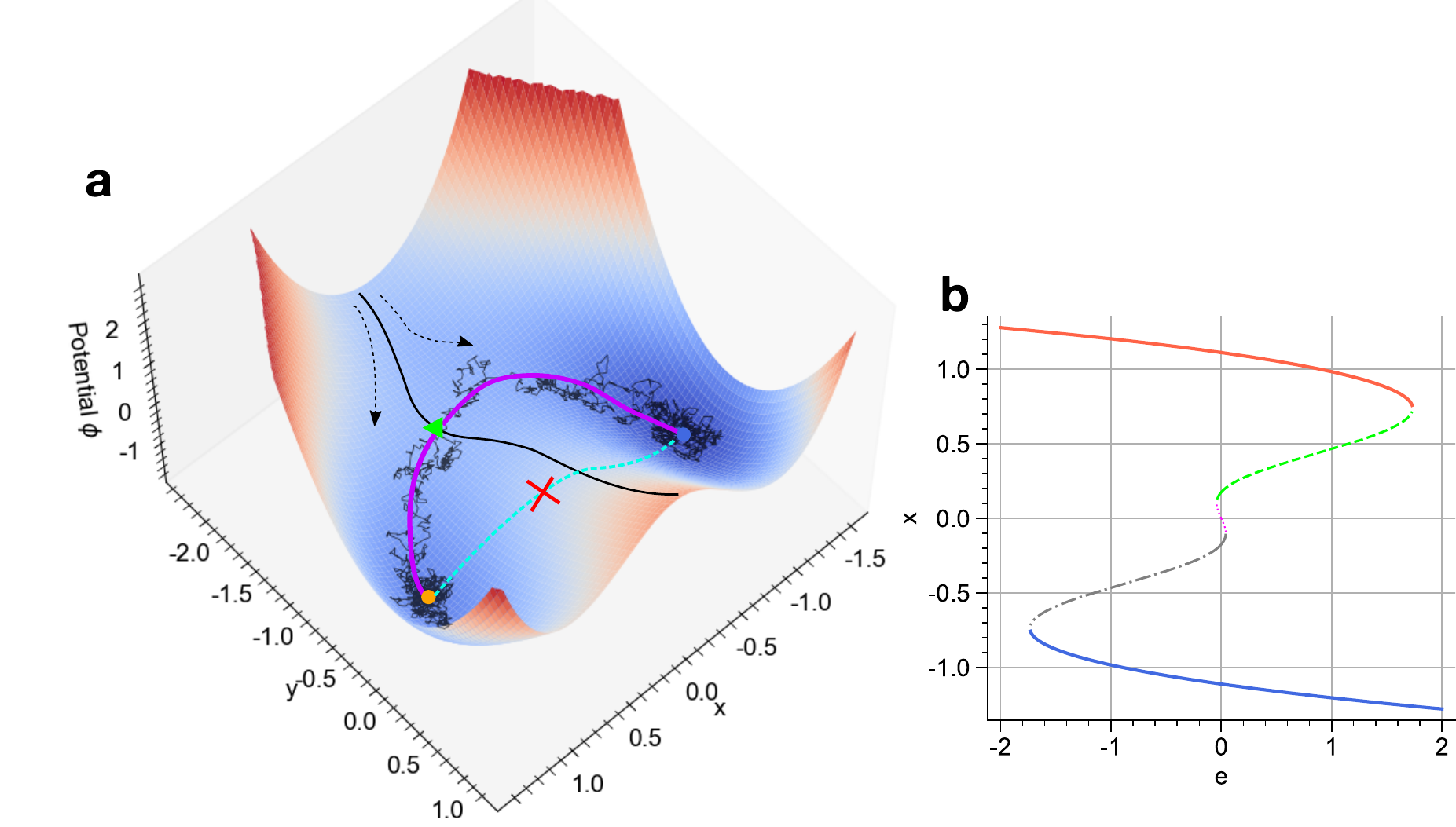}
\caption{\label{fig:potential2D} 
{\bf a} Potential defined by Eq.~\ref{eq:potential} with $e=0.2$, as well as one realization of a noise-induced transition between the two stable fixed points using $\sigma_x = \sigma_y = 0.1$. Also shown is the basin boundary (black) and the instanton (purple), as well as a hypothetical more ``direct'' transition path of shorter length in phase space (dashed blue line). 
{\bf b} Bifurcation diagram of the system described by Eq.~\ref{eq:gradient}
projected onto the variable $x$. The dashed line is the saddle. 
}
\end{figure}

We first consider a gradient system of two variables $x$ and $y$, which is defined by the potential 
\begin{equation}
\label{eq:potential}
\phi(x,y) = x^2(x^2 + y^2 - a) + y\frac{cy+d}{x^2 + b} + ex. 
\end{equation}
Adding Gaussian white noise independently to both variables yields the system of stochastic differential equations

%\begin{equation}
%\label{eq:gradient}
% \begin{aligned}
% \frac{dx}{dt} &= -\frac{\partial \phi}{\partial x} = - 4x^3 + 2ax -2xy^2 + \frac{2x %(cy^2+dy)}{(x^2 + b)^2} - e \\
% \frac{dy}{dt} &= -\frac{\partial \phi}{\partial y} = -2x^2y - \frac{2cy + d}{x^2 + b}.
%\end{aligned}
%\end{equation}

\begin{equation}
\label{eq:gradient}
\begin{pmatrix}
dx_t\\[\jot]
dy_t
\end{pmatrix}=\begin{pmatrix}
 -\frac{\partial \phi}{\partial x}\\[\jot]-\frac{\partial \phi}{\partial y}
\end{pmatrix} dt + \begin{pmatrix}
\sigma_x \, dW_{x,t}\\[\jot]\sigma_y \, dW_{y,t} ,
\end{pmatrix}
\end{equation}
where $dW_{x,t}$ and $dW_{y,t}$ are standard Wiener processes. Fixed values $a=2.5$, $b=0.5$, $c=0.2$, $d=0.5$ are used, and $e$ is the bifurcation parameter. For small $e$, there are two stable fixed points and one saddle point (Fig.~\ref{fig:potential2D}) in the deterministic system. The saddle, i.e., the edge state, is the point of lowest potential on the basin boundary, which is its stable manifold. Thus, in the low-noise limit, a noise-induced transition will cross the basin boundary in the vicinity of the saddle, instead of following a path of shorter length in phase space \cite{FRE84, Graham1987, Graham1991}. Typical transition trajectories are concentrated closely around a path that minimizes the Freidlin-Wentzell action, the so-called instanton. 
% Consider some equations?
In Fig.~\ref{fig:potential2D}a the instanton is shown by the purple line, which we computed here following the method and PyRitz {\it python} package presented in \cite{KIK20}. Since this is a gradient system, the instanton is made up by the unstable manifold of the saddle, and it is the same path for a transition going in either direction. 

\begin{figure}%[floatfix]%!htb
\includegraphics[width=0.99\textwidth]{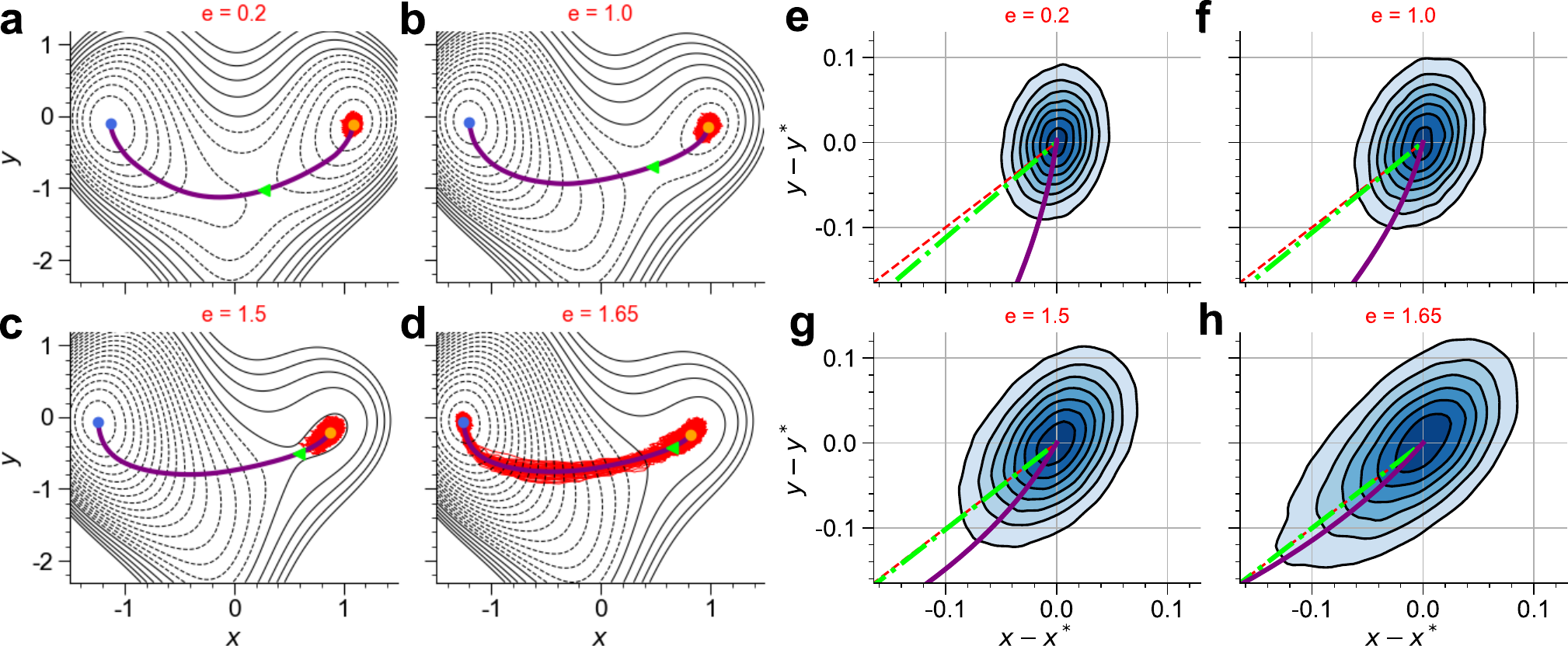}
\caption{\label{fig:gradient2D_results} 
{\bf a-d} Potential isolines of Eq.~\ref{eq:potential}, along with the fixed points (colored symbols) of Eq.~\ref{eq:gradient} for four values of the bifurcation parameter $e$ in increasing order until shortly before the bifurcation point. The green triangle is the edge state. Also shown in each panel is an ensemble of 100 simulations of Eq.~\ref{eq:gradient} with additive Gaussian white noise (red trajectory), and the instanton (purple) computed by the method in \cite{KIK20}. The duration of the simulations is $t=1500$. %simulated for 300,000 time steps, i.e., T=1500
{\bf e-h} Probability densities of residuals around the fixed point ($x^*$, $y^*$) aggregated from an ensemble of simulations with $N=1000$ realizations (fixed simulation time $t=1500$) of the model in Eq.~\ref{eq:gradient} with added Gaussian white noise and $\sigma_x = \sigma_y = 0.1$. Each realization is initialized from the fixed point ($x^*$, $y^*$) with $x>0$, and simulations have been performed for four different values of the control parameter $e$. For the largest $e$ ({\bf d}), which is close to the bifurcation, noise-induced transitions occur, the effect of which on the residuals is removed by cutting the time series at the time when the $x$-value of the saddle is crossed for the last time before tipping to the other fixed point. The red dashed line is the identity line, associated with the observable $x+y$, and the green line is the vector pointing from the attractor to the edge state. 
}
\end{figure}

A bifurcation occurs at $e_c \approx 1.73$, where the saddle and the node with $x>0$ collide. Level sets of the potential for four values of $e$ are shown in Fig.~\ref{fig:gradient2D_results}a-d. 
We simulate the noisy system with $\sigma_x = \sigma_y = 0.1$ using the Euler-Maruyama scheme with a time step of $dt = 0.005$, initialized at the fixed point ($x^*$, $y^*$) with $x>0$. When far away from the bifurcation, this noise strength is too small to induce a transition to the other attractor even during long simulations (Fig.~\ref{fig:gradient2D_results}a,b). Here, the fluctuations do not evolve far enough along the instanton (purple line) to bring the system close to the saddle. When increasing $e$ towards the bifurcation point, the saddle moves closer to the node, and the fluctuations eventually reach the saddle and cross the basin boundary (Fig.~\ref{fig:gradient2D_results}c,d). This is the scenario of so-called N-tipping \cite{ASH12}.

The fluctuations away from the node are increasingly directed towards the saddle as the bifurcation is approached. This is seen by collecting the residuals (deviations from the node) of repeated simulations with a fixed duration of $t=1500$ for different values of $e$ (Fig.~\ref{fig:gradient2D_results}e-h), while discarding any parts of a trajectory that crossed the basin boundary and tipped to the other fixed point. For our choice of $t$, tipping only happens when $e$ is close to the bifurcation. 
The probability densities of the fluctuations (Fig.~\ref{fig:gradient2D_results}e-h) become wider and more eccentric, with the major axes pointing in the direction of the instanton (purple line), which becomes increasingly aligned with the direction defined by the vector from the node to the saddle (green dashed lines).

\begin{figure}%[floatfix]%!htb
\includegraphics[width=0.99\textwidth]{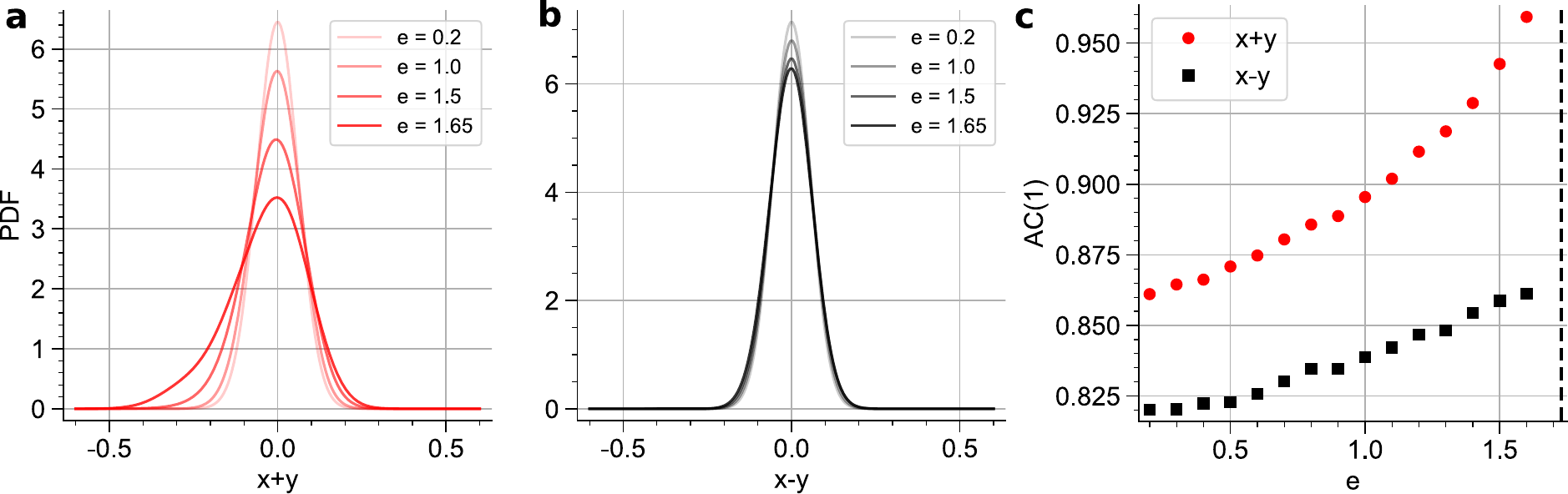}
\caption{\label{fig:fluctuations_2D_observables} 
From the same simulations as in Fig.~\ref{fig:gradient2D_results}e-h we extract the distribution of fluctuations (around the mean value) of the observables $x+y$ ({\bf a}) and $x-y$ ({\bf b}) for four values of the control parameter $e$. 
Panel ({\bf c}) shows the evolution of the autocorrelation at lag 1 for the two observables towards the bifurcation, which has been averaged over all realizations of ensembles for a range of values of $e$. By ``lag 1'' we mean to indicate that the correlation of subsequent time series samples is calculated, where a sample spacing of 0.05 time units, i.e., every tenth time step of the numerical integration, is used.  
The vertical dashed line is the bifurcation point. 
}
\end{figure}

This means that the system displays CSD as expected, but the degree to which it can be measured depends on which observables are available as time series of finite length. While an observable could be any function of a subset of state variables, we consider the simplest case of a linear combination $\alpha x + \beta y$ of the state variables. For spatio-temporal data as often obtained in climate science, important cases such as spatial averages or gradients would fall into this category. The observable $x+y$, i.e., a projection of the residuals onto a vector parallel to the identity line (red line in Fig.~\ref{fig:gradient2D_results}e-h), is almost parallel to an observable defined by the direction of the saddle (green line). Using the time series defined by this observable we find that $x+y$ shows clear increases in variance (Fig.~\ref{fig:fluctuations_2D_observables}a), and the expected tendence of the autocorrelation coefficient towards 1 (Fig.~\ref{fig:fluctuations_2D_observables}c), 
In contrast, the observable $x-y$ is almost perpendicular to the direction of the edge state. It shows almost no increase in variance (Fig.~\ref{fig:fluctuations_2D_observables}b) and only a modest increase in autocorrelation (Fig.~\ref{fig:fluctuations_2D_observables}c). 

As the bifurcation is approached, there is an increase in asymmetry between the lower and the upper tails of the density of the observable $x+y$, with large negative fluctuations becoming relatively more likely, in agreement with \cite{FAR14}. Indeed, increasing skewness is an often-cited indicator of CSD in systems reduced to one dimension \cite{GUT08}, and here we see that in the multi-dimensional case it holds (only) for observables that that have a strong projection onto the vector pointing from the base attractor to the edge state, or onto the instanton path. 

\subsubsection{Stommel model}
\label{sec:stommel}

Next, we demonstrate applicability of the principle to non-gradient systems with the example of Stommel's well-known box model of the AMOC \cite{STO61} with added Gaussian white noise:

\begin{equation}
\label{eq:stommel}
\begin{pmatrix}
dT_t\\[\jot]
dS_t
\end{pmatrix}=\begin{pmatrix}
 \eta_1 - T -| T-S| T \\[\jot]\eta_2 - \eta_3 S -| T-S| S
\end{pmatrix} dt + \begin{pmatrix}
\sigma_T \, dW_{T,t}\\[\jot]\sigma_S \, dW_{S,t}, 
\end{pmatrix}
\end{equation}
where $dW_{S,t}$ and $dW_{T,t}$ are independent Wiener processes. 
$T = \alpha_T (T_e - T_p)$ and $S = \alpha_S(S_e - S_p)$ are the dimensionless meridional difference in temperature and salinity of an equatorial and a polar ocean basin. The circulation strength $q$  is defined as proportional to the meridional density gradient $\rho_p - \rho_e$, where density is defined by the equation of state
\begin{equation}
 \rho_{e,p} = \rho_0 \left[1 - \alpha_T (T_{e,p}-T_0) + \alpha_S (S_{e,p}-S_0)\right ],
\end{equation}
with reference density $\rho_0$, temperature $T_0$ and salinity $S_0$. Thus, the strength of the overturning circulation is $q = (\rho_p - \rho_e)/\rho_0 = T - S$.

%NOTE SHORTEN much of the following.
The non-dimensional form of the model we consider here has three parameters. $\eta_1$ ($\eta_2$) represents the meridional gradient of the atmospheric temperature (salinity) forcing, and is furthermore proportional to $\alpha_T$ ($\alpha_S$). $\eta_3$ is the ratio of the time scales of relaxation to atmospheric forcing by the ocean salinity with respect to temperature,  
where a low value is required for bistability (we use $\eta_3 = 0.3$). For more details see \cite{DIJ08}. Changing either $\eta_1$ or $\eta_2$ can decrease the meridional density gradient, and thus weaken the AMOC. Here we use $\eta_1$ as bifurcation parameter and keep $\eta_2=1.0$ fixed, but equivalent results are obtained when varying $\eta_2$. 

\begin{figure}%[floatfix]%!htb
\includegraphics[width=0.95\textwidth]{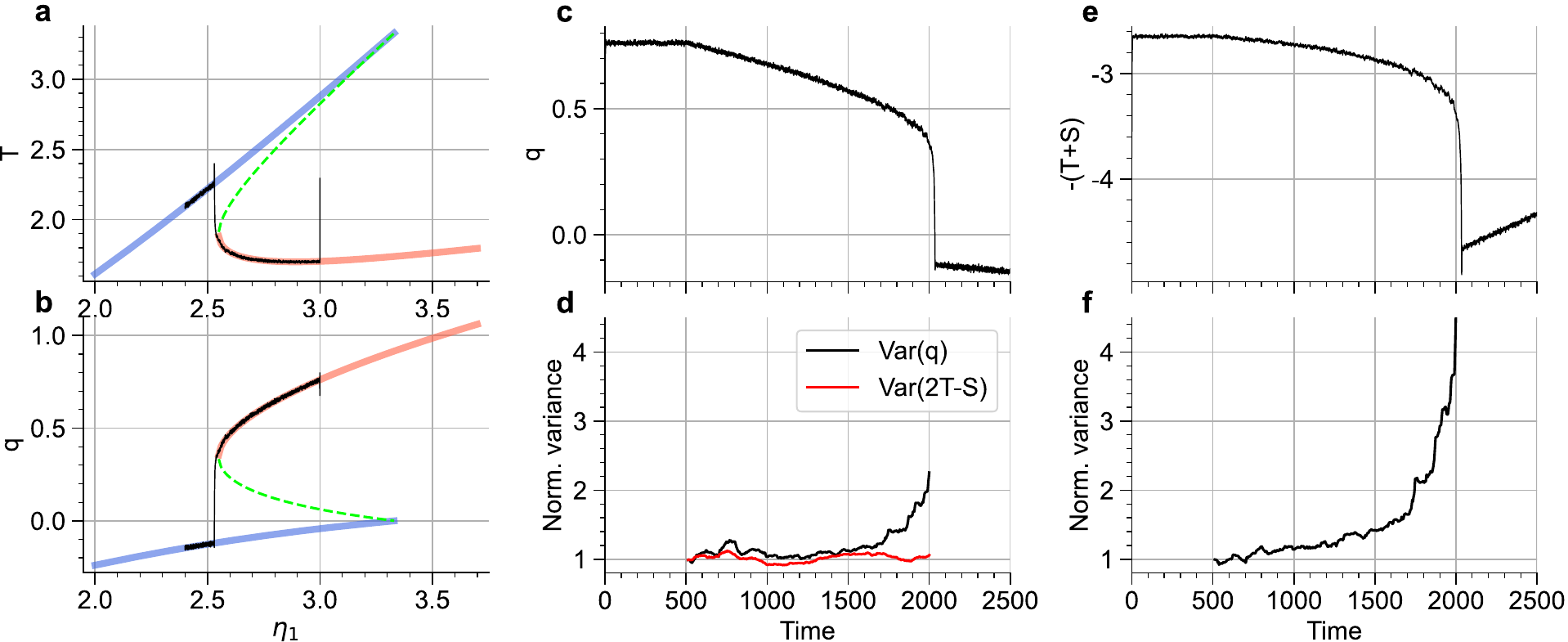}
\caption{\label{fig:stommel_bif_ews} 
{\bf a,b} Bifurcation diagram of the Stommel model, projected on the variable $T$ ({\bf a}) as well as the observable $q=T-S$ ({\bf b}). The 'ON' ('OFF') state is shown in red (blue), and the saddle by the green dashed line. 
{\bf c,e} Time series of $q$ and the observable $T+S$ for a simulation where $\eta_1$ is ramped up linearly across the bifurcation. 
{\bf d,f} Early-warning signal of increased variance in a sliding window for the time series in ({\bf c,e}). The variance is normalized by the variance in the first 500 years, where $\eta_1$ was held fixed. The time points are the endpoints of the sliding windows. Panel {\bf d} shows the variance for $q$ as well as the observable $2T-S$ (red), and {\bf f} shows the variance for the observable $T+S$. 
}
\end{figure}

The model features bistability, where a strong AMOC `ON' state and a weak, reversed circulation (`OFF' state) co-exist (Fig.~\ref{fig:stommel_bif_ews}a,b). When decreasing $\eta_1$, the ON state collides with a saddle point and disappears. A  simulation of Eq. \ref{eq:stommel} initialized in the 'ON' state with $\sigma_T = \sigma_S = 0.005$, where $\eta_1$ is ramped slowly and linearly from 3.0 to 2.4 from $t=500$ to $t=2500$, is shown by the black trajectory in Fig.~\ref{fig:stommel_bif_ews}a,b. In Fig.~\ref{fig:stommel_bif_ews}c,e are corresponding time series for two different observables. For these observables, we compute the evolution of the variance over time, using a sliding window of length $\tau = 500$ and cutting the timeseries at $t=2000$, which is before tipping. The time series in each window is first detrended with a cubic spline. While $q$ shows a doubling of the variance leading up to the tipping point (Fig.~\ref{fig:stommel_bif_ews}d), the observable $T+S$ has a much stronger, and more consistent trend in the variance (Fig.~\ref{fig:stommel_bif_ews}f). One may also find observables, such as $2T - S$ (red line in Fig.~\ref{fig:stommel_bif_ews}d), which show virtually no increase in variance.

\begin{figure}%[floatfix]%!htb
\includegraphics[width=0.99\textwidth]{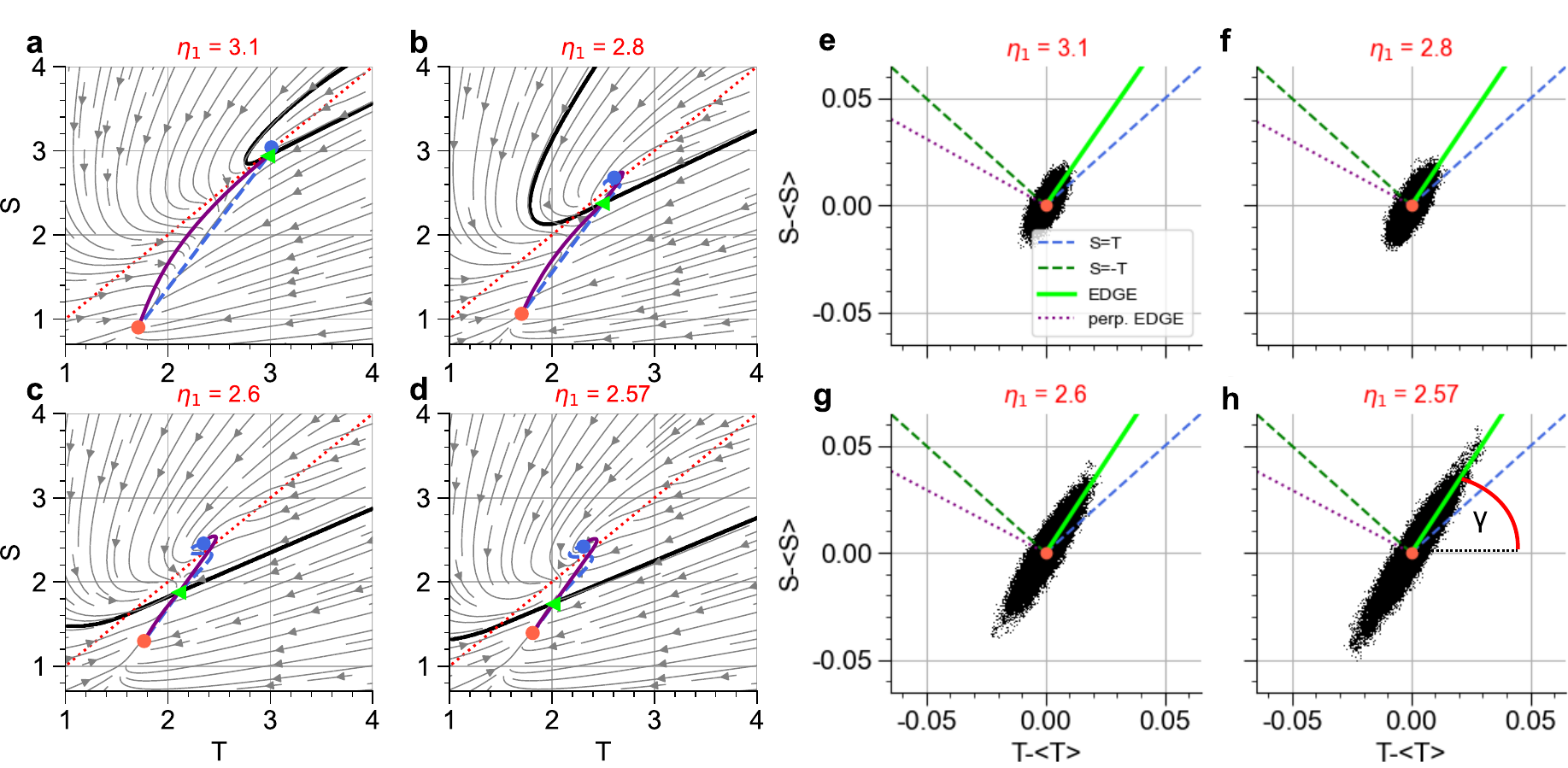}
\caption{\label{fig:stommel_results} 
{\bf a-d} Phase portrait of the Stommel model for four values of $\eta_1$ progressing towards the bifurcation where the ON state loses stability. The two stable fixed points are shown by the red (ON) and blue (OFF) dots, and the saddle point is the green triangle. The basin boundary is in black. The red dashed line is the identity line $T=S$. The instanton from ON to OFF is in purple, and the one from OFF to ON is the blue dashed line. Both are computed by the method in \cite{KIK20}. 
{\bf e-h} Residuals with respect to the ON fixed point obtained by an ensemble of long simulations for four values of $e$ (same as in panels a-d).
The different straight lines correspond to vectors defining observables discussed in the main text. The green solid line indicates the direction of the edge state. 
}
\end{figure}

%%% SHORTEN!
To see more clearly which observable shows strongest CSD, we investigate the system at four fixed values of $\eta_1$ approaching the bifurcation. The flow, fixed points, instantons and basin boundary of the system at these parameter values are shown in Fig.~\ref{fig:stommel_results}a-d. 
Since it is a non-gradient system, the instanton is not equivalent to the unstable manifold of the saddle in the deterministic system, and the paths in opposite directions are not the same, i.e., the relaxation and fluctuation dynamics are different. 
As the TP is approached, the instanton path from ON to OFF (purple) becomes very accurately aligned with the vector pointing from the ON state to the saddle. As a result, fluctuations driven by noise generally venture furthest away from the fixed point along the direction of the saddle. Figure~\ref{fig:stommel_results}e-h shows scatterplots for the residuals of $T$ and $S$ (with respect to the ON fixed point) for long simulations initialized at the ON state with low noise $\sigma_T = \sigma_S = 0.005$. During the simulations there is no tipping towards the OFF state. As the bifurcation is approached, the residuals form an increasingly eccentric ellipse, with a major axis aligned to the direction of the saddle (solid green line). 
Thus, an observable defined by a scalar projection of the residuals onto a vector aligned with the major axis will show high increases in variance. Observables that are directed along a projection close to the minor axis of the ellipse show almost no increase in variance. 

\begin{figure}%[floatfix]%!htb
\includegraphics[width=0.92\textwidth]{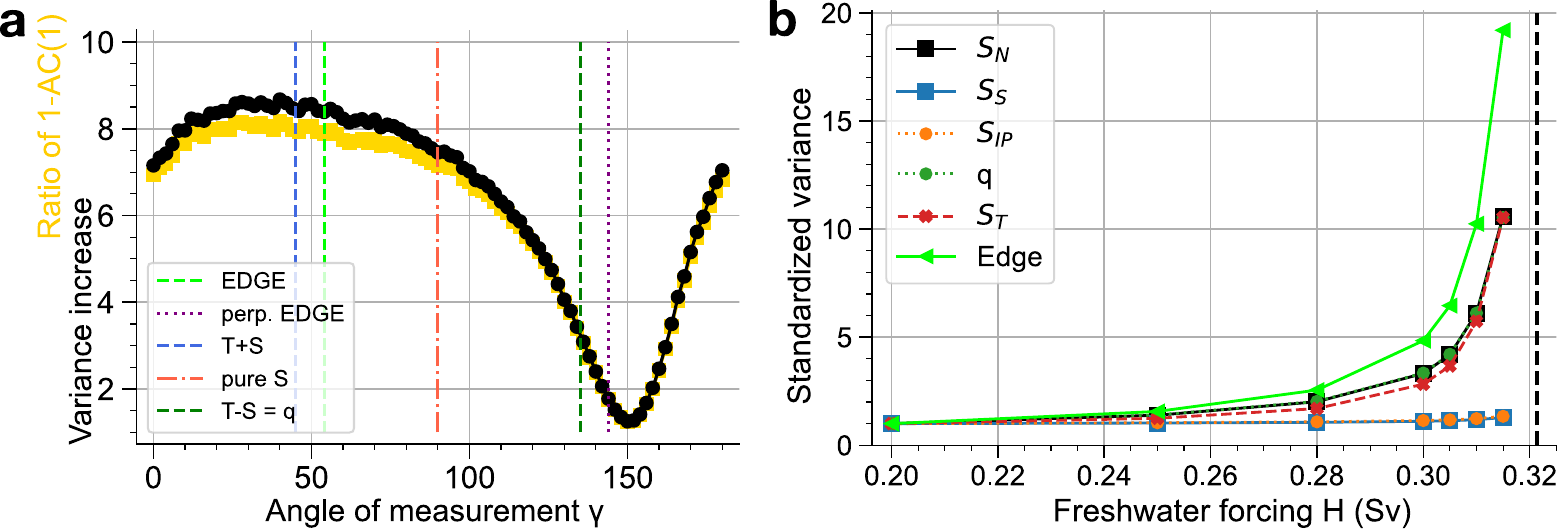}
\caption{\label{fig:stommel_var_increase} 
{\bf a} Variance of linear observables (defined by the angle of measurement $\gamma$, see main text) for long simulations of the Stommel model at the parameter value $\eta_1 = 2.57$ 
divided by the variance at the parameter value $\eta_1 = 3.1$, which is further away from the TP. The various vertical lines indicate the values of $\gamma$ that define different, specific observables. 
Shown in yellow is the ratio of $1-AC(1)$ at $\eta_1 = 2.57$ divided by the same quantity at $\eta_1 = 3.1$. Here, $AC(1)$ is the autocorrelation coefficient at lag 1 of the observable time series, with a sample spacing of 0.05 time units. 
{\bf b} Variance in the fluctuations around the ON fixed point in different variables and observables of the 5-box AMOC model, averaged over large ensembles of simulations at $\sigma = 10^{-6}$ at different values of the control parameter $H$, increasing until shortly before the bifurcation point (vertical dashed line). For each variable/observable the variance is normalized to be equal 1 at $H=0.2$~Sv. The observable marked ``edge'' is defined by $S_T - 1.0841\cdot S_N$.
}
\end{figure}

We define an ``angle of measurement'' $\gamma$ with respect to the axis $\tilde{S}=0$ (Fig.~\ref{fig:stommel_results}h), and an observable by the scalar projection of ($\tilde{T}$, $\tilde{S}$) onto the vector ($\cos \gamma$, $\sin \gamma$)$^\top$. The observable $T+S$ then corresponds to $\cos \gamma = \sin \gamma$, i.e. an angle of 45 degrees. $q = T-S$ corresponds to 135 degrees. We show the increase in variance going from $\eta_1 = 3.1$ to $\eta_1 = 2.57$ averaged over a large number of long simulations (discarding any portions of trajectories that include a noise-induced transition) as a function of $\gamma$ in Fig.~\ref{fig:stommel_var_increase}a. Also shown is a measure of the increase in autocorrelation, which produces almost exactly the same values as the increase in variance. 
%NOTE I need to state at what parameter value EDGE was defined here. 
The AMOC strength $q$ is close to a minimum corresponding to an angle where there is almost zero variance increase (ratio of 1), and thus shows only small CSD. 
In contrast, there is a broad maximum of the variance increase for angles in the quadrant of the direction of the saddle (green dashed line), where the latter has been defined at $\eta_1 = 3.1$ and does not change much as $\eta_1$ is decreased towards the bifurcation (see Fig.~\ref{fig:stommel_results}). 

Knowing the edge state allows us to find a physically meaningful observable with almost maximal variance increase. The gradient of ``spiciness'' $T+S$, i.e., a projection onto the line $T=S$ (red dotted line in Fig.~\ref{fig:stommel_results}a-d, and blue dashed line in Fig.~\ref{fig:stommel_results}e-h), is only at a small angle to the instanton and the edge vector and thus enjoys near-maximum variance increase. 
For reference, the oceanographic term spiciness is defined by a sum of temperature and salinity of a parcel of water, such that combined changes in $T$ and $S$ leave the density of the parcel unchanged. It follows that the perpendicular observable of the density gradient $T-S$, which is proportional to the AMOC strength, shows almost minimal increase in variance.

%%% give numbers for the edge vector etc. (at what parameter value?)

% consider whether this is precise enough? Can anything be perpendicular without a 
% reference scale?
Projections onto the single state variables $T$ or $S$ also yield relatively strong CSD. In higher-dimensional systems there may, however, be many variables that are almost perpendicular to the direction of the saddle, i.e., where the saddle is not much different from the attractors when projected onto these variables. 

%%% deviation from exact edge direction may be because it is really instanton direction that counts, and fuzziness with regards to what parameter value the edge direction should
%%% be estimated from. 

% also note that we talk about INCREASE in variance and not just variance. 
% -> Variance (at given eta1) has much clearer peak in edge direction. (should be obvious from the scatter plots). 

\subsubsection{Five-box model of the AMOC}
\label{sec:five_box}

%%% THIRD case: higher dimensions. 
The models discussed so far were two-dimensional examples, where the critical dynamics projected significantly onto both variables. Next, we want to consider a higher-dimensional conceptual model of the AMOC, which should illustrate the more ``typical'' case of a complex system with many degrees of freedom not directly involved in the critical dynamics. In this case, there can be a larger contrast of EWS derived from state variables compared to more optimized observables. In this model, the global ocean is divided into five boxes $X$ \cite{WOO19} of volume $V_X$, where the subscripts $X$ denote the following: $S$ is the Southern Ocean near-surface water; $T$ is the Tropical Atlantic thermocline; $N$ is the North Atlantic Deep Water (NADW) box; $B$ refers to the ``bottom'' waters and represents the southward moving NADW, as well as the deep water in the Pacific. $IP$ represents the Indo-Pacific from the surface down to the thermocline. 

The model variables are the box-averaged salinities $S_X$, which are coupled unidirectionally by the thermohaline overturning circulation, and bi-directionally by the wind-driven circulation (see Fig.~S11 for a schematic). All boxes except $B$ are forced by an atmospheric freshwater flux $F_X$ multiplied by the reference salinity $S_0 = 0.035$, which is then modulated by the additional contribution $H A_X$ to emulate the effect of climate change, where $H$ is the control parameter. The closed-loop coupling implies that the total salt content of the ocean is a constant $C$ for all times \cite{ALK19}, given that the atmospheric fluxes cancel $\sum_X F_X = 0$, which is the case here. Thus, the system can be reduced to four variables, expressing $S_B$ by the other variables: 
\begin{equation}
\label{eq:sb_solve}
S_B = (C - V_N S_N - V_T S_T - V_S S_S - V_{IP} S_{IP})/V_B,
\end{equation}
where we fix $C=4.4466 \times 10^{16}$. The varying strength of the overturning flow $q$ is assumed proportional to the density difference in the northern and southern boxes \cite{WOO19}:
\begin{equation}
    q = \lambda ( \alpha (T_{\rm{S}} - T_{\rm{N}}) + \beta (S_{\rm{N}} - S_{\rm{S}})).
    \label{eq:q_1}
\end{equation}
Here, $\alpha$ and $\beta$ define a linear equation of state for the density. The temperatures $T_X$ are fixed everywhere except the northern box, where it is assumed $T_N = \mu q + T_0$, with a global reference temperature $T_0$. It thus follows

\begin{equation}
    q =  \lambda\frac{  \alpha (T_{\rm{S}} - T_{\rm{0}}) + \beta (S_{\rm{N}} - S_{\rm{S}})}{1 + \lambda \alpha \mu }.
    \label{eq:q_2}
\end{equation}

As in the Stommel model, $q>0$ corresponds to an AMOC `ON' state, and it is assumed that in case of a reversed circulation $q<0$ the unidirectional coupling by the overturning flow is reversed. This yields different dynamics for postive and negative $q$, and a non-smooth system of four ODEs, using the Heaviside function $\Theta (\cdot)$:

\begin{subequations}  \label{eq:5box}
\begin{flalign}
   & V_N \frac{dS_N}{dt} =  |q| \left[ \Theta(q) (S_T - S_B) + S_B - S_N \right] + K_N(S_T - S_N) - (F_N + H A_N) S_0 \\
   & V_T \frac{dS_T}{dt} = |q| \left[ \Theta(q) (\gamma S_S + (1-\gamma) S_{IP} -S_N) +S_N -S_T \right] + K_S(S_S - S_T)+ \notag \\ 
     & \qquad \qquad \qquad K_N(S_N - S_T) - (F_T + H A_T)  S_0  \\
    &V_S \frac{dS_S}{dt} =  \gamma |q| \left[\Theta(q)(S_B - S_T) + S_T - S_S \right] + K_{IP}(S_{IP} - S_S) + K_S(S_T - S_S) + \notag \\    
     & \qquad \qquad \qquad  \eta (S_B - S_S) - (F_S + H A_S) S_0 \\
    &V_{IP} \frac{dS_{IP}}{dt} = (1-\gamma) |q| \left[ \Theta(q) (S_B - S_T) + S_T - S_{IP} \right] + K_{IP}(S_S-S_{IP}) - (F_{IP} + H A_{IP}) S_0. 
   %& V_B \frac{dS_B}{dt} = |q| \left[ \Theta(q) (S_N - S_B - \gamma S_S - (1-\gamma) S_{IP}) + \gamma S_S + (1-\gamma)S_{IP} - S_B \right]   + \eta (S_S - S_B).
\end{flalign}
\end{subequations}

Here, $K_S$, $K_N$ and $K_{IP}$ describe the constant coupling by the wind-driven circulation, and $\eta$ is the mixing strength between the Southern Ocean and Bottom waters. $\gamma$ represents the proportion of AMOC return flow via a cold water path, % correct?
where NADW up-wells in the Southern Ocean. We use parameters as in \cite{WOO19}, slightly adjusted to remove a Hopf bifurcation for simplicity. Time is re-scaled by $\tau_Y = 3.15 \times 10^{7}$ to go from seconds to years, and the remaining parameter values are in Table~S1. 
%NOTE we also use H in terms of m3/s!!!
For these parameter values and a range of $H>0$ (increased freshwater input in the North due to glacier melt), the model has two stable fixed points with a vigorous and collapsed  AMOC, respectively. At $H\approx 0.3214$~Sv the vigorous AMOC state collides with the edge state at a saddle-node bifurcation and vanishes (see bifurcation diagrams in Fig.~S12 and S13). To study the changes in fluctuations around the ON fixed point before the bifurcation, we add independent Gaussian white noise to each variable:
\begin{equation}
    dS_X = f_X (S_X,H) dt + \sigma_X dW_X,
    \label{eq:SDE}
\end{equation}
with $X \in \{ N,T,S,IP \}$, $\sigma_X = 10^{-6}$. $f_X$ represents the deterministic model, as given in Eq.~\ref{eq:5box}. $dW_X$ are standard independent Wiener processes. 

\begin{figure}%[floatfix]%!htb
\includegraphics[width=0.99\textwidth]{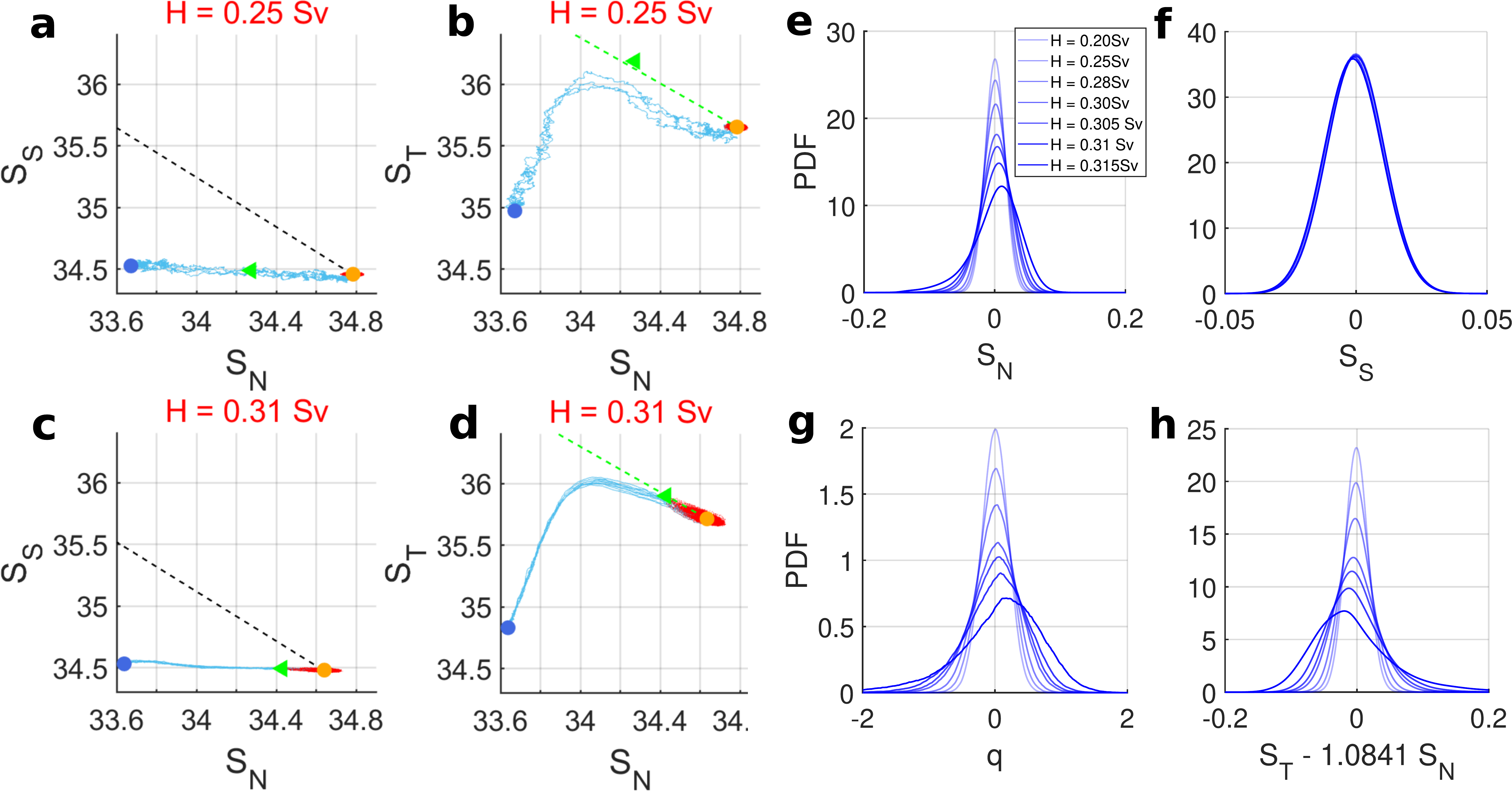}
\caption{\label{fig:5box_results} 
{\bf a-d} Phase space projections of the five-box AMOC model, showing fixed points (colored symbols) and ensembles of noise-driven trajectories (red and blue) for two values of the control parameter $H=0.25$~Sv (top row) and $H=0.31$~Sv (bottom row). 
Shown are projection onto the variables $S_S$ and $S_N$ ({\bf a,c}), as well as $S_T$ and $S_N$ ({\bf b,d}). The values $S_X$ are in PSU (practical salinity units), whereas in the dynamical equations the values are a factor of 1000 smaller, due to units in parts per million. 
Rare trajectories that reached the OFF state in a noise-induced tipping (only in {\bf c,d}, blue trajectories) were cut at the time they last crossed the saddle, and only the part shown in red is kept for statistical analysis. 
Additionally, we present a few transitions at $H=0.25$~Sv simulated with five times higher noise (blue in {\bf a,b}). The black line dashed in line in {\bf a,c} represents a vector aligned with the observable $q \propto (S_N - S_S)$, and the green dashed line in {\bf b,d} is a vector from the ON state to the edge state (calculated at $H=0.305$~Sv). 
{\bf e-h} Distributions of the fluctuations around the ON fixed point in an ensemble of simulations of Eq.'s \ref{eq:5box} and \ref{eq:SDE}, projected onto the state variables $S_N$ ({\bf e}) and $S_S$ ({\bf f}), as well as the observables $q=S_N-S_S$ ({\bf g}) and $S_T- 1.0841\cdot S_N$ ({\bf h}). Shown are distributions for seven values of the control parameter $H$.
}
\end{figure}

For fixed values of the control parameter, starting in the middle of the bistability regime at $H=0.2$ and increasing until $H=0.315$ shortly before the TP, we perform ensembles of simulations with the ON fixed point as initial condition. For $H$ closest to the bifurcation, there are cases of noise-induced tipping to the OFF state during the simulation time of $t=5000$~years. Here we remove any points after which the saddle had been crossed and the system had subsequently tipped. As $H$ reaches values close to the TP, the size of fluctuations in some variables increases, while it stays virtually unchanged in others (Fig.~\ref{fig:stommel_var_increase}b). 
This is because of the directionality in phase space of the emerging fluctuations towards the edge state (compare Fig.~\ref{fig:5box_results}a,b with Fig.~\ref{fig:5box_results}c,d). 
When projected onto the variables $S_S$ and $S_{IP}$, the three fixed points do not differ much (Fig.~\ref{fig:5box_results}a-d). Thus, the fluctuations towards the edge state are approximately perpendicular to the axes of the variables and they do not increase towards the TP (Fig.~\ref{fig:5box_results}f).
%NOTE ! variability in other variables now smaller, and separation of states
% is kind of comparable to variability...!
% REDO! In slow variables SS and SI, the separation of fixed points is very small in absolute terms, and of similar magnitude to the flucutations. 
In contrast, $S_N$ and $S_T$ show a clear increase in variance by a factor of 10 (Fig.~\ref{fig:stommel_var_increase}b, Fig.~\ref{fig:5box_results}e). 
This is inherited by $q$ (Fig.~\ref{fig:5box_results}g) via its dependence on $S_N$, being proportional to $S_N - S_S$ (Eq.~\ref{eq:q_2}), and the fluctuations in $S_S$ being small and unchanged (Fig.~\ref{fig:5box_results}f). Since $q$ does not point directly towards the edge state (see black dashed line in Fig.~\ref{fig:5box_results}a,c) it only partly captures the critical dynamics.
Corresponding figures for the remaining variables are in Fig.~S14. 

An observable with larger CSD is found from a projection onto $S_N$ and $S_T$ (Fig.~\ref{fig:5box_results}b,d), which are the only variables where the three fixed points differ significantly in absolute terms of salinity concentration and compared to the size of the noisy fluctuations in these variables. 
Choosing $H=0.305$ close to the TP, the observable obtained by projecting onto the vector pointing from ON to the edge state (-$1.0841\cdot S_N$, $S_T$)$^\top$ (green dashed line in Fig.~\ref{fig:5box_results}b,d) shows an increase in variance by a factor of almost 20 (Fig.~\ref{fig:5box_results}h and Fig.~\ref{fig:stommel_var_increase}b). 
%NOTE it does not matter at all what H we use to estimate the edge vector. 
In terms of this observable, the edge state has a value more extreme than the OFF state, which may be part of the reason of the large increase in variability.  
In particular, using this observable as metric, we may consider the difference of its values in the attractors as well as the size of the fluctuations as two natural scales. 
Then, initially (far from the bifurcation) the edge state is very far away from the ON state, whereas it necessarily needs to get close to the ON state as the bifurcation is approached. Thus, in the direction in phase space implied by this observable we can expect a large change in the underlying quasipotential. This in turn leads to a large change of the noise-driven variability. 
In contrast, observables with values that do not differ much between the edge state and the base attractor - even when still far from the bifurcation - are not expected to be associated with directions in phase space that experience much change. 

Using the same two-dimensional projection, the direction of the edge state furthermore motivates the very similar, but physically more meaningful observable $S_T - S_N$, i.e., the salinity gradient from the tropical to the northern Atlantic, which has larger CSD compared to the AMOC strength $q$. % give numbers for this observable too.
%need to explain more?

%%% check this with the edge vector. It could be for other variables too, but there
%%% it is marginal because the variable does not differ much for ON and OFF to begin with.
%%% (when compared to the scale of the fluctuations)

%%% NOTE there could be other reasons for why the increase in variance is highest for
%%% this particular projection.
%%% 1. the variables are the ``fastest''. 
%%% 2. signal-to-noise ratio (related)
%%% 3. the individual variables have largest CSD. 

\subsection{AMOC collapse in global ocean model with stochastic forcing}
\label{sec:veros}

\subsubsection{Stability landscape and edge state of the deterministic system}
\label{sec:stability_landscape}

\begin{figure}%[floatfix]%!htb
\includegraphics[width=0.65\textwidth]{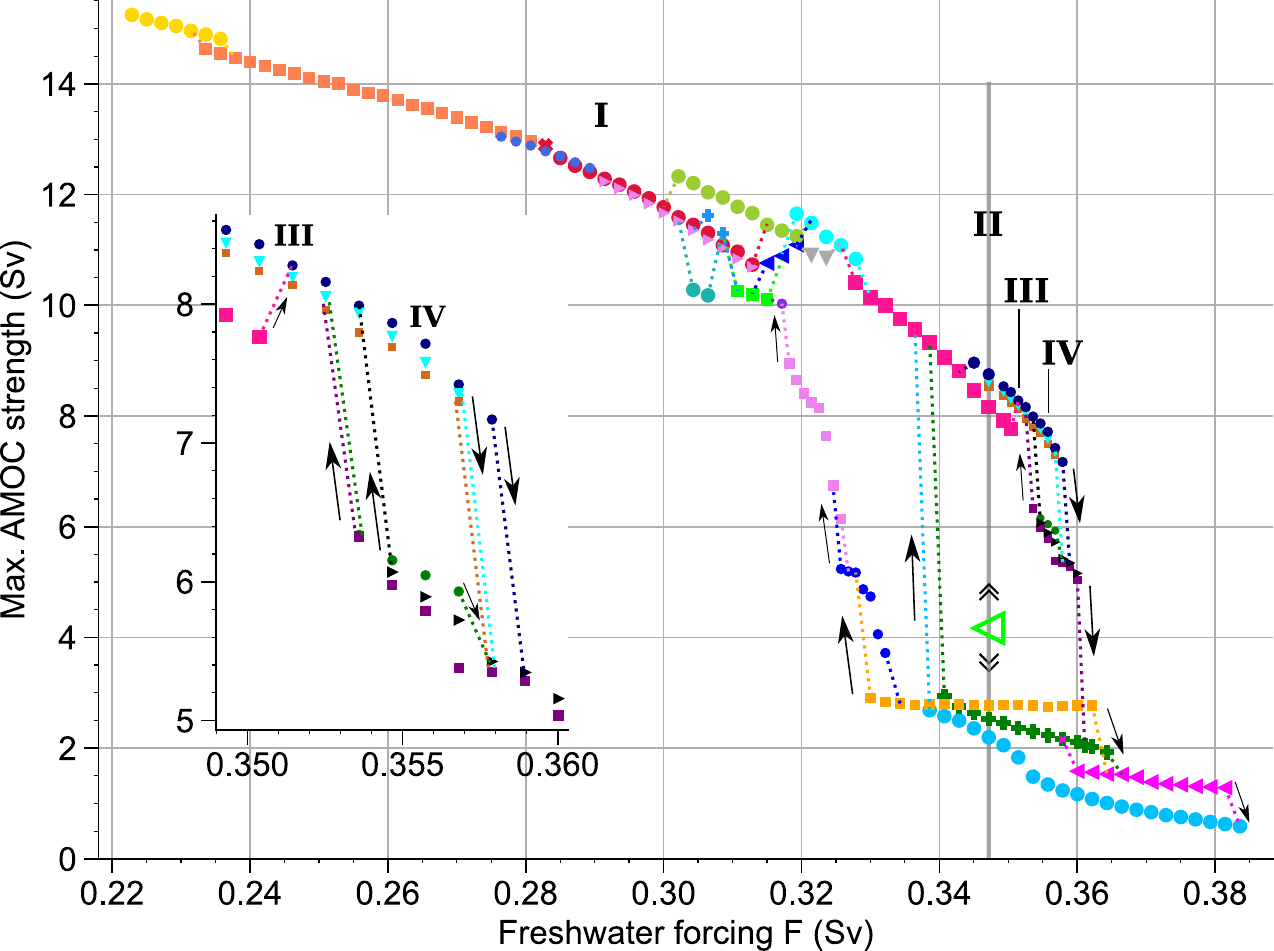}
\caption{\label{fig:veros_bif_diagram} 
Bifurcation diagram of the Veros global ocean model under change in NA freshwater forcing $F$ as obtained in \cite{LOH24}. The symbols of different color and shape correspond to different branches of attractors, which have been distinguished by the mean state and spatio-temporal variability of the AMOC and other relevant ocean properties. The colored dotted lines as well as the arrows indicate which branch of attractors collapses onto which other branch, as the stability of the former is lost. Marked with the open green triangle is the position of the edge state at the parameter value $F=0.3472$~Sv, which was computed in \cite{LOH24b}. This edge state lies on the basin boundary between two attractors denoted ON and OFF in the main text, which are the branches marked with dark blue dots and green crosses, respectively. The parameter values labeled I to IV are those used in Sec.~\ref{sec:veros_ews}. 
}
\end{figure}

In order to test the relevance of our approach to more complex dynamical environments, we consider the TP of an AMOC collapse in the global ocean model {\it Veros} \cite{HAE18,HAE21} (see Appendix~\ref{AppModel} and \cite{LOH21,LOH22,LOH24} for details on the model). 
The model in its usual form is deterministic, and its stability landscape subject to North Atlantic (NA) freshwater forcing has been shown previously \cite{LOH24} to feature a large number of co-existing stable states as the TP is approached (see bifurcation diagram in Fig.~\ref{fig:veros_bif_diagram}). These stable states are not fixed points, but chaotic attractors that display various forms of spatio-temporal variability \cite{LOH24}. 
The co-existing attractors are born and lose stability in rapid succession before the eventual AMOC collapse, leading to ``intermediate'' TPs at various critical values of the forcing, where the system undergoes transitions between attractors with different AMOC mean and variability. As a result, the overall AMOC collapse in the deterministic system comprises a series of step-wise changes in the spatio-temporal variability. 
This makes it hard for EWS based on CSD - which indicate loss of local stability of a given stable state at a single bifurcation - to show a reliable trend towards the AMOC collapse \cite{LOH24}. 
As discussed further below, depending on its strength, the addition of noise to emulate atmospheric fluctuations can to some degree alleviate this issue by an effective merging of co-existing attractors and by dominating the changes in deterministic variability.

\begin{figure}%[floatfix]%!htb
\includegraphics[width=0.99\textwidth]{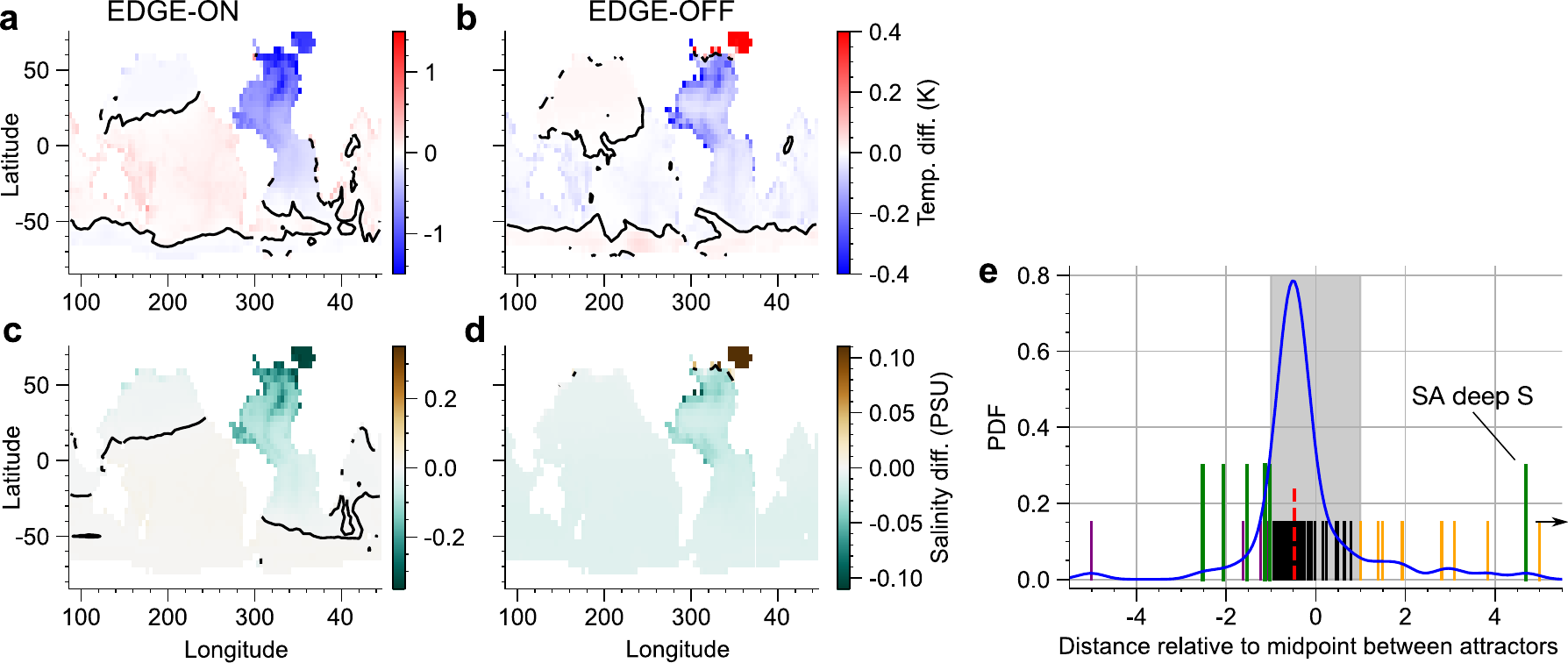}
\caption{\label{fig:global_deep_ocean} 
Deep ocean properties of the edge state at $F=0.3472$, marked ``II'' in Fig.~\ref{fig:veros_bif_diagram}, as computed in \cite{LOH24b}. Shown is the temperature ({\bf a,b}) and salinity ({\bf c,d}) anomaly of the edge state compared to long-term averages on the ON and OFF attractors, averaged over the ocean below 1000 meter depth. The contours at 0 are shown with black lines.
{\bf e} Distribution over an ensemble of rescaled observables (see main text) of the relative position of the edge state compared to the ON and OFF attractors. The values of the individual observables are given by the vertical lines. For observables with positive (negative) values, the edge state is closer to the ON (OFF) state. When the absolute value of the distance is larger than 1, the edge state lies outside of the range spanned by the ON and OFF states in the given observable. The red dashed line corresponds to the maximum AMOC strength as observable. 
}
\end{figure}

Recently, an unstable edge state that separates the regimes of a vigorous and collapsed AMOC was computed for the model using an edge tracking algorithm \cite{LOH24b} (green triangle in Fig.~\ref{fig:veros_bif_diagram}). It was found that, while the climatological properties of the edge state lie between the ON and OFF regimes in terms of the upper ocean and the Atlantic circulation, the deep Atlantic shows a characteristic anomaly with cooler and fresher (less ``spicy'') water than in either ON or OFF (Fig.~\ref{fig:global_deep_ocean}a-d). 
This is an interesting parallel to the much simpler Stommel model, where the edge state was also oriented in the direction of spiciness.

% Definition of suite of observables.
To illustrate in terms of which properties the state of the ocean in the edge state differs strongly from the average state in the attractors, we define a suite of observables, consisting of temperature, salinity and density in the different ocean basins (north, tropical and southern Atlantic/Pacific, the Indian ocean, and the Southern ocean) averaged over the surface layer, the sub-surface (until 1000 meter depth) and the deep ocean (below 1000 meter). Additionally, from the barotropic streamfunction we derive the strengths of the Antarctic circumpolar current and the ocean gyres (subtropical in north and south Atlantic, Pacific and Indian ocean, as well as subpolar in north Atlantic/Pacific). Finally, the strength of the AMOC (measured both by the spatial maximum of the meridional streamfunction, and by its maximum at the equator below 500 meter) and Antarctic bottom water (AABW, minimum of streamfunction at equator below 2000 meter depth) are considered. This yields a total of 83 observables. 
%NOTE: define AMOC strength even more clearly!

For each of these observables we quantify the location of the edge state in relation to the two attractors. To this end, each observable is normalized so that the difference of the values in the two attractors is equal to 2, and then shifted such that 1 (-1) corresponds to the location of the ON (OFF) attractor. Then we consider the value of this scaled and shifted observable in the edge state. By definition, 0 corresponds to a situation where the edge state is exactly halfway between ON and OFF, and a value >1 (< -1) is attained when the edge state is outside the range of attractors and beyond the ON (OFF) state, as measured with this observable. 
In Fig.~\ref{fig:global_deep_ocean}e we show a distribution over the suite of rescaled observables evaluated at the edge state. Note that the values do not tell whether the ocean is, e.g., warmer/saltier/denser in ON or in OFF. It only illustrates the portion of observables where the edge state is more extreme than ON and OFF. 
%define ``more extreme''? more extreme than ON: Edge > ON > OFF or Edge < ON < OFF; more extreme than OFF: Edge > OFF > ON or Edge < OFF < ON.
In most observables the edge state lies between ON and OFF (gray shaded area). The temperature and salinity observables in the deep Atlantic ocean (green vertical lines) all show cooler and fresher conditions in the edge state compared to both ON and OFF. With the exception of the SA deep salinity, where the ON state is fresher than OFF, these observables are such that the OFF state is cooler and fresher than ON, and the edge state is cooler and fresher still (thus yielding values < -1). 
%Even more explicitly? I.e., OFF is a ``barrier'' between ON and EDGE. 
There are three other observables with this property (salinity in deep North Pacific, density in sub-surface South Pacific, and AABW), but in terms of these observables the distance of the ON and OFF attractors is only very small and thus we don't expect to see much change as the TP is approached. %%% IN ABSOLUTE / PHYSICAL TERMS.
%%% compared to the Atlantic variables. 
%%% MAKE SUPP FIGURE TO FULLY SUPPORT THIS. 
This suggests the deep Atlantic temperature and salinity as observables that would likely capture an approach to the TP via increased fluctuations, which is tested in the following sections by the addition of noise. 
%%% Comment on the observables with large positive values?

\subsubsection{AMOC dynamics with noise}
\label{sec:veros_noise}

The Veros model is usually forced by fixed climatological fields of temperature, salinity and wind stress derived from the ERA-40 re-analysis \cite{UPP05}. To excite fluctuations  away from the attractors, noise fields are added in surface temperature and salinity forcing with zero mean and negligible seasonality at every grid point. 
These are computed every month from an autoregressive model on the coefficients of empirical orthogonal functions (EOFs) derived from re-analysis data of sea surface temperature and salinity (see Appendix~B). 
A variety of simulations have been performed to probe the model's response to noise forcing, using the two components individually or jointly, and varying the noise strength by adjusting the temperature and salinity fields with scalar factors $\sigma_T$ and $\sigma_S$. Detailed results are given in the supplemental Sec.~S1, and here we only summarize the broad picture that is relevant for the study of EWS. 

\begin{figure}%[floatfix]%!htb
\includegraphics[width=0.99\textwidth]{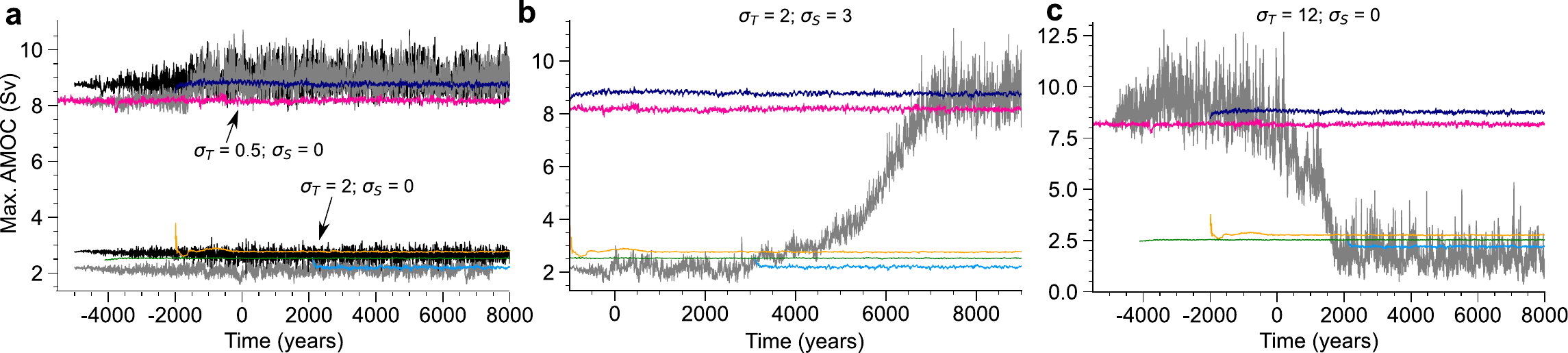}
\caption{\label{fig:veros_transitions} 
Time series of the maximum AMOC strength (5-year averages) for simulations with added noise at $F=0.3472$~Sv (gray and black trajectories) compared to steady-state simulations on the co-existing attractors of the deterministic system (colored trajectories). 
The colors of the deterministic time series correspond to the ones of the attractor branches given in Fig.~\ref{fig:veros_bif_diagram}.
All time series have been shifted to end at the same time, and some of the deterministic trajectories have not been simulated for the maximum duration, thus starting later and including short transients. The noise strength is ramped up linearly for 5000 years ({\bf a,c}) and 1000 years ({\bf b}) to allow for a slow and controlled shift of the system from an initial state on an attractor of the deterministic system to its new statistical steady state under noise forcing. The maximal noise strength is reached at time 0 and held constant thereafter. {\bf a} Four simulations with low noise (here temperature-only noise), initialized on two coexisting attractors each for the ON and OFF regimes. {\bf b} Noise-induced resurgences of the AMOC for medium noise level. 
{\bf c} At strong noise (here shown for temperature-only noise), a simulation initialized in an AMOC ON attractor undergoes a transition to a collapsed AMOC. 
}
\end{figure}

As we gradually increase the noise strength from zero to a finite level, transitions appear from certain attractors to others (Fig.~\ref{fig:veros_transitions}a). Often two close-by attractors appear to effectively merge. As a result, the phenomenology of intermediate TPs due to the close-by coexisting attractors gets blurred at a certain noise level, also because the noise forcing dominates specific spectral features (such as quasi-periodic decadal-scale AMOC variability) that distinguish the different attractor branches \cite{LOH24}. Thus, discontinuous variability changes under increasing freshwater forcing are less pronounced, which is good for the prospects of early-warning. 
There is also a non-trivial influence of the noise on the mean state of the system, such as the mean AMOC strength. This is due to a non-linear response of the system \cite{LUC12,ABR17} to the short-term buoyancy forcing, for instance by the model's convection scheme.

At the chosen parameter value $F=0.3472$~Sv, when further increasing the noise strength we first see transitions from a collapsed to a vigorous AMOC state (Fig.~\ref{fig:veros_transitions}b). The attained AMOC ON state appears stable on a time scale of at least 10,000 years. Interestingly, at much larger noise the reverse seems to be the case: The AMOC ON regime consistently disappears, and a state with collapsed AMOC seems essentially stable (Fig.~\ref{fig:veros_transitions}c). 
The fact that the AMOC seems to be harder to collapse may be due to the physics of the edge state, which requires a large change in the deep ocean that may be easier to accomplish with a short-term onset of convection (AMOC resurgence) rather than a shutdown. 

Due to the nature of our model we cannot say what noise level should be chosen such as to emulate a realistic forcing by the chaotic atmosphere. Thus, for our conceptual study we focus first noise as low as possible as to generate a transition between attractors. This is done to establish whether the instanton indeed points towards the edge state in the relevant observables, and whether a transition at non-infinitesimal noise strength indeed passes by the edge state. Then, in order to study EWS, in Sec.~\ref{sec:veros_ews} we employ intermediate noise where no transitions are observed. 

\subsubsection{Noise-induced transitions and passage of the edge state}
\label{sec:veros_transitions}

For a good approximation of the true instanton, we wish to obtain low-noise transitions that may be very rare, and for limited computing resources, only accessible by rare event algorithms (REA). Here we implement the Giardina-Kurchan-Lecomte-Tailleur (GKLT) algorithm \cite{GIA11}, where an ensemble of $N=50$ trajectories is propagated for a given time horizon, after which the most promising members (as measured by a score function) are cloned, while the least promising ones are killed (see Sec.~S2 for details). We choose $F=0.3515$ (marked III in Fig.~\ref{fig:veros_bif_diagram}), since here there are no competing vigorous or partially-collapsed states that would complicate the path to an AMOC collapse. Note, however, that there are still (at least) three very close-by attractors that differ by about 0.1~Sv in AMOC strength. 
Not knowing the instanton, and thus an optimal score function (committor function), we use the most parsimonious approach with the AMOC strength as the score function, killing trajectories that show the least AMOC decline. 

\begin{figure}%[floatfix]%!htb
\includegraphics[width=0.75\textwidth]{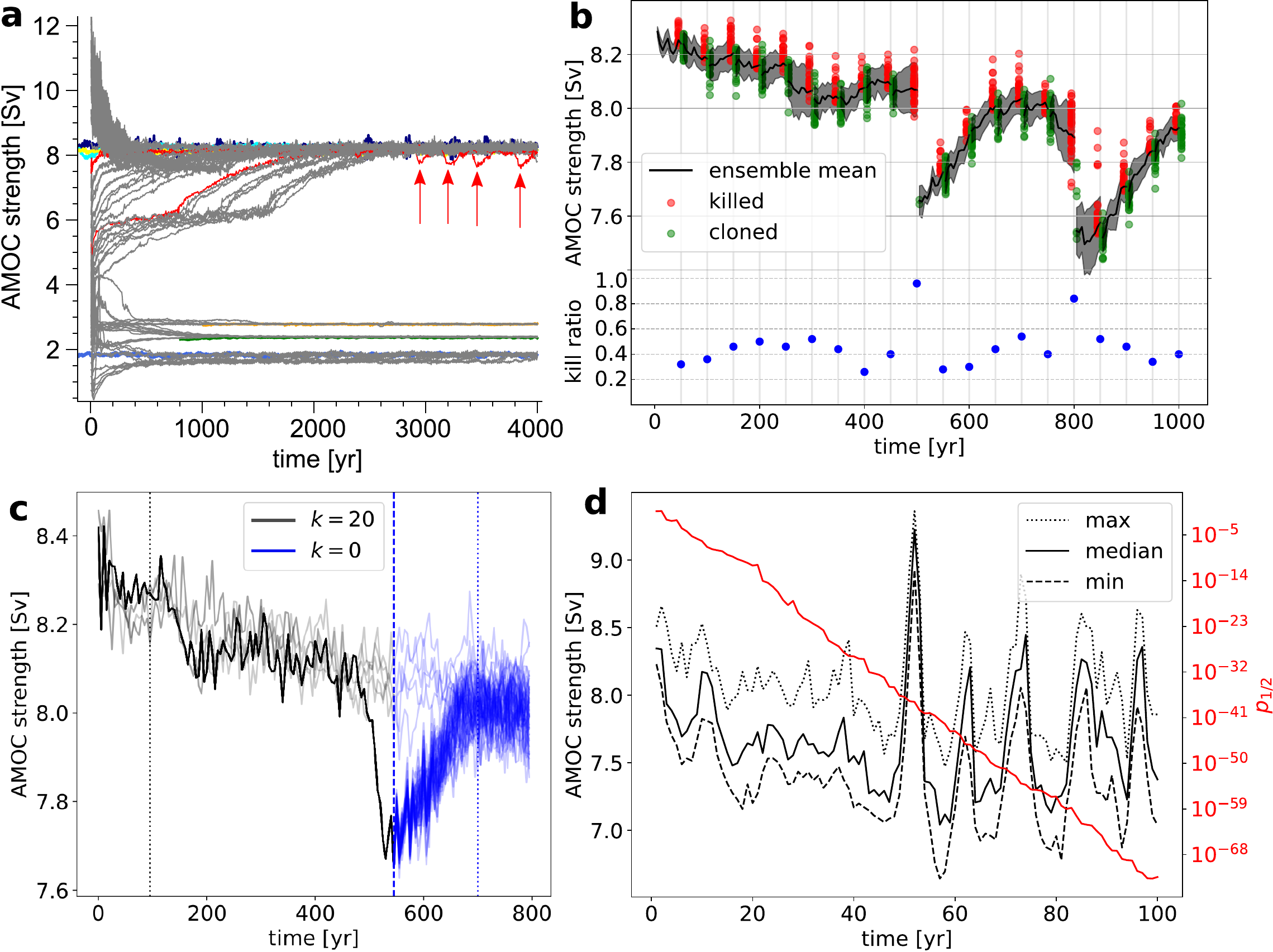}
\caption{\label{fig:rare_event} 
{\bf a} $N=64$ unforced simulations at $F=0.3515$~Sv with a wide range of initial conditions (gray) converging to different attractors (colored timeseries). Four realizations (red) that converge to the AMOC ON regime show abrupt, temporary AMOC weakenings (red arrows).
{\bf b} Evolution of REA ensemble with selection strength $k=20$~Sv$^{-1}$ and resampling time $\tau=50$~years (see Sec.~S2). The shaded area corresponds to one standard deviation. In red are the ensemble members killed and in green the ones that survived to the next iteration. On the bottom is the fraction of members killed at each resampling step. At $t=500$~years ($t=800$~years) there are abrupt AMOC weakenings, and only 2 (8) out of the 50 members survive, resulting in a ``near-collapse'' of the ensemble.
{\bf c} Transition of another ensemble to a different stable attractor, using same parameters as in {\bf b}. In black are backward reconstructed trajectories of the 5-year average AMOC strength after 10 REA iterations. The run terminates with an ensemble collapse similar to the one in panel {\bf b}, with the darkest trajectory consisting of 45 identical clones. In blue the 50 ensemble members are propagated forward for 250 years, with no REA sampling. All ensemble members recover from the temporary fluctuation in 150 years, and then stabilize around a different value of the AMOC strength. 
{\bf d} REA run with added atmospheric noise, $\tau=1$~year and $k=20$~Sv$^{-1}$. In black (left y-axis) is the minimum, median and maximum values of the 1-year averaged AMOC strength before each resampling step. In red (right y-axis) is the unbiased probability $p_{1/2}$ (Eq.~4 in Sec.~S2) of a natural fluctuation of the system being below the ensemble median. 
}
\end{figure}

% leave out ...? or perhaps just summarize and then Supp figure?
% perhaps all could be summarized very briefly: there were several obstacles, including 
% centennial-scale AMOC reductions, that do not lead to collapse, as well as 
% decadal-scale oscillations. 
% as a result there were no transitions generated using the naive score function 
We first drive the REA without additional noise, but only with an infinitesimal perturbation to the state at the cloning step. 
This is done because while very long control simulations do not show a spontaneous AMOC collapse \cite{LOH24}, the system dynamics on the vigorous attractors in the multistable regime feature rare AMOC weakenings of up to 1~Sv that occur abruptly and subsequently decay in 100-200 years (red arrows in Fig.~\ref{fig:rare_event}a). 
The weakenings occur roughly once per 20,000 years (note they occur only in four out of 64 realizations in Fig.~\ref{fig:rare_event}a), and it is interesting to test whether these could, in even rarer cases, lead to an actual collapse. 
%%% results for 50 year resampling time: 
Indeed, the REA quickly explores these abrupt weakenings, but the ensemble always recovers to stronger AMOC states (Fig.~\ref{fig:rare_event}b). The weakenings are still rare given the ensemble size of $N=50$, and thus their occurrence leads to a ``collapse'' of the ensemble (Sec.~S2), which drastically reduces the diversity of the ensemble and thus limits our ability to explore trajectories leading to a full AMOC collapse. Nevertheless, less catastrophic, yet technically critical, noise-induced transitions are sampled, i.e., transitions between the three very close-by vigorous AMOC attractors (see inset Fig.~\ref{fig:veros_bif_diagram}). After recovering from a temporary weakening, the ensemble can be found in a basin of attraction that does not belong to the attractor the ensemble was initialized in. This can be seen by stopping the REA sampling, and observing that the ensemble relaxes to an attractor with weaker AMOC (Fig.~\ref{fig:rare_event}c).

We attempt to mitigate the effect of temporary AMOC weakenings by providing alternative pathways in phase space with the addition of temperature noise, and a lower resampling time in order to avoid a relaxation back from the weakening between consecutive sampling steps. %much shorter resampling time, also because now we are not constrained by lyapunov time as much. 
A simplified noise model is used that only consists of a temperature anomaly field in the Atlantic, again derived from EOFs of reanalysis sea surface temperature data. This is to isolate perturbations that would plausibly lead to more frequent anomalies driving the system towards an AMOC collapse (in the Veros model temperature and salinity anomalies act in similar ways on the surface buoyancy).

Here, another issue arises that prevents sampling an AMOC collapse. For this noise implementation and value of $F$, a decadal-scale oscillation in the AMOC strength is excited (Fig.~S15). The REA repeatedly samples this oscillation, where during the downward slope it acts to amplify the oscillation, selecting members for which the AMOC strength decreases faster, while on the upward slope it has a dampening effect but cannot escape the oscillation altogether (Fig.~\ref{fig:rare_event}d). Since the algorithm keeps selecting rare trajectories with relatively weak AMOC, the sampled probabilities keep decreasing drastically (red line in Fig.~\ref{fig:rare_event}d). But by the end of a decadal oscillation the system loses memory and thus no qualitative long term progress towards lower AMOC strength is observed. %is this a correct interpretation?
%%% larger resampling times were tried, but did not yield transitions. 

The failure to observe an AMOC collapse could indicate that the instanton is not directed towards lower AMOC strength and thus evades our score function, or at least that it is directed along a much more specific path, whereas typical (more frequent) fluctuations towards lower AMOC are not along this path. 
Indeed, at least for the conceptual 5-box AMOC model discussed in Sec.~\ref{sec:five_box}, it turns out that the instanton from the ON to the OFF state begins with a phase of increasing AMOC strength \cite{SOO24}. %Physically this could mean a more complicated path to AMOC collapse than often assumed, e.g., not only a surface buoyancy anomaly that is amplified by the salt-advection feedback. 
Thus, we focus on brute force simulations at the lowest noise level that resulted in a transition between the AMOC ON and OFF regime, which in our case were AMOC resurgences (Fig.~\ref{fig:veros_transitions}b). 

\begin{figure}%[floatfix]%!htb
\includegraphics[width=0.99\textwidth]{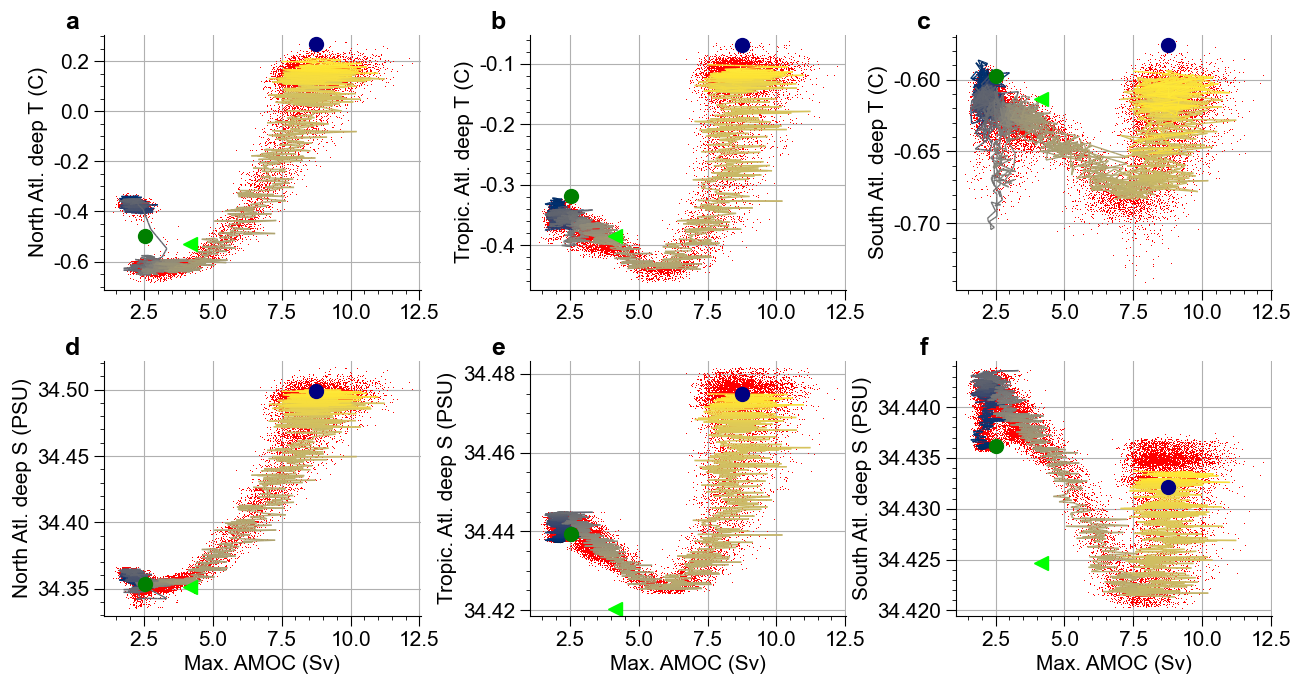}
\caption{\label{fig:edge_passage} 
Trajectory of a noise-induced resurgence of the AMOC in six projections onto different observables. The color code from blue to yellow indicates time of a 10,000 year simulation, where the corresponding time series of the AMOC strength is in Fig.~\ref{fig:veros_transitions}b. The symbols indicate the locations of the deterministic ON and OFF attractors, as well as the edge state (green triangle).
The clouds of red points in the background correspond to the trajectories of a total of 10 realizations of a noise-induced transition using the same parameter values and initial conditions. 
}
\end{figure}

Figure~\ref{fig:edge_passage} shows trajectories of AMOC resurgences with $\sigma_T = 2$ and $\sigma_S = 3$ projected onto the AMOC strength and six observables as key signatures of the edge state, namely the average temperatures and salinities in three sectors of the Atlantic below 1000 meter depth. Indeed, starting at OFF, there is a decrease in temperature and salinity in all sectors of the deep Atlantic, before these increase again as the system relaxes to the ON state. 
Thus, the initial segments of the instanton point at least qualitatively to the edge state in the relevant observables. 
%NOTE comment here on multiple realizations. and the fact that for larger noise
% there is expected quite a bit of spread, and indivdual realizations never exactly
% pass edge state. 
While the values of the edge state are passed neither precisely nor synchronously in all observables, this indicates that in order to achieve a transition the system needs to attain a state similar to the edge state along its trajectory and overcome the according potential barrier \cite{LOH24b}. 

A non-exact passage of the edge state could be due to the finite noise amplitude. Here, the instanton may not be tracked precisely and the basin boundary may be crossed elsewhere first after which an attraction by the stable manifold of the edge state precedes the ejection towards the ON state. We find that in all cases the transition is initiated by a rapid onset of NA convection (Fig.~S16) lasting 10-20 years. 
%due to a buoyancy anomaly as a result of a surface noise anomaly + some preconditioning
%%% perhaps: supp figure with some noise field. but hard to show, since hard to 
%%% align the realizations. 
This leads to a sharp transition in the NA deep ocean, which cools by 0.3~K within 25 years and also shows a smaller decrease in salinity (Fig.~S17). 
%NOTE: this may constitute the basin boundary crossing. 
%therefore overall increase in deep density.
Thereafter, the deep tropical and southern Atlantic slowly become colder and fresher, until the trends reverse and the system evolves towards ON. Transitions with temperature- or salinity-only noise evolve similarly (Fig.~S18 and S19). 
%NOTE this is an important point, since noise structure of course matters in theory!
% emphasize this? -> Some slight evidence for robustness of edge state and instanton. 
%It could also be that the edge state of the deterministic system is not crossed exactly because of a general shift in the steady states as noise is added. 
%Indeed, also the mean values of noisy trajectories before and after the transitions are slightly shifted with respect to the mean values of the deterministic states.
The reverse transitions from the ON state to an OFF state for very large noise levels also show some signatures of cooling and freshening of the deep Atlantic, but in this case the mean ocean state is shifted so much from the deterministic case that it is not meaningful to directly compare it to the edge state.

The chosen noise structure can also affect the dynamics and the path/likelihood of a transition. On a fundamental level, the noise is degenerate, as it acts only on a small subset of the degrees of freedom (namely, the surface temperature and salinity fields). Hence, there is no direct stochastic forcing acting on the velocity field at all levels nor on the subsurface temperature and salinity fields. Still, the impact of stochastic forcing propagates to all degrees of freedom by advection, diffusion, and the coupling between buoyancy and velocity fields. This allows us to circumvent the mathematical difficulty of considering degenerate noise, because we can assume that we deal with a so-called hypoelliptic diffusion (in stochastic, not physical terms) process \cite{BEL04}. As a result (see discussion in \cite{LUC20,MAR21}) it is reasonable to assume that the notion of quasi-potential is still useful to describe the noise-induced transitions of the system between competing states. 
On a physical level, in our mixed noise model the temperature and salinity fields are not correlated, as would be the case in reality. Thus, for instance, a strong warm and fresh anomaly that could lead to an AMOC collapse has to occur purely by chance in our model, whereas in reality it may occur more likely for physical reasons. For example ENSO leads to correlated temperature and precipitation anomalies. Furthermore, fluctuations in wind stress are missing, which could contribute to the dynamics that promote a noise-induced transition, as has been observed for spontaneous AMOC collapses in a coupled model using a rare-event algorithm \cite{CIN24}.

\subsubsection{Early-warning signals}
% Fourth: check for increases in fluctations of observables, and compare with 
% characteristic observables for edge state. 
\label{sec:veros_ews}

While the true low-noise instanton remains unknown, the previous section gave some evidence for an alignment of the instanton and the direction of the edge state in terms of the relevant observables, and for a passage via the edge state even in the large noise regime and for different structures of the noise model. Reverting to lower noise levels, where no transitions from the ON state are observed, we test which observables show changes in variability around the ON attractor during steady-state simulations for different fixed values of $F$.

\begin{figure}%[floatfix]%!htb
\includegraphics[width=0.85\textwidth]{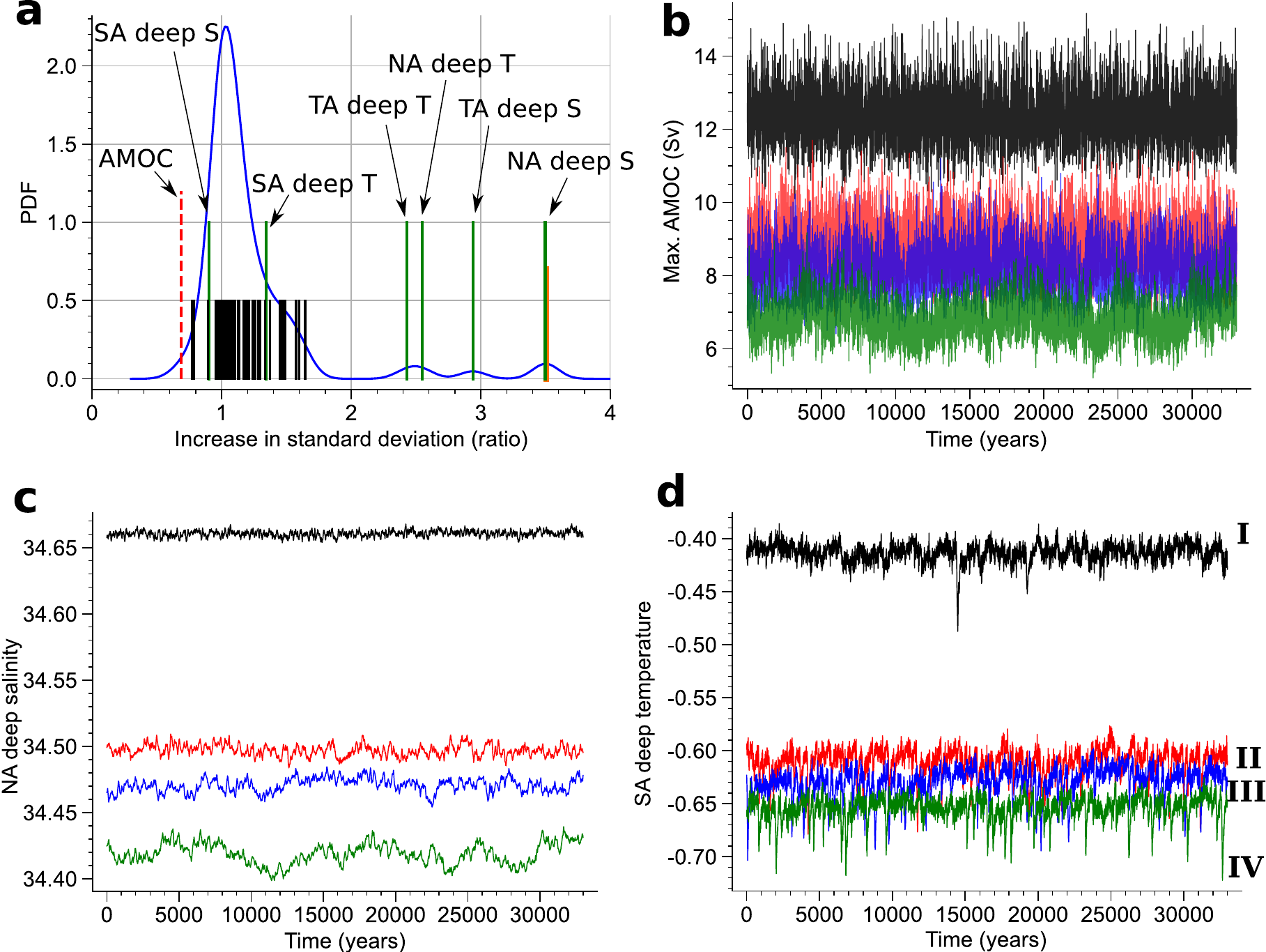}
\caption{\label{fig:veros_ews} 
{\bf a} Distribution across the suite of observables of the increases in standard deviation (SD) towards the AMOC collapse tipping point. The increase in SD is defined by the ratio of the SD of an observable for a long simulation with $\sigma_T = \sigma_S = 2$ close to the tipping points at $F=0.2957$~Sv divided by the SD of a simulation far from the tipping point at $F=0.3557$~Sv. In blue is a Gaussian Kernel estimate of the distribution, and in black are the values for the individual observables. The values for the northern, tropical, and southern Atlantic deep ocean salinity and temperature are shown in green, and the value for the AMOC strength is given by the red dashed line. 
{\bf b-d} Corresponding time series of the simulations at the parameter values I, II, III and IV (see {\bf d} for color coding) for three observables.
}
\end{figure}

To this end, we compare the amplitude of fluctuations in the collection of observables defined in Sec.~\ref{sec:stability_landscape} for four values of $F$ (marked I to IV in Fig.~\ref{fig:veros_bif_diagram}). The ratio of the standard deviation (SD) of an observable for the largest versus the smallest $F$ (IV versus I) is shown in Fig.~\ref{fig:veros_ews}a. The majority of observables do see an increase in SD, likely due to being influenced by critical slowing down. 
There are some observables with decreasing variability, and, curiously, the AMOC maximum is the observable with the largest (albeit still mild) decrease (see time series in Fig.~\ref{fig:veros_ews}b). 
Importantly, from a statistical perspective, in most observables the change in variability is small and thus unlikely to be detected as significant in a practical setting. 

The only observables in the far tail of the distribution (blue line Fig.~\ref{fig:veros_ews}a), which show a large variability increase with at least a doubling of the standard deviation are the deep ocean salinities and temperatures in the tropical and northern Atlantic, which we previously linked to the edge state (see Fig.~\ref{fig:global_deep_ocean}e). For all of these, the edge state lies outside the range of ON and OFF and is more extreme than OFF (as opposed to, for instance, the SA deep ocean salinity). 
Figure~\ref{fig:veros_ews}c shows time series for the deep northern Atlantic salinity, 
which features the strongest increase in variance and the emergence of low-frequency variability. 
Note that also the observable capturing NA subsurface salinity shows an equivalent increase in variability (orange vertical line in Fig.~\ref{fig:veros_ews}a), which is related to the same phenomenon, and our division of deep and sub-surface ocean is somewhat arbitrary. 

\begin{figure}%[floatfix]%!htb
\includegraphics[width=0.99\textwidth]{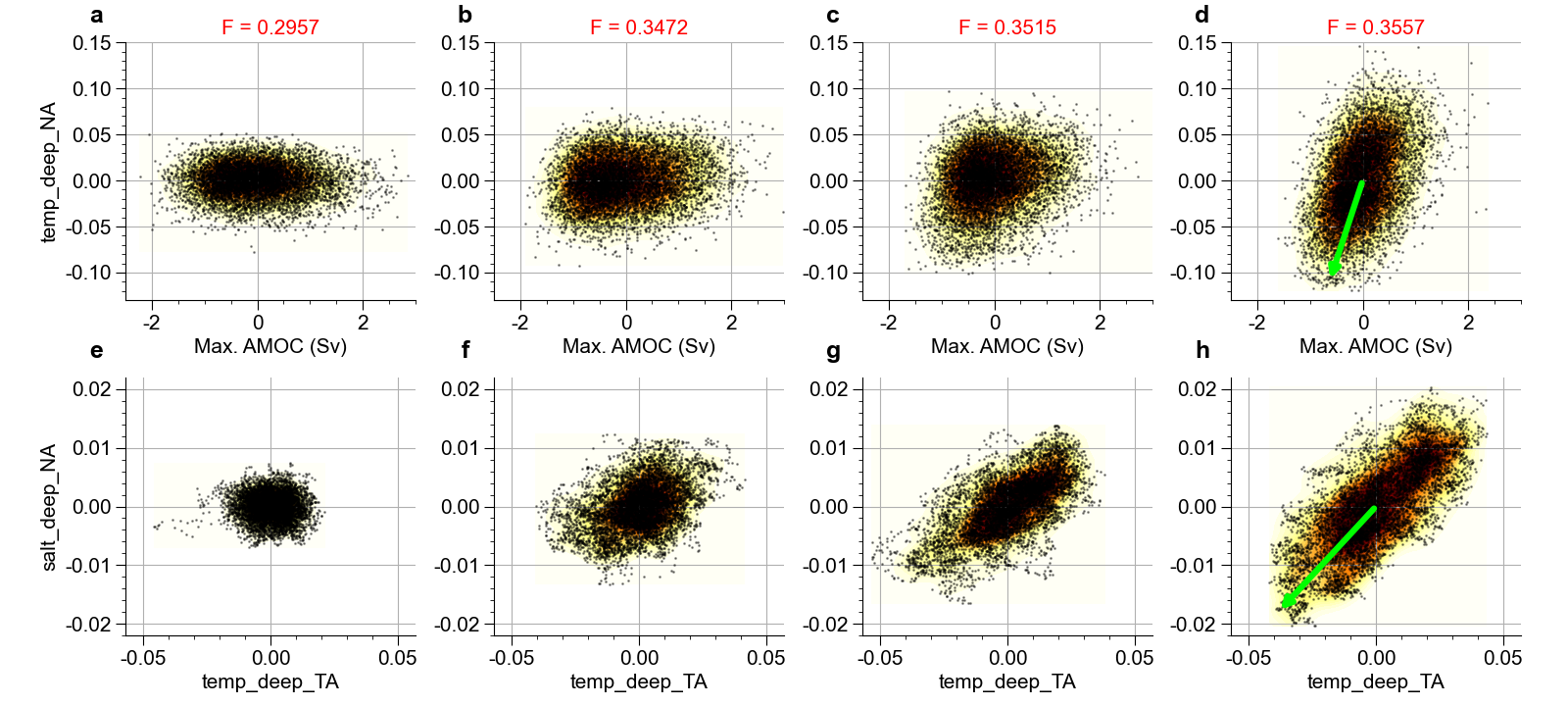}
\caption{\label{fig:veros_fluctuations} 
Residuals of the same simulations as in Fig.~\ref{fig:veros_ews} projected onto the NA deep ocean temperature versus the maximum AMOC strength ({\bf a-d}), as well as the NA deep ocean salinity versus the tropical Atlantic deep ocean temperature ({\bf e-h}), for the four values I, II, III and IV of the control parameter $F$. For the parameter value closest to the tipping point ({\bf d,h}) the green arrow indicates the direction of the edge state, estimated from the deterministic system at $F=0.3472$~Sv.
}
\end{figure}

The deep southern Atlantic is less affected by the increasing fluctuations towards the edge state, which mirrors our observation that noise-induced transitions are initiated  in the north. It is, however, affected by another mode of variability that gets increasingly excited as the TP is approached. These are rapid excursions towards colder conditions that originate in the Southern Ocean and spread to the deep southern Atlantic (Fig.~\ref{fig:veros_ews}d). 
The anomaly does not propagate to the tropical or north Atlantic, and in fact tends to occur when salinities and temperatures in the deep Atlantic are higher than usual. Thus, these fluctuations seem to evolve in a direction opposite to the edge state. 
The deep ocean temperature, salinity and density anomalies during these noise-induced fluctuations are very similar to the rare, temporary AMOC weakenings encountered in the deterministic system and during the REA simulations (Fig.~\ref{fig:rare_event}a). The noise-induced excursions also show a slight AMOC weakening, which may be caused by a disturbance of the Southern Ocean upwelling due to the cold anomaly, but it is masked by other noise-driven variability. If these are indeed excursions away from the edge state, 
it would explain why the rare-event algorithm could not find noise-induced AMOC collapses, and it would motivate the need of a more specific score function. 

%edge_vector = {'amoc_max': -4.590344087, 'temp_deep_NA': -0.7991384496141118, 'temp_deep_TA': -0.315678035936083, 'temp_deep_SA': -0.03848311601438503, 'salt_deep_NA': -0.14826214647282399, 'salt_deep_TA': -0.05477043103248036, 'salt_deep_SA': -0.007534909522306066}

As the AMOC collapse is approached the fluctuations in the deep ocean become increasingly aligned along the ``spiciness'' direction (cold and fresh or warm and saline anomalies) that is characteristic of the edge state, as can be quantified by a vector pointing from the mean state on the ON attractor to the edge state (Fig.~\ref{fig:veros_fluctuations}). We define a projection of this ``edge'' vector ${\bf v_e}$ onto a lower-dimensional observable sub-space by taking the difference between the mean values at the edge and the ON state of several key observables (see Fig.~S20). This yields
${\bf v_e} = (-4.590, -0.799, -0.316, -0.038, -0.148, -0.055, -0.008)^\top$, where the first component is the AMOC strength anomaly (in Sv), and the remaining components are the temperature anomalies in north, tropical, and south Atlantic (in K), followed by the corresponding salinity anomalies (in PSU). In Fig.~\ref{fig:veros_fluctuations}d,h we plot normalized vectors projected further onto two observables. As the parameter value is increased towards the TP, an almost perfect alignment of the fluctuations and the edge vector emerges. This is despite the edge state having been estimated at a lower $F$, indicating that the direction of the edge state does not change much towards the TP, as was also the case for the conceptual models studied in Sec.~\ref{sec:conceptual}. 

%%% NOTE: another way to reason about the alignment of the fluctuations is that, for a given observable, the size of the fluctuations (as response to noise) is 
%%% proportional to the (initial) distance to the edge state.
%%% this is not really a statement about the amount of INCREASE in variance though.

Hence, it is plausible that the increases of variability in the identified observables is due to the approach of the edge state to the ON regime as the TP is reached, and a corresponding ``softening'' of the quasi-potential in that direction in phase space (i.e. CSD). It is worth mentioning, however, that in contrast to the conceptual models introduced earlier, the Veros model has a much more complex stability landscape such that there are not only two co-existing attractors and one edge state in between. 
We cannot fully exclude that the increasing number of attractors with a non-collapsed AMOC towards the TP known from the deterministic system, which become accessible by the noise forcing, are part of the reason for increases in fluctuations in certain variables. 
Some of these co-exisiting attractors indeed span a non-negligible range of deep Atlantic salinities and temperatures, and thus there could be increases in fluctuations as these  are mixed by the stochastic forcing. %(as they are indeed). %%% SUPP FIGURE for F=8.3
But these co-existing attractors also differ in many other variables, and thus it does not explain why specifically the observables with significance to the edge state show increased variability. Furthermore, in the case of state III ($F=0.3515$~Sv, Fig.~\ref{fig:veros_fluctuations}c,g) there are no co-existing attractors to be mixed besides the three extremely close-by states (inset Fig.~\ref{fig:veros_bif_diagram}).

\section{Discussion}
\label{sec:conclusions}

Here we propose that knowledge of the edge state separating a (desired) base attractor and an (undesired) alternative attractor is useful for finding observables that carry efficient EWS of a TP from the base attractor. 
Using conceptual models as well as a global ocean model, we illustrate that ``canonical'' variables or observables (as the AMOC strength in the case of the AMOC collapse TP), which a priori may best summarize the state of the system, can display little or no CSD as the TP is approached. 
% refer to other studies that show no CSD for AMOC strength. 
Instead, observables that take into account the location of the edge state with respect to the base attractor should be used, because from large-deviation theory it is known that noise-driven fluctuations in the low-noise limit will be directed along the instanton path towards the edge state. Such fluctuations will be explored by the system more frequently as the TP is approached and the potential barrier becomes lower. 

The models investigated here suggest that observables carrying good EWS are those where the edge state lies outside the range of the base and alternative attractors, and specifially at values beyond the alternative attractor and thus very far from the base attractor when the system is not near the TP yet. 
%%% some heuristic why this is plausible? 
In the case of the AMOC collapse in the global ocean model, this criterion pointed towards the salinity and temperature in the deep Atlantic, which indeed turned out to be the observables with largest increase in variability as the TP is approached (Fig.~\ref{fig:veros_ews}). 
This is not a sufficient or necessary criterion for observing significant CSD, and a better theoretical understanding needs to be developed to see whether it can be generalized. Nevertheless, a projection onto several of such observables indeed showed an accurate alignment to the direction of the edge state in the ocean model (Fig.~\ref{fig:veros_fluctuations}). Detecting this emerging directionality of fluctuations may be quite general as EWS, since for a given (low) noise strength the system's trajectory can only sample the path towards the edge state when the system is close to the TP. 

%CONNECTION TO PREVIOUS WORK: WHAT IS NEW IN THIS WORK?
The problem of detecting EWS for TPs in higher dimensional systems is known, and it has been pointed out that there can be state variables that carry no EWS even in simple low-dimensional systems \cite{BOE13}. A proposed remedy is data-driven methods for dimensionality reduction to find the relevant degrees of freedom, such as EOFs or principal oscillation patterns \cite{HEL04,BAT13,KWA18,PRE19}. 
% and then EWS on the principle component time series. 
In the case of an ocean circulation TP, this would entail creating three-dimensional, multi-variate EOFs from the entire set of state variables. A comparison of our results with such a technique is beyond the scope of this paper. 
Similar data-driven techniques include searching the direction in phase space of lowest resilience via autocorrelation functions \cite{WEIN19} and reconstructions of the Jacobian \cite{WIL15}. %%% NOTE and langevin approach Morr 
%fits time series to multivariate AR process, and then diagonalizes. 
% BUT we want modes/directions/observables that show largest INCREASE in variability. 
% this has been discussed too by prettyman et al 2019; they argue it is likely still the first EOF that is relevant for EWS: the first EOF should project the system on a loading vector very similar to the first eigenvector of the linearized system (the one where the eigenvalue becomes 1). thus it should capture the EWS. 

% Faranda et al 2014; They explicitly talk about fluctuations towards the unstable
% state, and measure this using extreme value laws. Related to skewness of course.
% but they dont consider how to find the relevant observables. 
%NOTE they also focus on the quantitative estimate of the bifurcation point!
% could say that this would only work when one has the correct observable?

%%% Maybe also talk about EWS in spatially extended systems. including Amazon papers. 
% Similarly to this: 
% Dylewski Sci Rep 2024: finding the low-dimensional embedding of the bifurcation. 
% they just look at spatial patterns of EWS (obtained from machine learning technique), assuming each location to be independent, and then say the spatial pattern of EWS gives clue about the embedding of the bifurcation...

The novelty in this work is the proposition to obtain the critical modes carrying EWS by an approach that is global (instead of local) in phase space, namely the estimation of the edge state (for a parameter value reasonably close to the TP) or the instanton. By requiring knowledge of an unstable state that cannot be observed directly in data, the method is system-specific and it sacrifices the ``universality'' of EWS, which is one of its main appeals. But it is crucial in cases where not all state variables are available by measurements to obtain the correct critical modes for EWS using the above-mentioned empirical methods. 
It allows one to know beforehand what observables would correctly warn of an impending tipping - and whether the required data is available - not least in order to avoid false positives when ``data-mining'' for variables that show an increase in variance by chance, or for reasons other than CSD.
The method can be a useful complement to EWS identified in complex models based on insights into the physics of the TP \cite{VWE24}, since it offers a general guiding principle to check whether the identified critical degrees of freedom are indeed compatible with what would be expected from dynamical systems theory. 

Another approach using edge states to determine the TP has been presented recently for the case of magnetohydrodynamic turbulence \cite{BRY24}. Here, a continuation of the edge state is performed in order to estimate at what value of the control parameter the TP occurs, based on extrapolation of the normal form of the saddle-node bifurcation. This is relevant in cases where the base state is not an attractor but a metastable, chaotic transient, and thus more difficult to continue along the control parameter. 

In practice, it can be difficult to derive a reliable approximation of the edge state of a system using models of sufficient complexity. In the case of climate, it would arguably require a number of different state-of-the-art models where an edge state is computed and its properties agree qualitatively. 
Furthermore, even if the crucial observables can be identified, these may turn out not to be measureable practically, or may yield low statistical significance as EWS in a given situation with finite data. This happens especially with observables evolving on long time scales, where long data horizons are required for robust statistics of the variability, which is further exaggerated as the dynamics slow down towards the TP. Which observable will show highest statistical significance in practise can thus depend on the length of available data, which is a statistical issue that deserves further study. 
The case of the AMOC TP in the global ocean model studied here highlights this: Temperature and salinity in the deep Atlantic, being the key characteristics of the edge state and indeed the only observables with substantial increase in variability towards the TP, have not been monitored with good spatio-temporal coverage in past and present. They evolve on very long timescales, which would require a long data horizon. This reduces the usefulness of the EWS derived from the edge state in this particular case. Note, however, that the characteristics of the edge state may change if one would estimate it from a coupled atmosphere-ocean GCM, which features more processes on faster time scales.

% NOTE general connection to the fact that there should only be one unstable direction
% of the saddle, since boundary has to be at most co-dimension 1
% -> what would be a reference here?

More work is needed to give the proposed EWS based on the edge state a theoretical underpinning. We showed with examples of different complexity that observables which best represent the relative location of the edge state with respect to the attractors are clearly among the (often few) ones that stand out most in terms of CSD. 
%other variables that seem a priori important might not show CSD.
But a definitive criterion how exactly these observables should be constructed is lacking. Specifically, a connection needs to be made between the distance (and change thereof) in a given observable subspace of the edge state relative to the base attractor and the expected increase in variability. 
% this may also be viewed as a statistical issue of finding the precise observable in a sub-space that would give . 
The statistical issue of finding an observable subspace that shows CSD of highest statistical significance is non-trivial. This is because the directionality of the edge state (and instanton) % with respect to base attractor
could change drastically % say non-monotonically?
during a parameter shift, making the result dependent on the range of the control parameter where the EWS are to be extracted. 

We remark that the choice of the noise law is critical in defining the geometrical properties of the quasipotential and the instantonic trajectories linking the attractors and the edge state, see the explicit example in a climate-relevant problem in \cite{LUC20}. As a result, the signal-to-noise ratio of an observable for EWS can depend on which state variables are influenced by the noise. If noise applied to a subset of variables does not induce fluctuations in the right direction \cite{BOE13}, or if the noise processes driving the state variables are correlated \cite{MOR23}, EWS in some observables can be masked. 
% Regarding noise model:
In the Veros model simulations, we chose noise in the surface layer in temperature and salinity only. Both constitute buoyancy forcing and may lead to an inhibition of NA convection as commonly presumed pathway to AMOC collapse. 
It will be interesting to investigate whether other noise structures yield qualitatively different results. For instance, wind stress noise could be implemented, which may affect the AMOC by other processes, such as by changes in upwelling in the Southern Ocean or by anomalies in NA Ekman transport as observed in \cite{CIN24}.

The principle of CSD is typically formulated for systems with fixed points and stochastic forcing. Our specific approach to CSD was motivated via examples of fixed point dynamics too (Sec.~\ref{sec:conceptual}), but then transferred to a much more complex model with chaotic attractors (Sec.~\ref{sec:veros}). Does this imply we expect the approach to work for any noisy system with limit cycles or chaotic attractors? The present work is not sufficient to prove this, but the proposed reasoning in terms of instantons and quasipotentials gives some merit that the concept can be transferred in many cases to non-fixed point situations, provided the following observations hold: 

\begin{enumerate}
 \item We expect any base attractor to get closer to the basin boundary before colliding at the TP with an unstable invariant object, which could be an unstable periodic orbit or a chaotic saddle (also referred to as melancholia state \cite{LUC19}). This object takes on the role of the edge state discussed here. 
 \item Under the addition of noise, Graham's quasipotential \cite{Graham1987} is a concept introduced for general nonequilibrium systems, i.e., systems not restricted to fixed points
 \item At the approach of the TP there is (plausibly) a reduction in the quasipotential barrier (defined by the value at its lowest point on the basin boundary)
 \item This would naturally lead to a flattening of the quasipotential in a particular direction, implying increased fluctuations
 \item This direction can be estimated by knowledge of the edge state or (more generally) a minimum action path, and translated into observables useful for EWS
\end{enumerate}

Regarding the last point, using scalar projections onto vectors pointing towards the edge state as observables is just the simplest approach. Generalizations are required, especially in the abovementioned case of chaotic attractors and saddles, and an associated complicated phase space structure. 
First, it is really the initial portion of the instanton connecting to the base attractor that would determine the directionality of fluctuations in the low-noise limit. The instanton is generally not a straight line, and its initial segment may be directed entirely differently compared to the straight line between the mean states of base attractor and edge state. This motivates further studies on how to best design non-linear observables for EWS based on the instanton path. 
While this discrepancy is likely to become smaller as the TP is approached, it may nevertheless be a problem in practice since for a given system it may only be possible to estimate the edge state when still reasonably far from the TP due to long transients. This causes issues when the edge state itself moves non-linearly in phase space as the control parameter is varied towards the bifurcation. This results in estimates of observables for EWS that are not accurate close to the TP. 

We further remark that there can be multiple edge states per basin boundary \cite{KHA13}, in which case it is not a priori clear which edge state is preferably visited by large fluctuations. The dominant transitions paths in the weak-noise limit will then in general vary when different noise laws are considered because of changes in the value of the quasipotential \cite{LUC20,MAR21}. Hence, again the actual minimum action path for a realistic noise law would be required for EWS. 

Another generalization concerns the assumption that the edge state, i.e., the dynamical saddle of the deterministic system, is also the set of minimum quasipotential on the basin boundary, and consequently that noise-driven fluctuations will (in the low-noise limit) drive the system towards the dynamical saddle. This can be proven in some cases \cite{MAI97, BOU16}, and is also made plausible in more complex cases by the study of edge states and transitions in turbulence \cite{LOZ12, KHA16, GOM22}, as well as our analysis of the global ocean model. 
But there may be certain systems, for instance with multiscale dynamics, where for non-infinitesimal, but weak noise the dynamical saddle is in fact avoided by noise-induced transitions \cite{NOR83, BER89, LUC99, BOE24}. Here the Freidlin-Wentzell instanton is not accurate and the target needs to be a more general minimum action path, such as the minimizer of the Onsager-Machlup action. 
Another case is degenerate noise, where it is not obvious that the invariant measure of the system is smooth with respect to Lebesgue. This is guaranteed in the case of hypoelliptic diffusion, whereby the noise is propagated to all degrees of freedom through the drift term \cite{BEL04}. Keeping in mind that the quasipotential reflects the minimal action required to transition between states along admissible paths, noise degeneracy implies that not all directions in the state space are accessible. This makes finding the minimizers of the action functional more challenging. Recent results indicate that in the case of hypoelliptic diffusion such a minimizer can be found \cite{INA24}. As a result of the constraints, it is not obvious that such a minimiser, which clearly crosses the basin boundary separating the competing states, will go through the dynamical saddle. The crossing point realised in this case might not be the edge state, but coincide with a so-called mediator state \cite{ROL24}. 

%NOTE: maybe also generalizations for Levy noise component?

It would be interesting in future studies to examine how the ideas presented here can be connected to purely deterministic cases of a crisis where an attractor collides with a chaotic saddle, where generalized EWS have been proposed \cite{TAN18, TAN20}. 
Further progress could be made by exploring the connections to previous work on critical transitions from the viewpoint of linear response theory, which investigated decay rates of Koopman modes \cite{GUT22}, and the related issue of finding ``reaction coordinates'', i.e., sets of observables %that are thermodynamic variables 
%giving a low-dimensional description of the macroscopic state 
that capture %time-dependent properties such as 
transitions between metastable states
%and optimal observable to capture the 
and critical resonances in the linear response \cite{ZAG24}. A promising method of obtaining a low-dimensional space of reaction coordinates that reproduces the global dynamics including the edge state and its stable manifold is spectral submanifold reduction \cite{KAS24}. 

%NOTE: ABOUT BOUNDARY CRISIS not hitting the edge state. 
Finally, it is worth investigating whether exceptions need to be made when in a boundary crisis \cite{OTT93} the base attractor does not collide with the edge state, as would be expected in analogy to the saddle-node bifurcation and as indeed is observed in some complex flows \cite{ZAM15, BRY24}, but instead with a different part of the basin boundary. This may again require abandoning the usage of the edge state, and to tackle the more difficult problem of estimating a general minimum action path. 

\section{Conclusion}
Summarizing, here we describe how EWS for TPs based on the idea of CSD can be improved by deriving suitable observables from the properties of an edge state that lies between a reference and an alternative attractor. This is useful in cases where EWS measured on ``canonical'' observables fail, as shown here for the case of a collapse of the AMOC with the AMOC strength as observable, and it is essential for attempts at quantitative prediction of forthcoming TPs based on observed time series data. 
The proposed methodology is heuristic, but based in large-deviation theory, which states that the edge state is typically on the minimum action path (instanton) of a noise-induced transition between the attractors. 
%though there can be exceptions, and it depends on noise structure, and finite time!!
When nearing the TP during a parameter shift, stochastic fluctuations along the instanton towards the edge state become more likely. Hence, the variability in observables that characterize this direction in phase space become larger, which serves as efficient EWS. 
% any caveats or improvements?
% could say that theory needs to be further developed. 

\textbf{Acknowledgements} We thank R. Nuterman and the Danish Center for Climate Computing for supporting the simulations with the Veros ocean model. JL has received support from Danmarks Frie Forskningsfond under grant no. 2032-00346B, and JL and RC were supported by a research grant (VIL59164) from VILLUM FONDEN. VL has received support from the EU Horizon Europe project ClimTIP (Grant No. 101137601), and from the EPSRC project LINK (Grant No. EP/Y026675/1). VL and AL have received funding from the Marie Curie Action CRITICALEARTH (Grant No. 956170).

%%%%%%%%%% Insert bibliography here %%%%%%%%%%%%%%

\bibliographystyle{apsrev4-1}
\bibliography{refs}

\appendix 

\section{Details on the Veros ocean model}\label{AppModel}

The primitive equation finite-difference global ocean model {\it Veros} is a direct translation of the Fortran backend of the ocean model {\it PyOM2} \cite{EDE14} into Python/JAX \cite{HAE18,HAE21}. Mesoscale turbulence is represented using the \cite{RED82} and \cite{GEN90} parameterization for isopycnal and thickness diffusion (diffusivity of 1~000 m$^2$/s). We use the second order turbulence closure of \cite{GAS90} to account for diapycnal mixing. A background diffusivity of 10$^{-5}$ m$^2$/s is employed, and static instabilities are removed by a gradual increase of the vertical diffusivity as the instability is approached. This corresponds to a smoothed variant of convective adjustment. 

The heat exchange boundary condition is expressed by a first-order Taylor expansion of the heat flux as a function of the anomaly of the modeled surface temperature with respect to a fixed surface temperature climatology \cite{BAR98}. For this, we use ERA-40 climatologies \cite{UPP05} of surface temperature and heat flux, as well as a climatology of the derivative of the heat flux with respect to changes in surface ocean temperature. The latter is derived from ERA-40 following \cite{BAR95} and has a global yearly average 30.26 $WK^{-1} m^{-2}$. Wind stress forcing is also taken from the ERA-40 data. Freshwater exchanges with the atmosphere are modeled by boundary conditions under which the sea surface salinity is relaxed towards a present-day ERA-40 climatological field within a given relaxation timescale. By choosing a long timescale of 2 years, oceanic salinity anomalies are less efficiently damped by the atmospheric forcing compared to temperature anomalies. This enables the positive salt advection feedback, which can lead to AMOC multi-stability and tipping. 

The bathymetry is obtained by smoothing the ETOPO1 global relief model \cite{AMA09} with a Gaussian filter to match the grid resolution. The model domain ends at 80$^o$N and thereby does not feature an Arctic connection of ocean basins. The horizontal grid contains 90 longitudinal and 40 latitudinal cells, where the latitudinal resolution increases from 5.3$^o$ at the poles to 2.1$^o$ at the Equator, as well as 40 vertical layers, which increase in thickness from 23 m at the surface to 274 m at the bottom. For further model details, see our previous studies that investigated AMOC tipping in this ocean model \cite{LOH21,LOH22,LOH24}.

In addition to the ERA-40 forcings, a freshwater anomaly is introduced in the North Atlantic, in order to drive the system closer to the tipping point of a collapsed AMOC.
To this end, a salinity flux anomaly $\tilde{\phi}$ is applied at the surface in the grid cells between 296$^o$E to 0$^o$E and 50$^o$N to 75$^o$N, corresponding to an area of about $A= 1.5 \, \text{mio.} \, km^2$. Here we use the equivalent total freshwater forcing $F = \tilde{\phi} A S_{ref}^{-1}$ as control parameter, with the reference salinity $S_{ref} = 35 \, g\cdot kg^{-1}$.

\section{Surface noise model}\label{AppNoise}

% 8. choose cutoff number of EOFs based on 0.95 explained variance. 
% 9. For each principle component (EOF coefficient) time series, we determine a maximum lag beyond which there is no significant partial autocorrelation anymore. for significance level there is a parameter alpha. 
% 10. Use yule-walker method to fit AR parameter for each PC timeseries. -> Done. 

The noise model is derived from the sea surface temperature dataset from Berkeley Earth \cite{ROH20} spanning the period from 1880 to 2022 and the ORAS5 sea surface salinity data set from 1958 to 2014 \cite{ORAS5}, on a 1 degree grid. From the data we construct a spatio-temporal noise model via EOF analysis and subsequent autoregressive modeling of the principal components (PCs). The raw data is preprocessed by subtracting the mean at all grid points. Next, missing monthly values for a given grid cell are filled using the neighboring grid cells via nearest-neighbor interpolation. This is also done for the permanently missing land grid cells in the sea surface salinity data, in order to allow better interpolation to the grid of {\it Veros}. In this case, we subsequently apply a spatial smoothing at each time step with a Gaussian kernel filter (standard deviation of 6 grid cells) to remove non-smoothness in the data as a result of shallow coastal areas. 

The data are made approximately stationary by removing the climatology at every grid point in a 10-year sliding window. Concretely, for a given monthly time point, the average value of the same calendar month in the five years before and after are subtracted. %handling of edges: only use 5 years after/before. 
This removes long term trends (anthropogenic climate change), as well as seasonality (see Fig.~S21 and S22 for examples at two grid points). By using a sliding window approach we can also to some degree remove effects of a change in seasonality over time. %is it actually clear what is meant by that?
% the point is that seasonality changes, and thus we have to remove seasonality 
% ``locally'' in time. 
Note that the choice of a 10-year window removes any multidecadal variability. 
% trade-off: good detrending and removal of changing seasonality (short window). 
% long-timescale variability (large window); full removal of seasonality (long window, but only if it is stationary). 
The final preprocessing step is linear interpolation from the $180\times360$ grid of the reanalysis data to the $40\times90$ grid of the Veros model. The preprocessed data is then decomposed in EOFs,  
%\cite{hannachiEmpiricalOrthogonalFunctions2007}, 
i.e., the spatial modes $e^{(j)}$ that diagonalize the covariance matrix of the spatio-temporal field $X$ (salinity or temperature). Then, for each month $n$ (at time $t_n$) at grid point $i$, we can write
\begin{equation} \label{eq:veros:eof}
    X_i(t_n) = \sum_{j=1}^{J} \tilde{c}_n^{(j)} e^{(j)}_i .
\end{equation}

%leave this out..?
In the decomposition of Eq.~\ref{eq:veros:eof}, we choose to save the information about the explained variance in the norm of the EOF $e^{(j)}$, so that $\tilde{c}_n^{(j)}$ have zero mean and unit variance.
We then reduce the dimensionality by truncating the above series and keeping only the first $J_0$ modes that cover 95\% of the variance. For temperature and salinity this yields $J_0 = 198$ and $J_0 = 70$, respectively.
%refer again to Supp figure with first EOFs?
The error made by this truncation is very low (Fig.~S23). Subsequently, the $J_0$ principal components $\tilde{c}_n^{(j)}$ are modeled as autoregressive processes of rank $R^{(j)}$:
\begin{equation}\label{eq:veros:autoregressive-process}
    \tilde{c}_n^{(j)} \approx c_n^{(j)} = \sigma^{(j)} \xi_n^{(j)} + \sum_{r=1}^{R^{(j)}} \rho_r^{(j)} c_{n-r}^{(j)} ,
\end{equation}
where $\xi_n^{(j)}$ is Gaussian white noise.

The maximum lag $R^{(j)}$ is computed following the procedure by Box and Jenkins \cite{boxTimeSeriesAnalysis2015}.
This involves computing the partial autocorrelation function (PACF), which at lag $l$ is the standard autocorrelation function from which the effect of all lags $l^\prime < l$ has been removed. For a true autoregressive process of rank $R$, the PACF is equal to 0 $\forall l > R$. We choose $R^{(j)}$ as the maximum lag $l$ at which the PACF is significantly different from 0 with a significance p-value of $\alpha = 10^{-4}$.
%NOTE check this value. correct for salinity, but 0.001 for T it seems. 
The maximal lag used across all PCs was $R^{(j)} = 240$ for the temperature data, and $R^{(j)} = 133$ for salinity. Finally, the autoregressive coefficients $\sigma^{(j)}$ and $\rho_r^{(j)}$ are fitted from the time series of $\tilde{c}_n^{(j)}$ with the Yule-Walker equations. %\cite{yuleVIIMethodInvestigating1927,walkerPeriodicitySeriesRelated1931}.

%NOTE: add info on what the ``typical'' maximal lag is? 
% from Alfred: the maximal lag used across the AR models will be 240 months for the second PC, and there is a minimum of 60 months.
% for salinity? max lag 
% min lag was 2 for just one random PC (almost last PC). Others around 60 or 48. 
% -> this has something to do with the fact that seasonality is suppressed beyond
% what would happen by chance (red noise). 

To run the noise model alongside Veros, we initialize it setting $c_n^{(j)} = 0 \, \forall n < 0$. Then, at the beginning of every month $n \geq 0$, we compute the next coefficients $c_{n+1}^{(j)}$ according to Eq.~\ref{eq:veros:autoregressive-process}, and combine them into the noise field
\begin{equation}
    \epsilon^T_i (t_{n+1}) = \sum_{j=1}^{J_0} c_{n+1}^{(j)} e^{(j)}_i .
\end{equation}
Finally, for each time step $t$ within month $n$, the actual noise field $\epsilon^T_i (t)$ is a linear interpolation between $\epsilon^T_i (t_{n})$ and $\epsilon^T_i (t_{n+1})$, and then added to the ERA-40 climatology that forces the sea surface salinity and temperature variables of the Veros model. 

The resulting noise models show multi-annual to decadal variability and inherit the most prominent spatio-temporal structures found in the reanalysis data, such as the El-Ni\~no-Southern-Oscillation (ENSO), the Pacific Decadal Oscillation and the North Atlantic Oscillation (see Fig.~S24 and S25).

\end{document}

\title{Supplemental Material to: The role of edge states for early-warning of tipping points} 

\author{Johannes Lohmann}
\email{johannes.lohmann@nbi.ku.dk}
\affiliation{Physics of Ice, Climate and Earth, Niels Bohr Institute, University of Copenhagen, Denmark}
\author{Alfred B. Hansen}
\affiliation{Physics of Ice, Climate and Earth, Niels Bohr Institute, University of Copenhagen, Denmark}
\author{Alessandro Lovo}
\affiliation{ENS de Lyon, CNRS, Laboratoire de Physique, F-69342 Lyon, France}
\author{Ruth Chapman}
\affiliation{Physics of Ice, Climate and Earth, Niels Bohr Institute, University of Copenhagen, Denmark}
\author{Freddy Bouchet}
\affiliation{ENS de Lyon, CNRS, Laboratoire de Physique, F-69342 Lyon, France}\author{Valerio Lucarini}
\affiliation{School of Computing and Mathematical Sciences, University of Leicester, Leicester, UK}

\date{}

%%%%%%%%%%%%%%%%% END OF PREAMBLE %%%%%%%%%%%%%%%%

\newcommand{\beginsupplement}{%
        \setcounter{table}{0}
        \renewcommand{\thetable}{S\arabic{table}}%
        \setcounter{figure}{0}
        \renewcommand{\thefigure}{S\arabic{figure}}%
        \setcounter{section}{0}
        \renewcommand{\thesection}{S\arabic{section}}%
}


\baselineskip24pt


\maketitle 

\beginsupplement

\section{Description of dynamics of the Veros model with added surface noise}\label{NoiseVeros}

\begin{figure}%[floatfix]%!htb
\includegraphics[width=0.99\textwidth]{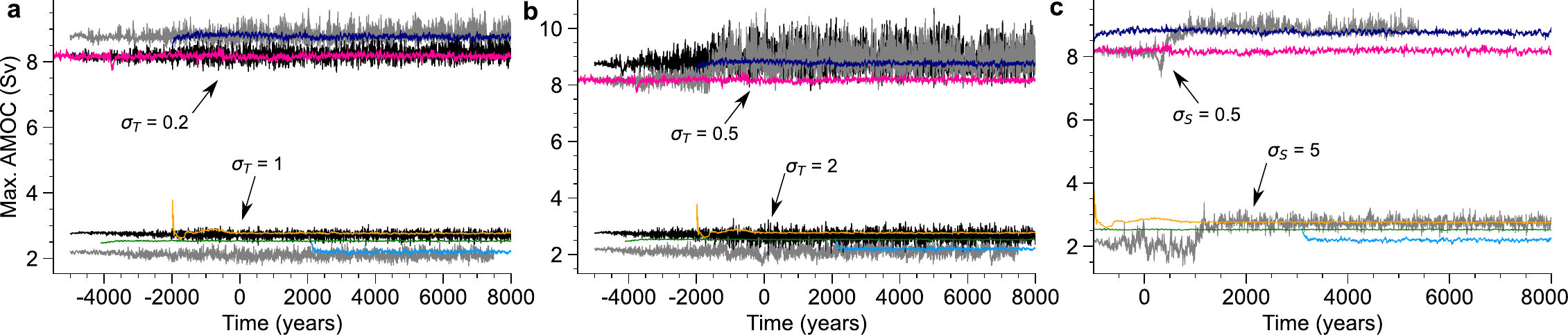}
\caption{\label{fig:low_noise} 
Time series of the maximum AMOC strength (5-year averages) for simulations with added noise at $F=0.3472$~Sv (gray and black trajectories) compared to steady-state simulations on the co-existing attractors of the deterministic system (colored trajectories). 
The colors of the deterministic time series correspond to the ones of the attractor branches given in Fig.~8.
All time series have been shifted to end at the same time $t=8000$~years, and some of the deterministic trajectories have not been simulated for the maximum duration, thus starting later and including short transients. The noise strength is ramped up linearly for 5000 years in ({\bf a,b}) and 1000 years in ({\bf c}) to allow for a slow and controlled shift of the system from an initial state on an attractor of the deterministic system to its new statistical steady state under noise forcing. The maximal noise strength is reached at time 0 and held constant thereafter. {\bf a} Four noisy simulations, initialized on two coexisting attractors each for the ON and OFF regimes. Shown are simulations where temperature-only noise was used with $\sigma_T = 0.2$ and $\sigma_T =1$ for the simulations in the ON and OFF branches, respectively. {\bf b} Same as {\bf a}, but noise strength is $\sigma_T = 0.5$ and $\sigma_T = 2$ for the simulations in the ON and OFF branches, respectively. 
{\bf c} Salinity-only noise is used for simulations initialized in one attractor each for the ON ($\sigma_S = 0.5$) and OFF ($\sigma_S = 5$) branches. 
}
\end{figure}

\begin{figure}[!htb]
\includegraphics[width=0.75\textwidth]{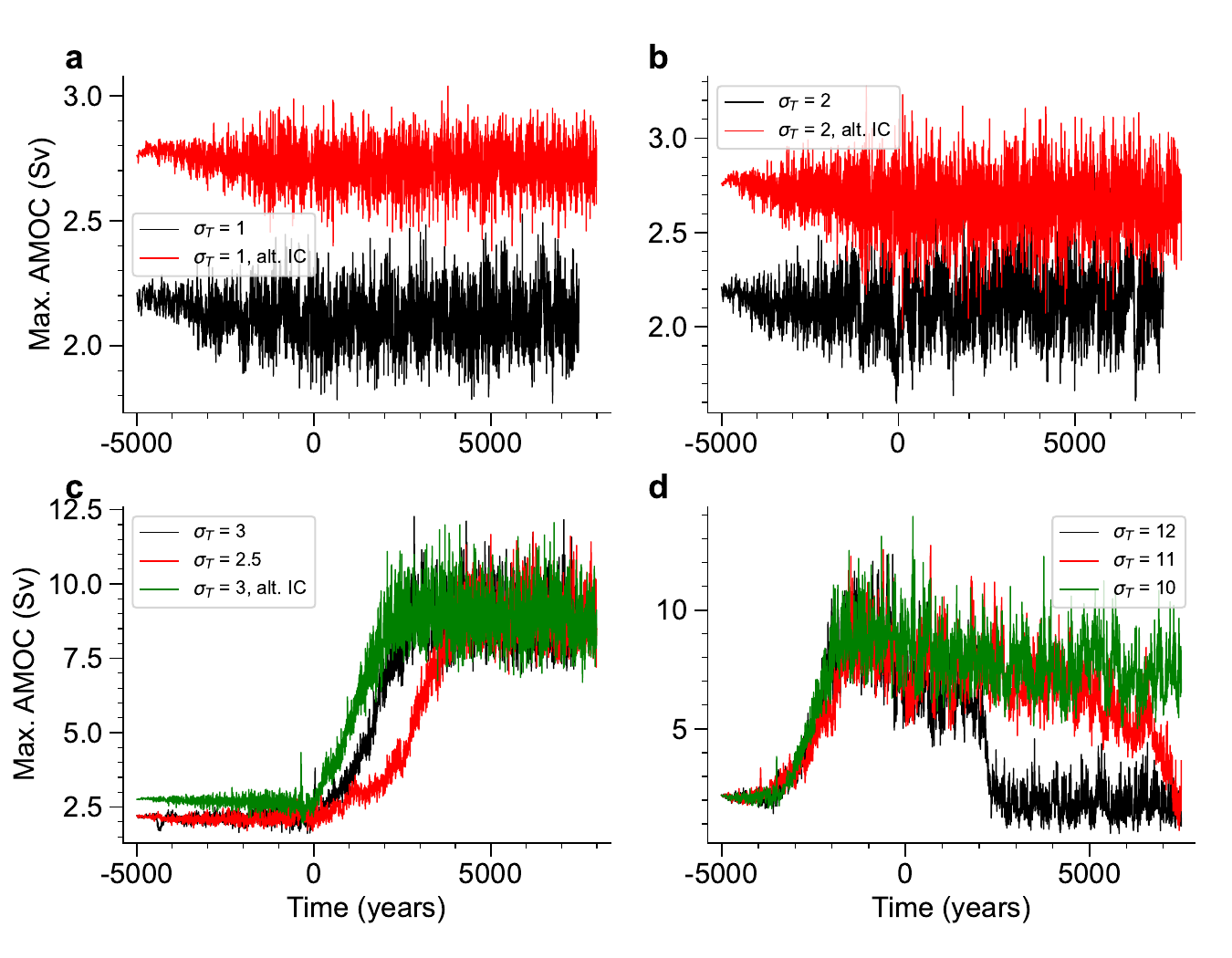}
\caption{\label{fig:timeseries_noise_temp_lower} 
Simulations of the Veros model under temperature-only noise forcing at the control parameter value $F=0.3472$~Sv. Shown are simulations initilized in different AMOC OFF states, and for different noise strengths $\sigma_T$ (see figure legends). The noise strength is ramped up linearly during the first 5000 years, and kept constant after year 0 in the figures. For low noise, there are no noise-induced transitions between the different AMOC OFF states ({\bf a,b}) or from an OFF state to an AMOC ON state. For intermediate noise strength we observe transitions from the OFF to the ON state ({\bf c}). At these noise levels and this value of $F$, we have not observed any transitions in the opposite direction (from ON to OFF). The same holds for intermediate noise levels using salinity-only or mixed noise forcing. For larger forcing, there are transitions from ON to OFF. This is seen in panel {\bf d}, where during the ramp-up of the noise strength there is a transition from OFF to ON, but after the noise strength has stabilized at a sufficiently high level of about $\sigma_T = 11$, there is a transition back to an OFF state. 
}
\end{figure}

Here we give a detailed account of how the surface noise forcing affects the dynamics of the Veros model. Simulations with a large variety of combinations of $\sigma_T$ and $\sigma_S$ have been performed to explore the dynamics. 
$\sigma_T = \sigma_S = 1$ corresponds to noise fields as derived directly from the data. 
We focus on the regime with co-existing AMOC ON and OFF states, and first consider the parameter value $F=0.3472$~Sv. For low noise strength, the dynamics evolve closely around the attractors from which the system was initialized in, and no transitions are observed (Fig.~\ref{fig:low_noise}a and Fig.~\ref{fig:timeseries_noise_temp_lower}a,b). 
For slightly increased noise strength transitions occur within the AMOC ON or OFF regimes from certain attractors onto others (Fig.~\ref{fig:low_noise}b,c). 
The transitions appear irreversible, in the sense that during the simulation time there were no transitions in the other direction. This may be partly because, at this value of the control parameter, the relative stability of the attractors in question may be very different, yielding a much longer residence time in one of the metastable states. 
Similarly, the residence time in one or both of the states may have become so short that the attractors appear effectively merged by the noisy dynamics on centennial and millennial time scales.
% explain further?
Alternatively, it may be that added noise above a certain strength, despite being of mean zero, changes the dynamical landscape of the system such that certain states that were deterministically stable cease to exist. %speculate the physics that would do this?

\begin{figure}%[floatfix]%!htb
\includegraphics[width=0.7\textwidth]{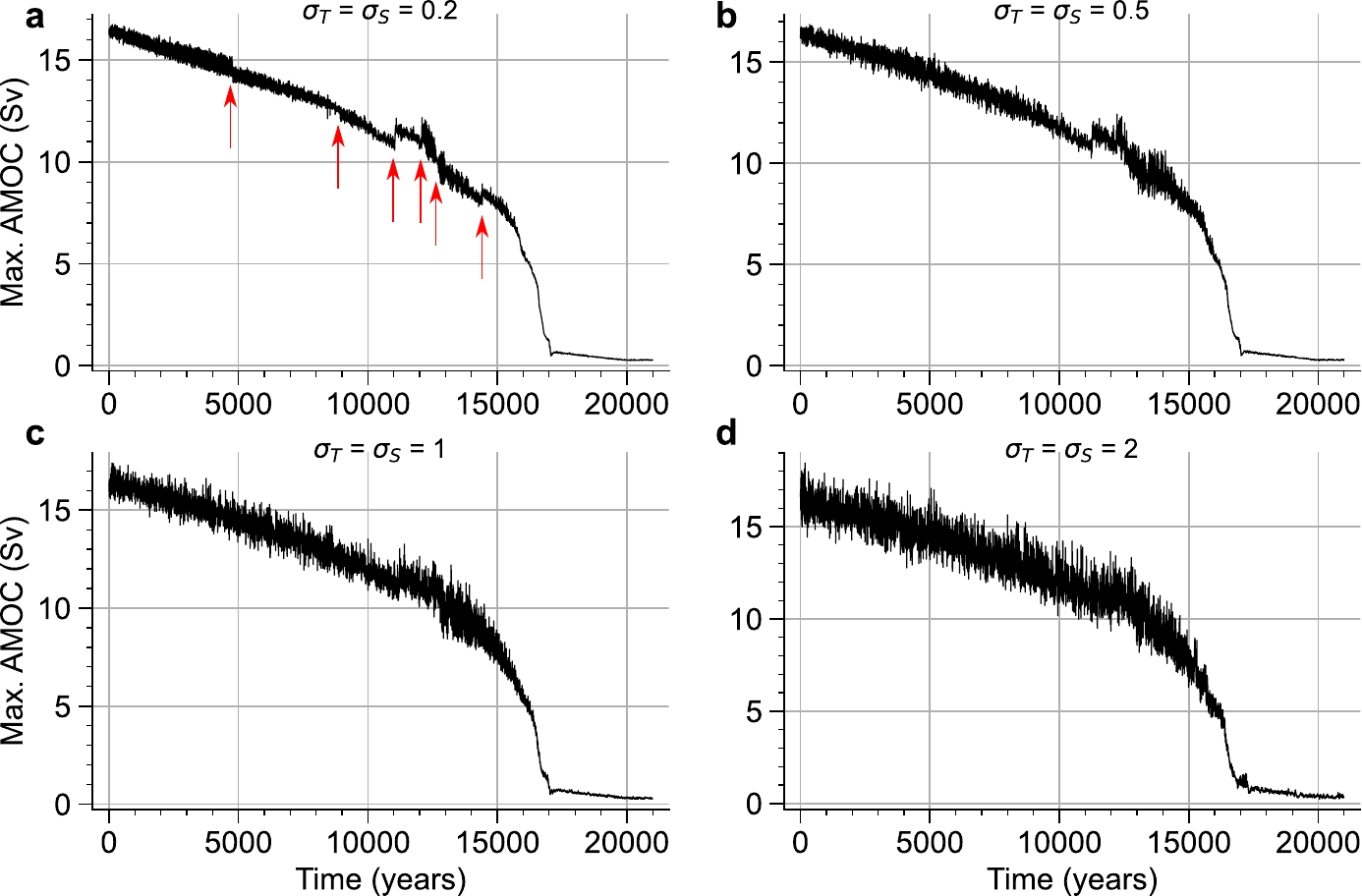}
\caption{\label{fig:transient_collapse} 
Transient simulations with linear freshwater forcing increase for four different noise strengths in a model with both temperature and salinity noise forcing. Shown are time series of the 5-year average AMOC strength. The simulations are started from a 10 kyr equilibrium simulation at $F = 0.184$~Sv. Thereafter (at time 0), the
freshwater forcing is increased linearly from $F = 0.184$~Sv to $F = 0.407$~Sv over a time period of 20,000 years. Afterwards it is held constant for 1000 years at $F = 0.407$~Sv. A number of intermediate tipping points known from the deterministic system \cite{LOH24} are shown in {\bf a} by the red arrows. 
}
\end{figure}

% transient runs:
By merging or loss of stability of attractors that were very close-by in phase space of the determinisitic system, some fine structure in the stability landscape is lost. As a result, some intermediate TPs are masked in time series of the AMOC strength (Fig.~\ref{fig:transient_collapse}). The amplitude of AMOC variability appears to evolve more gradually in the noisy case as the TP is approached, and for large noise levels the disappearance of intermediate TPs may recover our ability to design robust EWS. 

\begin{figure}%[floatfix]%!htb
\includegraphics[width=0.8\textwidth]{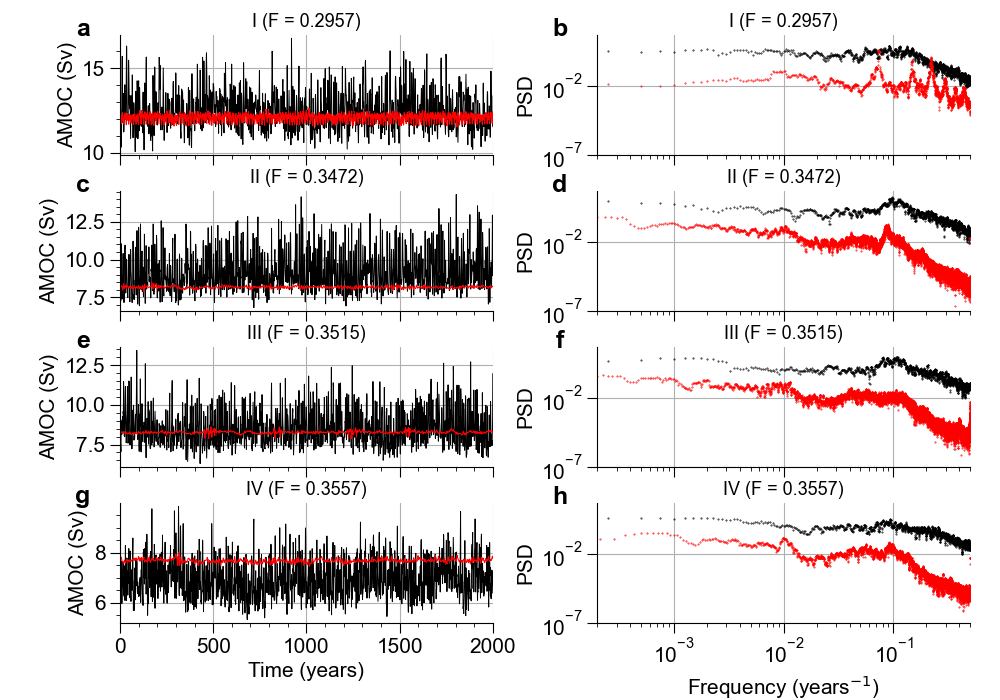}
\caption{\label{fig:timeseries_spectra} 
Annual average time series of the AMOC strength along with the power spectrum, for simulations with added temperature and salinity noise (black) using $\sigma_T = \sigma_S = 2$,  along with corresponding simulations without noise forcing (red) performed at the same values of the control parameter $F$. The simulations in black are 2,000 year long continuations of the much longer simulations shown in Fig.~13b-d, thus indicating that noise-induced AMOC collapses using this noise structure and strength are very rare. 
}
\end{figure}

The AMOC collapse is not advanced noteably by moderate noise forcing (Fig.~\ref{fig:transient_collapse}), as could be expected if there is a noise-induced transition in the multistable regime. Indeed, noise levels that induce considerable AMOC variability can be applied even very close to the tipping point of the deterministic system without observing a transition during long simulations. Fig.~\ref{fig:timeseries_spectra} shows simulations with $\sigma_T = \sigma_S = 2$ at four values of $F$, the largest of which is very close to the loss of stability of the ON regime. At any $F$ this noise level increases the AMOC variability by an order of magnitude (see spectra in Fig.~\ref{fig:timeseries_spectra}b,d,f,h). For $F$ further away from the TP, the quasi-periodic decadal-scale variability (Fig.~\ref{fig:timeseries_spectra}b,d), which is related to a thermal Rossby wave in the NA \cite{LOH24}, is dominated by a more broadband variability under noise forcing. 
% Already mention? -> Visually no evidence for increased 
% correlation time of fluctuations and increased variance (neither in noisy nor in 
% deterministic runs)
% Probably just mention later, referring back to the spectra. -> no spectral reddening. 

There are also shifts in the mean AMOC strength compared to the attractor the system was initialized on (red time series in Fig.~\ref{fig:timeseries_spectra}a,c,e,g). This may be partly a result of the merging of close-by attractors. For $F$ as in Fig.~\ref{fig:timeseries_spectra}c, there is another vigorous AMOC attractor with slightly stronger AMOC (see Fig.~\ref{fig:low_noise}b), and the two attractors effectively merge at this noise level. For $F$ as in Fig.~\ref{fig:timeseries_spectra}g, there are states with a partially-collapsed AMOC (see lower branches of states in inset of Fig.~8) that merge with the vigorous attractor leading to a lower mean AMOC strength. 
%%% NOTE EXPLAIN partially-collapsed? And refer to SUPP FIGURE with simulation starting from the PC state. 
%%% see Fig.~SX for longer time series of longer simulations? or coming later 
%%% for EWS results. 
But also when there are no known co-existing vigorous attractors with substantially different mean AMOC strength, there are clear shifts in the AMOC mean as noise is added (Fig.~\ref{fig:timeseries_spectra}a,e).

\begin{figure}%[floatfix]%!htb
\includegraphics[width=0.99\textwidth]{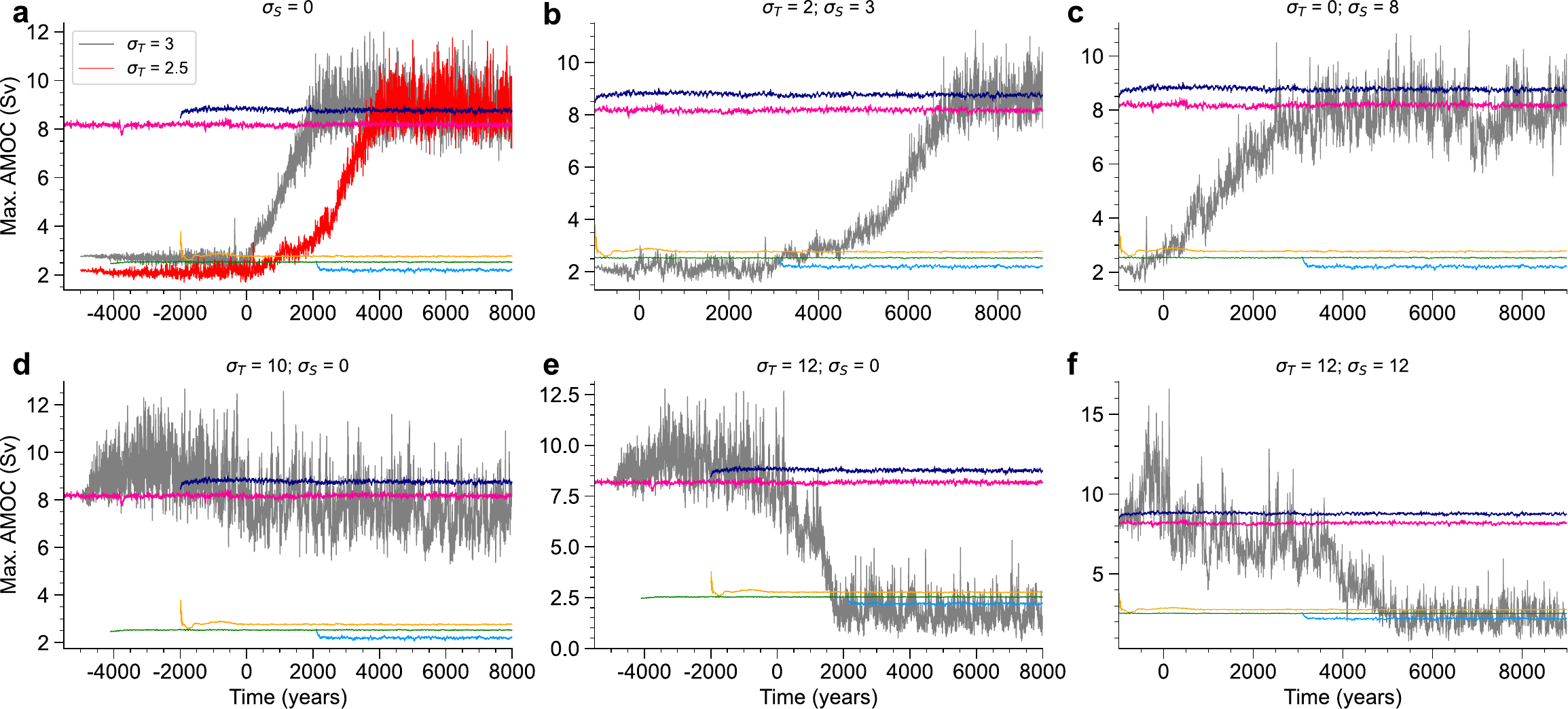}
\caption{\label{fig:veros_transitions_all} 
Examples of noise-induced transitions in the Veros model, using different noise strengths and combinations of the temperature and salinity noise components. 
{\bf a-c} Noise-induced resurgences of the AMOC for temperature-only ({\bf a}), mixed ({\bf b}), and salinity-only ({\bf c}) noise at $F=0.3472$~Sv. As in Fig.~\ref{fig:low_noise}, the noise strength is ramped up linearly until the time 0. 
{\bf d-f} Increasing the noise strength further, noise-induced collapses of the AMOC are found in temperature-only ({\bf e}) and mixed noise ({\bf f}) simulations. 
}
\end{figure}

\begin{figure}[!htb]
\includegraphics[width=0.85\textwidth]{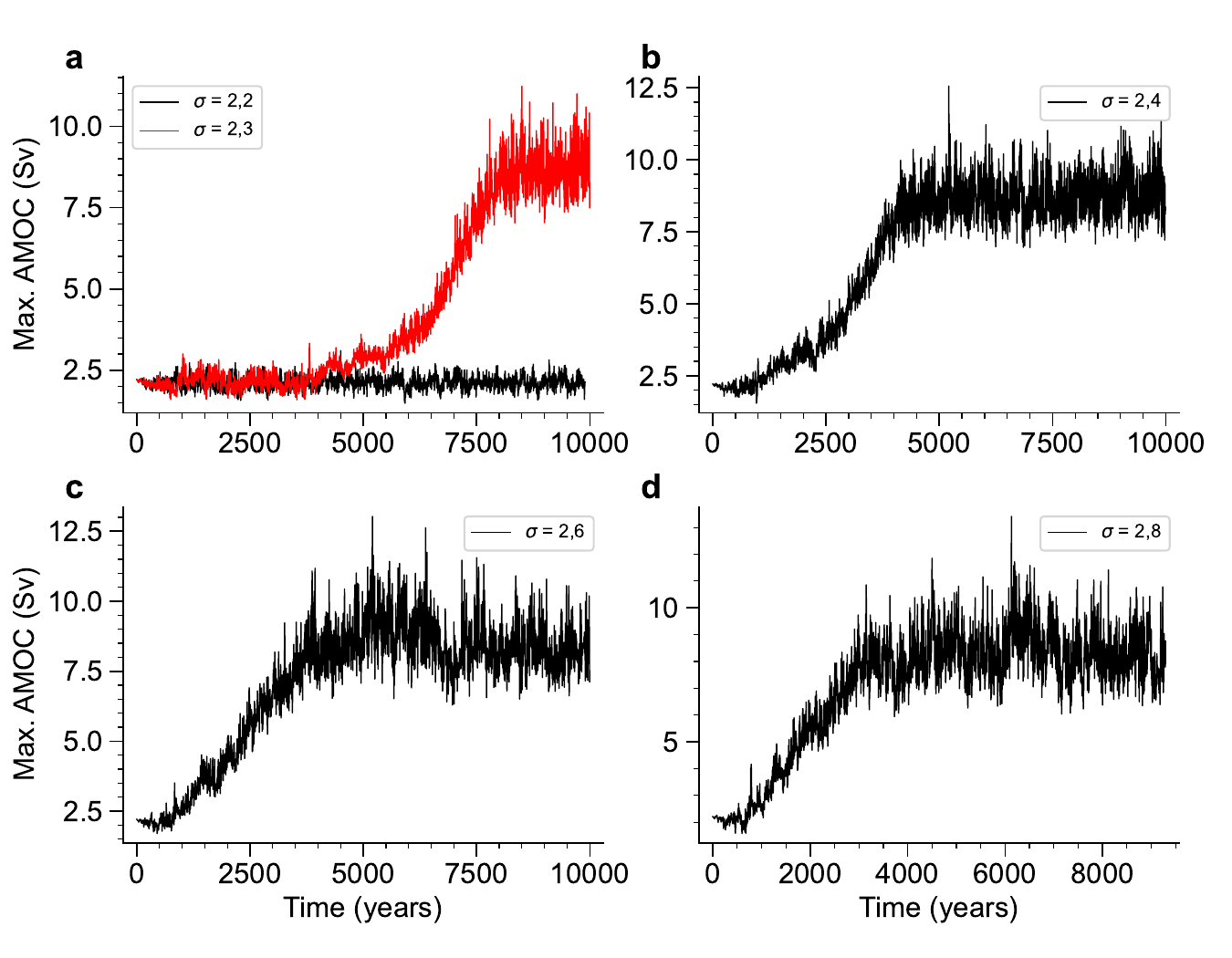}
\caption{\label{fig:timeseries_mixed_lower} 
Simulations initialized in an AMOC OFF state with the Veros ocean model forced by a noise model with both temperature and salinity components at $F=0.3472$~Sv. Both noise strengths are ramped up linearly during the first 1000 years, and kept constant thereafter. The simulations all feature the temperature noise strength $\sigma_T = 2$, whereas the strength of the salinity noise field $\sigma_S$ is varied (see panel legends indicating $\sigma = (\sigma_T, \sigma_S)$). 
}
\end{figure}

\begin{figure}[!htb]
\includegraphics[width=0.99\textwidth]{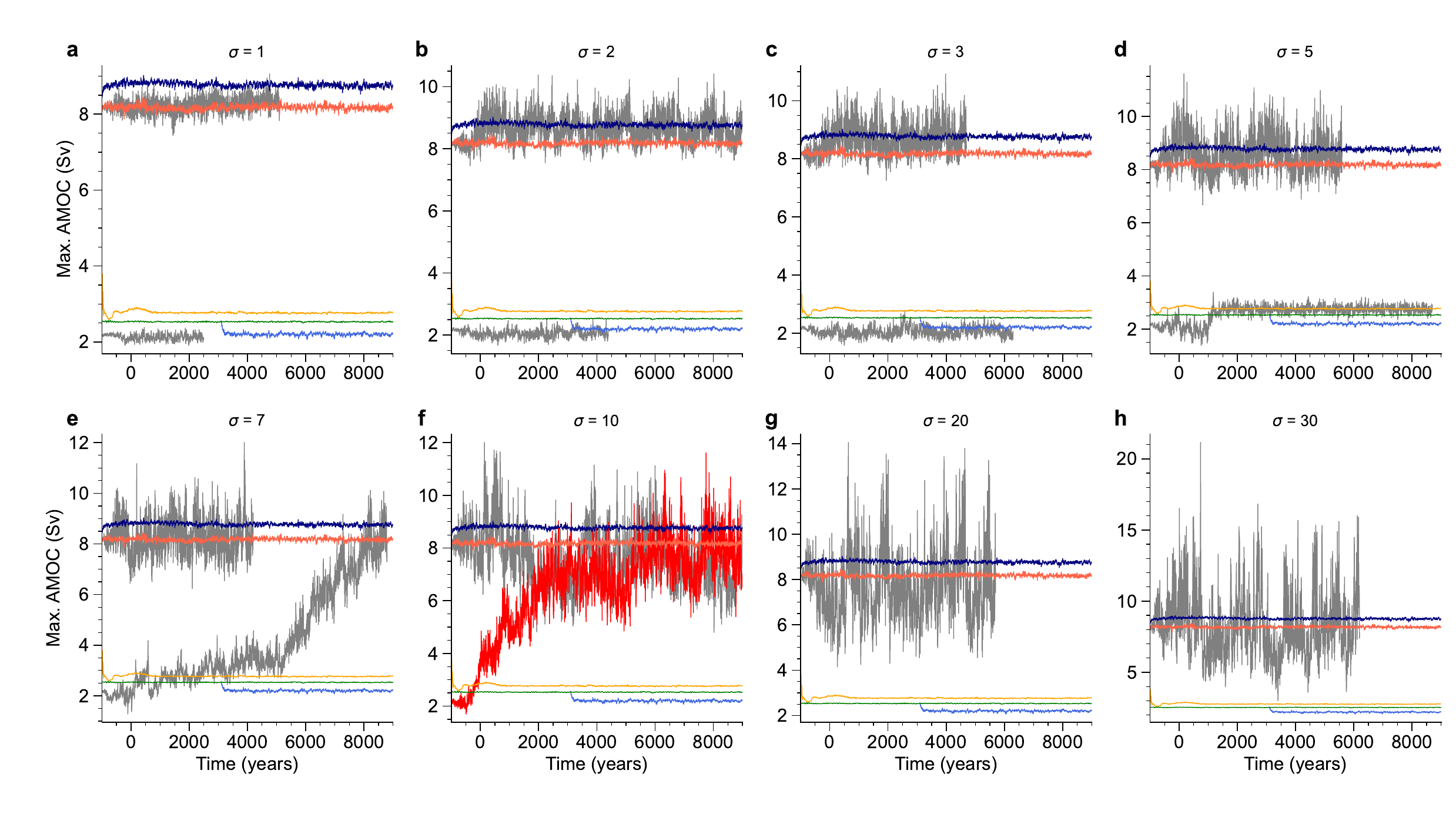}
\caption{\label{fig:salinity_noise_timeseries} 
Simulations with the Veros ocean model forced by a noise model with salinity component only at the control parameter value $F=0.3472$~Sv. Note here that, as opposed to the simulations with temperature-only forcing (Fig.~\ref{fig:timeseries_noise_temp_lower}), the noise strength is ramped up over  the first 1000 years only. 
The panels are ordered according to increasing strength of the salinity noise field. Steady-state simulations on the deterministic attractors are colored, and the noisy trajectories are in gray, except for one realization at $\sigma_S=10$ (panel {\bf f}) that performs a transitions from the AMOC OFF to the AMOC ON regime, which is shown in red. We have not seen any clear noise-induced AMOC collapse using the salinity-only forcing. If the noise strength is further increased, the numerical model eventually becomes unstable (not shown here). 
}
\end{figure}

\begin{figure}[!htb]
\includegraphics[width=0.85\textwidth]{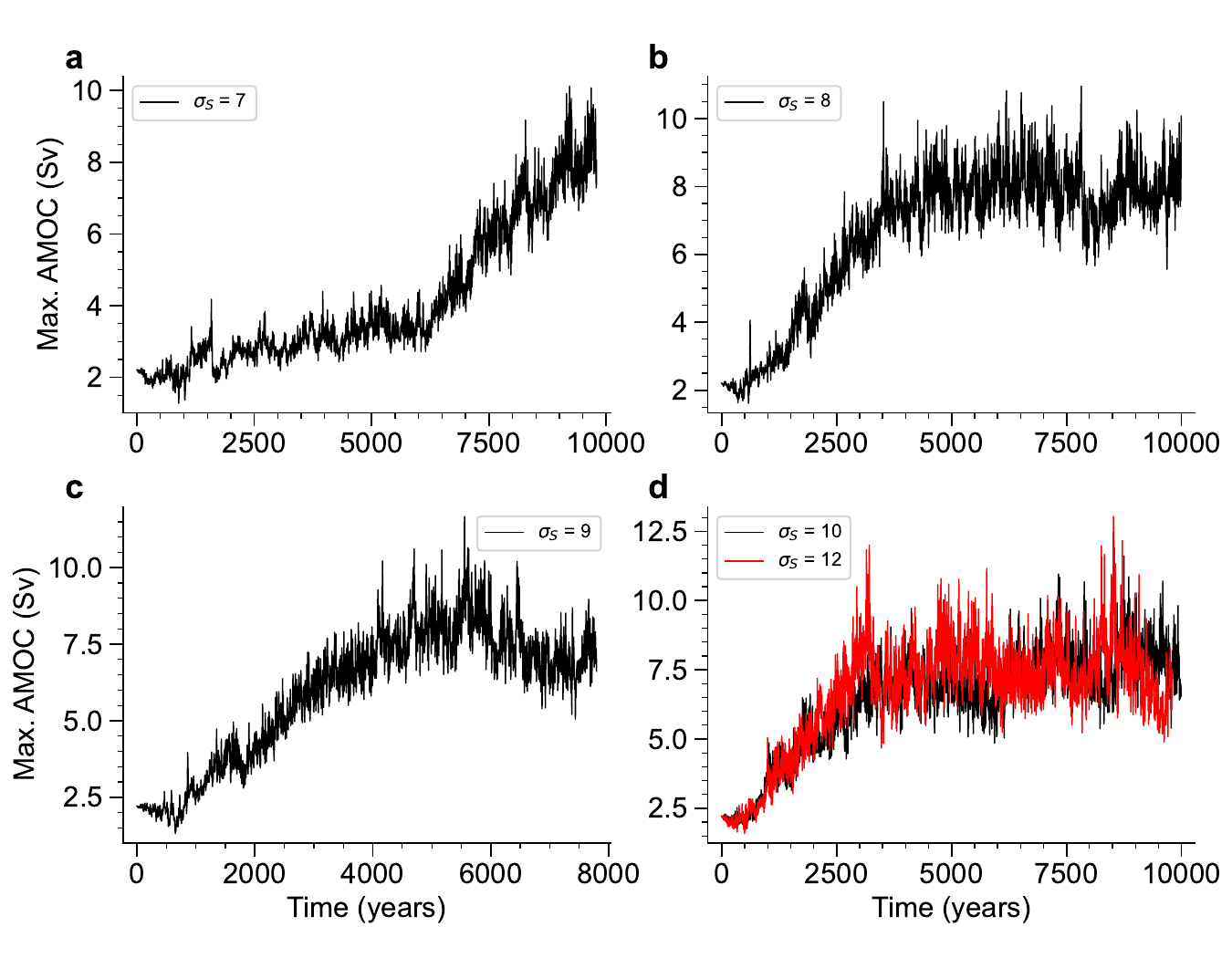}
\caption{\label{fig:salinity_noise_timeseries_lower} 
Simulations with the Veros ocean model forced by a noise model with salinity component only at $F=0.3472$~Sv. The panels are ordered according to increasing strength of the salinity noise field. The noise strength is ramped up linearly during the first 1000 years, and kept constant thereafter. Shown are simulations initialized in an AMOC OFF state. 
}
\end{figure}

When increasing $\sigma_T$ or $\sigma_S$, transitions from the OFF regime to the ON regime appear (Fig.~\ref{fig:veros_transitions_all}a-c and Fig.~\ref{fig:timeseries_mixed_lower}-~\ref{fig:salinity_noise_timeseries_lower}). 
The transitions are not abrupt in the context of anthropogenic climate change, taking at least 1,000 years from when there is a first increase in AMOC strength until the system reaches a state with a vigorous AMOC. As discussed further below, they are, however, initiated by an abrupt convective event. At these noise levels, none of our simulations show a noise-induced transition back to the OFF regime. 

\begin{figure}[!htb]
\includegraphics[width=0.85\textwidth]{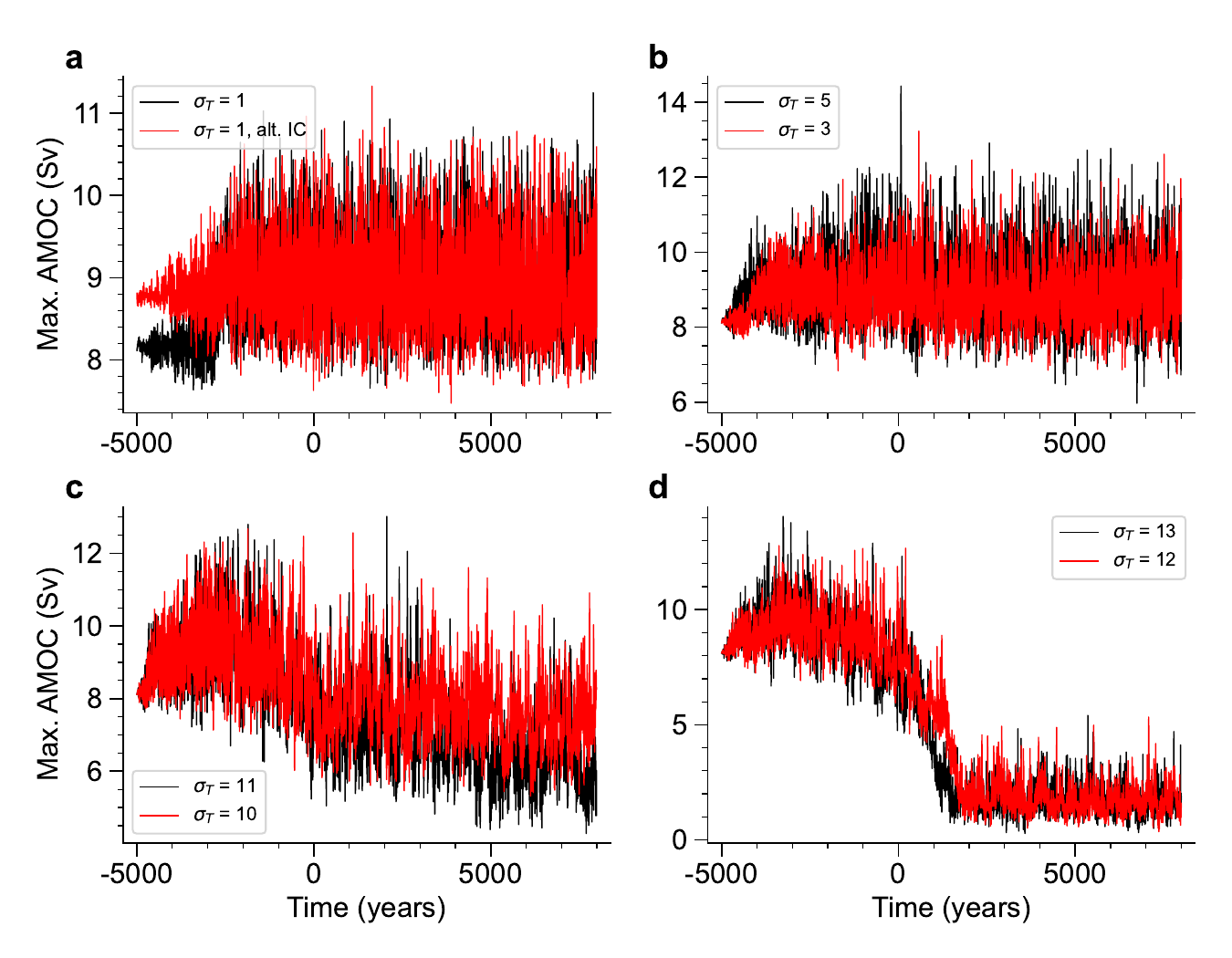}
\caption{\label{fig:timeseries_noise_temp_upper} 
Simulations of the Veros model under temperature-only noise forcing at the control parameter value $F=0.3472$~Sv. Shown are simulations initilized in different AMOC ON states, and for different noise strengths $\sigma_T$ (see figure legends). The noise strength is ramped up linearly during the first 5000 years, and kept constant after year 0 in the figures. 
}
\end{figure}

\begin{figure}[!htb]
\includegraphics[width=0.85\textwidth]{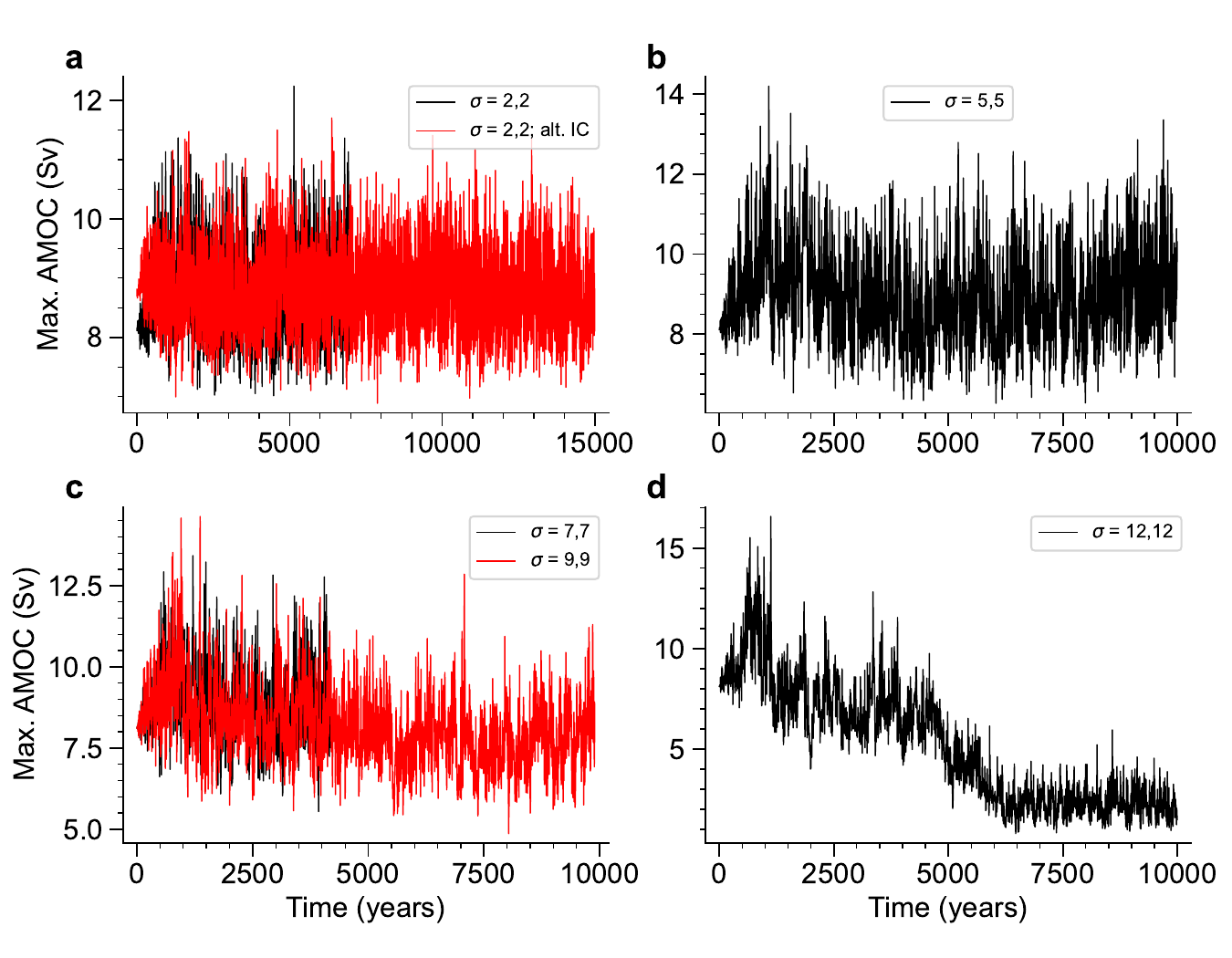}
\caption{\label{fig:timeseries_mixed_upper} 
Simulations initialized in different AMOC ON states with the Veros ocean model forced by a noise model with both temperature and salinity components at $F=0.3472$~Sv. Both noise strengths are ramped up linearly during the first 1000 years, and kept constant thereafter. The varying noise levels are given in the panel legends, as $\sigma = (\sigma_T, \sigma_S)$. 
}
\end{figure}

When further increasing the noise level we eventually find transitions where the AMOC collapses (Fig.~\ref{fig:veros_transitions_all}d-f and Fig.~\ref{fig:timeseries_noise_temp_upper}-\ref{fig:timeseries_mixed_upper}), except in the salinity-only case (Fig.~\ref{fig:salinity_noise_timeseries}).
Again, the transitions are not abrupt and we have not observed transitions in the other direction back to an AMOC ON regime. In fact, initializing the system in the OFF regime and ramping up the noise strength linearly, there is first a transition from OFF to ON during the ramp up, after which the system transitions back to the OFF regime as the noise has stabilized at higher levels (Fig.~S7d). 
This non-monotonic noise influence on the dynamics for fixed control parameter
is counterintuitive, since one would not assume the additive noise to change the underlying dynamical landscape, i.e., the height of potential barriers between attractors. If one of two competing states has a much lower residence time at a given noise level, one would expect this to remain so as the noise strength is increased. Instead, it appears that the OFF regime is less stable for lower noise, while for larger noise the ON state decreases in stability and eventually becomes less stable compared to OFF.

%%% 
The physical properties of the edge state might give clues as to why the ON state is more difficult to collapse. If a noise-induced transition needs to pass by the edge state, water that is both fresher and colder compared to both ON and OFF needs to be brought to the deep Atlantic. It seems plausible that this is more easily achieved by a rapid, noise-induced {\it onset} of convection, which can relatively quickly flush the deep Atlantic with water of different properties, as opposed to a temporary {\it shutoff} of convection due to a surface noise anomaly, which instead leaves the deep ocean properties intact at their current state. % until changed slowly by other processes.
A more detailed study is needed to unravel the physical mechanisms that explain the model's non-monotonic response to noise forcing.

%General comments on the effect of noise on the dynamics and realism in terms of the 
%coupled ocean-atmosphere system. 
Further work is needed to clarify by what mechanisms the noise affects the ocean circulation, and to draw conclusions with regards to the coupled ocean-atmosphere system. The simulations may suggest that the high multistability in the oceanic dynamical landscape (Fig.~8) would be lost in the presence of atmospheric fluctuations. However, while we derived the atmospheric forcing from reanalysis data, we cannot be sure what noise strength and structure should be considered realistic due to the boundary conditions of the ocean model. The temperature noise field at $\sigma_T = 1$ may represent the effect of atmospheric fluctuations on the ocean somewhat closely, since atmospheric temperature anomalies directly influence the surface ocean, which is inherited by the temperature boundary conditions. %refer to appendix. 
But a correct scaling of the salinity noise forcing remains unknown. First, this is because the noise field is estimated from the variability of sea surface salinity and not evaporation and precipitation. Second, the salinity boundary condition is implemented by a relaxation towards the climatology of sea surface salinity, including the noisy perturbation thereof. This relaxation involves a poorly constrained, parameterized time scale, which directly influences the size of the fluctuations induced by noise of a given strength. 

\section{Rare-event sampling with the Giardin\`a-Kurchan-Lecomte-Tailleur algorithm}\label{AppREA}

The general idea of the Giardin\`a-Kurchan-Lecomte-Tailleur algorithm \cite{GIA11} is to propagate an ensemble of trajectories forwards in time, iteratively cloning or killing them according to a metric of `goodness'. This way, the ensemble will gradually get `better' at sampling a particular rare event of interest. What follows is a detailed description of the algorithm.

Let us assume we are interested in a stochastic system $\text{d} X = f(X, t) \text{d} t + \sigma(X, t) \text{d} B(t)$ with $X \in \mathbb{R}^d$, and an observable $O(X(t)) \in \mathbb{R}$. We can then identify a region $\mathcal{A} \subset \mathbb{R}^d$ of the phase space that is not often visited by the system, and where the observable $O$ features large anomalies compared to it expectation value over the invariant measure.
We then define a score function $V(\{X\}, t)$ that will work as a proxy for the probability of trajectory $\{X\}$ reaching $\mathcal{A}$ \cite{CHR20}.

In our case, the observable is the AMOC strength itself $O = \mathrm{AMOC}$, possibly time-averaged. For simplicity here, we will use as score function the observable itself, $V({X}, t) = -\mathrm{AMOC}(X(t))$, which is a quite common choice in the literature (see for instance \cite{GKTL-Ly,Ragone24}). This choice makes the assumption of persistence to approximate probabilities, namely: ``the AMOC is more likely to collapse in the future if it is already weak right now''.

After we have defined the score function, we initialize $N$ ensemble members with trajectories $\{X^{(0)}_j\}, \, j = 1, \ldots, N$, each ending at time $t^{(0)}_j$. For simplicity and without loss of generality, we can assume $t^{(0)}_j = t_0 = 0$, and, since the dynamics is Markovian, we can ignore everything that happens before $t_0$.
Finally, to complete the initialization of our ensemble we compute the initial scores $V_j^{(0)} = V(\{X^{(0)}_j\})$, and assign to each trajectory an unbiasing weight $\pi_j^{(0)} = 1$. This weight will track the likelihood to observe trajectory $j$ when running the climate model with no REA, and it will be crucial to have access to the unbiased probabilities of events sampled by the biased trajectories.

Once we have our initial ensemble, we need to select a resampling time $\tau$ and a selection strength $k$. Then, for each iteration $i = 1, \ldots, I$, we will perform the following steps:

\begin{enumerate}
    \item Extend each trajectory integrating forward in time for one resampling step $\tau$, obtaining $\{\tilde{X}^{(i)}_j\}, \, j = 1, \ldots, N$ that run from $t_0$ to $t_i = i\tau$. Then compute for each trajectory its score function at the end of the resampling step $\tilde{V}_j^{(i)} = V(\{\tilde{X}_j^{i}\}, t_i)$.

    \item Compute for each trajectory a weight according to the improvement of its score in the last resampling step $\tilde{w}_j^{(i)} = \exp\left( k\left( \tilde{V}_j^{(i)} - V_j^{(i-1)} \right) \right) $, and then normalize the weights so that they sum up to $N$:
    $$w_j^{(i)} = \frac{1}{Z^{(i)}} \tilde{w}_j^{(i)}, \quad Z^{(i)} = \frac{1}{N}\sum_{j=1}^N \tilde{w}_j^{(i)}$$

    \item Assign to each trajectory a random number of clones $m_j^{(i)}$ with expectation $w_j^{(i)}$. This is achieved by drawing $N$ random numbers $u_j^{(i)} \sim \mathcal{U}(0, 1)$, uniformly distributed between $0$ and $1$, and then setting $m_j^{(i)} = \lfloor w_j^{(i)} + u_j^{(i)} \rfloor$. Trajectories with $m_j^{(i)} = 0$ will be discarded. In case $\Delta N^{(i)} = \sum_{j=1}^N m_j^{(i)} - N \neq 0$, $|\Delta N^{(i)}|$ additional trajectories are either cloned or killed to ensure that the ensemble remains of size $N$.

    \item Update the trajectories in the ensemble by creating the parent mapping function $p^{(i)}(j)$, which links each resampled trajectory $j$ to the one it was cloned from. Then, set
    $$ \{X^{(i)}_j\} = \{\tilde{X}^{(i)}_{p^{(i)}(j)}\} , \quad V_j^{(i)} = \tilde{V}^{(i)}_{p^{(i)}(j)} \quad \text{and} \quad \pi_j^{(i)} = \frac{\pi_{p^{(i)}(j)}^{(i-1)}}{w^{(i)}_{p^{(i)}(j)}} $$
\end{enumerate}

In the original paper by Giardina et al. \cite{GIA11}, the weights in point 2 are computed not as the improvement of the score function but as the score function itself at the end of the resampling step. However, in \cite{GAR06}, the authors suggest using the improvement of the score function.
% move to later.

We implemented the algorithm both for a fully deterministic ocean-only Veros configuration, as well as for a configuration with a temperature noise component. In the deterministic case, when cloning trajectories (step 4), we add a small random perturbation to the temperature $T$ and salinity $S$ fields, which allows them to diverge, thanks to the chaotic nature of the flow. The perturbation is defined as
\begin{equation}\label{eq:veros:perturbation}
    \begin{aligned}
        T &\mapsto T + T\epsilon_T \eta \\
        S &\mapsto S + \epsilon_S \eta ,
    \end{aligned}
\end{equation}
where $\eta$ is sampled from a normal distribution with mean 0 and variance 1, independently for each field and grid point. The noise amplitudes are chosen to be $\epsilon_T = 0.001$ and $\epsilon_S = 0.002 \,\, \text{g} \, \text{kg}^{-1}$. Typical salinity values are around $34.5 \,\, \text{g} \, \text{kg}^{-1}$ with fluctuations of the order of $0.2 \,\, \text{g} \, \text{kg}^{-1}$, while temperature values have a mean of 5 deg C with a standard deviation of the order of 10 deg C, though the temperature distribution is strongly skewed due to the heating from the top. Thus, in Eq.~\ref{eq:veros:perturbation} we employ a perturbation that is multiplicative in temperature, such that the deep ocean temperatures that are close to 0 degrees are perturbed less than the warm surface ocean.
% The perturbations in salinity are additive, since all values are very close on an absolute scale.

Now, once we reach the final iteration of the algorithm, we can use the unbiasing weights $\pi_j^{(I)}$ to compute the proper average of any observable $U(\{X\})$ over our biased ensemble:
\begin{equation}\label{eq:veros:unbiased-averaging}
    \bar{U}_N = \frac{1}{N}\sum_{j=1}^N \pi_j^{(I)} U(\{X^{(I)}_j\}) \sim_{N \to \infty} \mathbb{E} \left[ U(\{X\}) \right],
\end{equation}
where $\mathbb{E}$ is the expectation over the stationary measure. Since the trajectories are not independent, the central limit theorem doesn't apply, and thus the typical error can be larger than $1/\sqrt{N}$.

% remove?
% In practice, convergence will be faster for observables closely linked to the score function $V$ used for biasing the ensemble, while for others it may be more efficient to run a simple unbiased Monte Carlo estimation.

% Note that if we are interested in computing probabilities rather than averages, we can simply think of a probability as the expectation of an indicator function:
% \begin{equation}
%     \mathbb{P} \left(U(\{X\}) > u \right) \equiv \mathbb{E} \left( \mathbb{I}_{U(\{X\}) > u}  \right)
% \end{equation}
Then, the probability that the AMOC is weaker than a given threshold $a$ is simply given by
\begin{equation}\label{eq:veros:unbiased-probabilities}
    \mathbb{P} (O(X(T)) \leq a) \equiv \mathbb{E} \left( \mathbb{I}_{O(X(T)) \leq a} \right) \approx \frac{1}{N} \sum_{j=1}^{n} \pi_j^{(I)} \mathbb{I}_{O(X_j^{(I)}(T)) \leq a} .
\end{equation}

In the main text, we give results for $a$ chosen as the median of the current distribution of the ensemble, as this uses most of the ensemble members and thus has low statistical uncertainty. Given ordered realizations in the ensemble at iteration $i$, this yields the unbiased probability $p_{1/2}^{(i)}$ of being below the median $a^{(i)}_{1/2}$:

\begin{gather}\label{eq:veros:p0.5}
        p_{1/2}^{(i)} = \mathbb{P}(O(X(i\tau)) < a^{(i)}_{1/2}) = \frac{1}{N} \sum_{l=1}^{N/2} \pi_{j_l}^{(i)}, \quad O(X_{j_1}^{(i)}) \leq \ldots \leq O(X_{j_{l}}^{(i)}) \leq \ldots \leq O(X_{j_{N}}^{(i)}) .
\end{gather}

The algorithm has four `hyperparameters': the ensemble size $N$, the resampling time $\tau$, the selection strength $k$ and the score function $V$.

As already discussed, we chose to use as score function the amoc strength itself, and in particular its 5-year average for the deterministic model and its yearly average for the model with temperature noise.

The resampling time should be of the order of the Lyapunov time \cite{WOU16}.
Long enough to allow clones of the same ensemble member to separate sufficiently, but not so long that they relax back to the attractor. Namely, we don't want segments of trajectories to completely lose memory of their initial condition. By analyzing the autocorrelation function of the AMOC strength we decided to use $\tau = 50$~yr for the deterministic model and $\tau = 1$~yr for the model with atmospheric noise.

Concerning the selection strength $k$, it should be chosen based on how extreme the event of interest is: the more extreme the event, the higher the selection strength. \cite{PRI24} suggests an empirical way to get the right order of magnitude for $k$: if over the integration period $T = I\tau$ the AMOC strength fluctuates with a standard deviation $\sigma$, then to push the system $n$ standard deviations from the mean, we should set $k \sim n/\sigma$. In our case this gives a suggested $k$ of the order of 10~Sv$^{-1}$.

Finally, the choice of the ensemble size is a matter of compromise between accuracy and computational cost, and in this work we chose to use $N = 50$ trajectories.

The practical implementation of the GKTL algorithm and its application to the Veros model is available at \url{https://github.com/AlessandroLovo/REA-Veros}.

\bibliographystyle{RS}
\bibliography{refs}

\newpage

\begingroup
\renewcommand{\arraystretch}{0.8} % Default value: 1
\begin{table}[h]
  \caption{Parameter values used for the five-box model, adapted from the {\tt FAMOUSA1xCO2} calibration in \cite{WOO19}. $\alpha = 0.12$ (thermal coefficient) and $\beta = 790$ (haline coefficient) define a linear equation of state for the density of sea water. $V_i$ is box volume,  $F_i$ the freshwater fluxes, $T$ are temperatures, $K_i$ are wind fluxes and $A_i$ determine the distribution of freshwater forcing. $\eta$ is a mixing parameter between the S and B boxes, $\gamma$ determines the proportion of water which takes the cold-water path, $\lambda$ and $\mu$ are constants. Subscripts indicate box labels, $i \in \{ N, T, S, IP, B \}$, and $0$ indicates a global value.}
\label{tab:params}
\centering
%\begin{tabular}{| l || c |}
\begin{tabular}{ l | c || l | c }
%\hline
Parameter  & Value & Parameter & Value\\
\hline
\hline
$V_{\rm{N}} (m^3)$ &  3.683 $\times 10^{16}$ & $F_{\rm{N}} (m^3  s^{-1})$ &  0.375 $\times 10^{6}$\\ 
$V_{\rm{T}} (m^3)$  & 5.151 $\times 10^{16}$ & $F_{\rm{T}} (m^3  s^{-1})$ &  -0.723 $\times 10^{6}$\\ 
$V_{\rm{S}} (m^3)$ &  10.28 $\times 10^{16}$ & $F_{\rm{S}} (m^3  s^{-1})$ &  1.014 $\times 10^{6}$\\
$V_{\rm{IP}} (m^3)$  & 21.29 $\times 10^{16}$ & $F_{\rm{IP}} (m^3  s^{-1})$ &  -0.666 $\times 10^{6}$\\
$V_{\rm{B}} (m^3)$  & 88.12 $\times 10^{16}$ &  $F_{\rm{B}} (m^3  s^{-1})$  &  0  \\ 
\hline
$A_{\rm{N}}$  & 0.194 & $\eta (m^3  s^{-1})$ & 66.061 $\times 10^{6}$\\
$A_{\rm{T}}$ & 0.597 & $\gamma$  & 0.1 \\
$A_{\rm{S}}$ & -0.226 & $\lambda (m^6 kg^{-1} s^{-1})$ & 2.66 $\times 10^{7}$\\
$A_{\rm{IP}}$  & -0.565 & $\mu (^o C m^{-3}s)$ & 7.0 $\times 10^{-8}$\\ 
\hline
$K_{\rm{N}} (m^3  s^{-1})$ & 5.439 $\times 10^{6}$& $T_{\rm{S}} (^o C)$ &  5.571 \\
$K_{\rm{S}} (m^3  s^{-1})$  & 3.760 $\times 10^{6}$& $T_{\rm{0}} (^o C)$  & 3.26 \\ 
$K_{\rm{IP}} (m^3  s^{-1})$  & 89.778 $\times 10^{6}$& & \\
%\hline
\end{tabular}
\end{table}
\endgroup

\begin{figure}[!htb]
\includegraphics[width=0.85\textwidth]{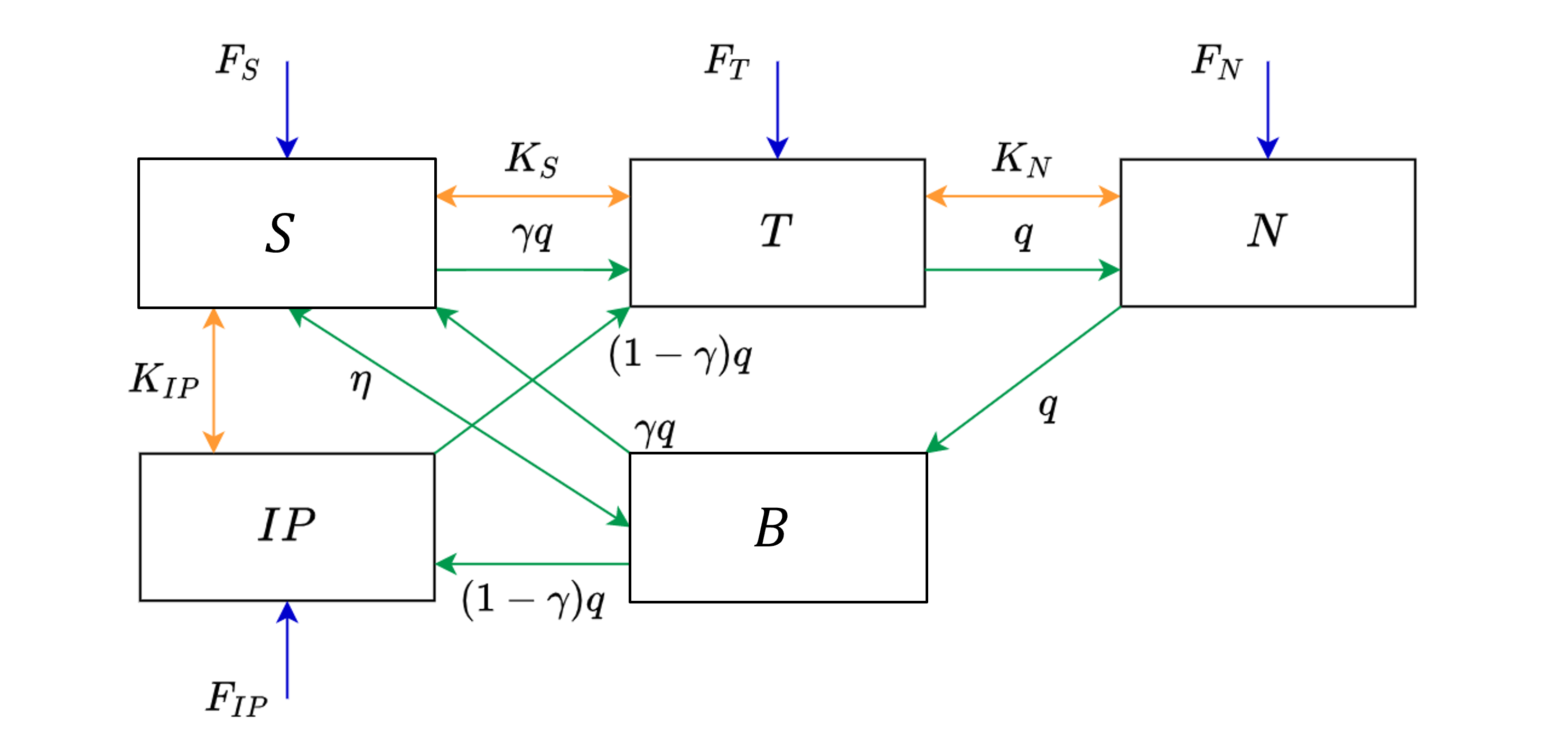}
\caption{\label{fig:FiveBoxModel} 
Schematic of the conceptual 5-box AMOC model by Wood et al. (2019) \cite{WOO19}, where the coupling of the different boxes by the overturning flow $q$, the wind-driven circulation $K_X$ and mixing $\eta$ are given by the arrows. While the wind-driven coupling is bidirectional, the coupling by the overturning flow is uni-directional and all arrows are reversed in the AMOC OFF state (negative $q$). This is reflected in the non-smooth dependence on $q$ of the corresponding differential equations (see main text). Shown here is the direction of coupling for the AMOC ON state with positive $q$. Also shown is the surface forcing $F_X$, acting on all boxes except the bottom waters $B$. 
}
\end{figure}

\begin{figure}[!htb]
\includegraphics[width=0.65\textwidth]{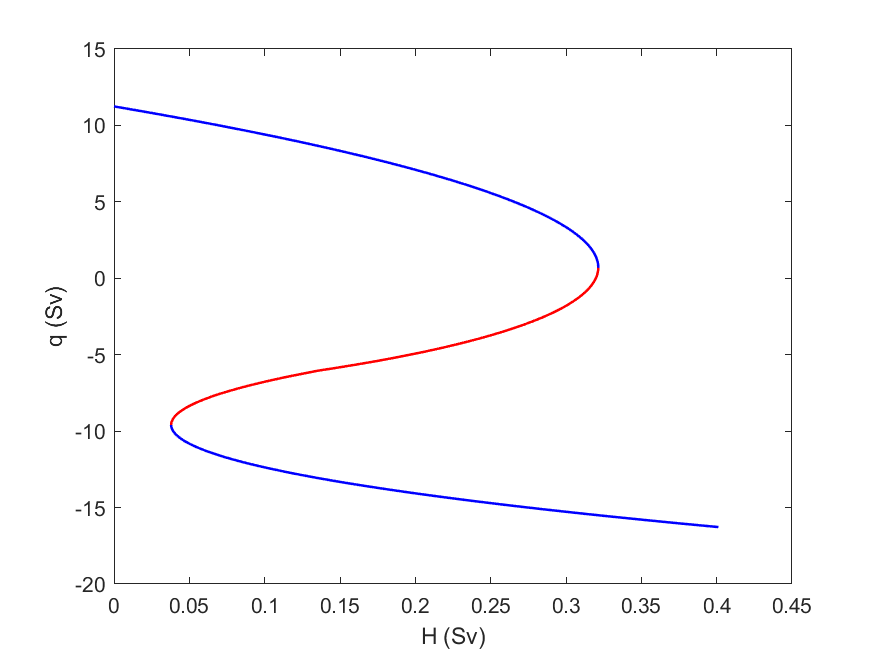}
\caption{\label{fig:FiveBoxBifQ} 
Bifurcation diagram of the 5-box AMOC model, projected onto the observable $q$ (see main text). The red line is the branch of unstable equilibria. 
}
\end{figure}

\begin{figure}[!htb]
\includegraphics[width=0.85\textwidth]{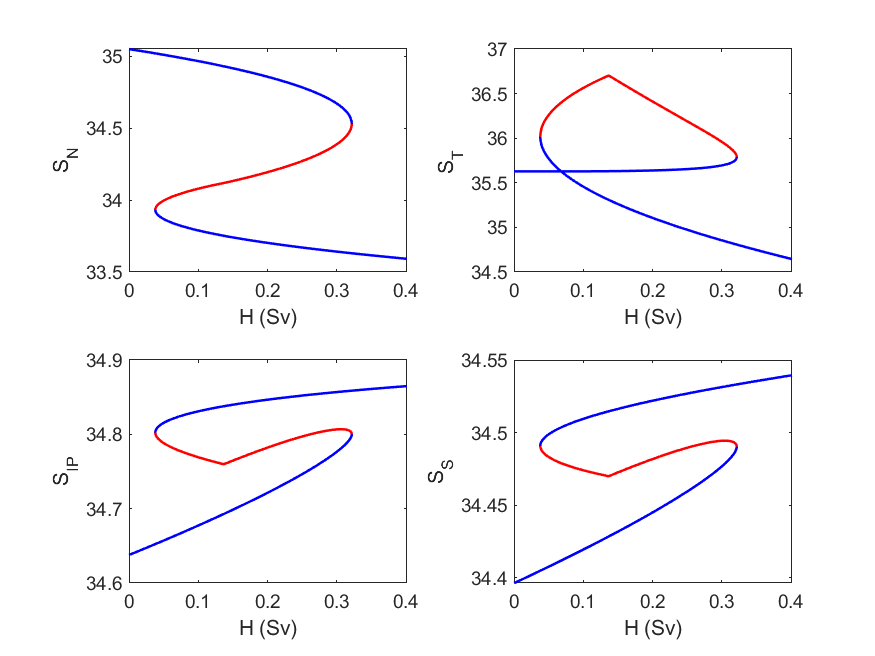}
\caption{\label{fig:FiveBoxBif} 
Corresponding bifurcation diagrams projected onto individual state variables. 
}
\end{figure}

\begin{figure}[!htb]
\includegraphics[width=0.7\textwidth]{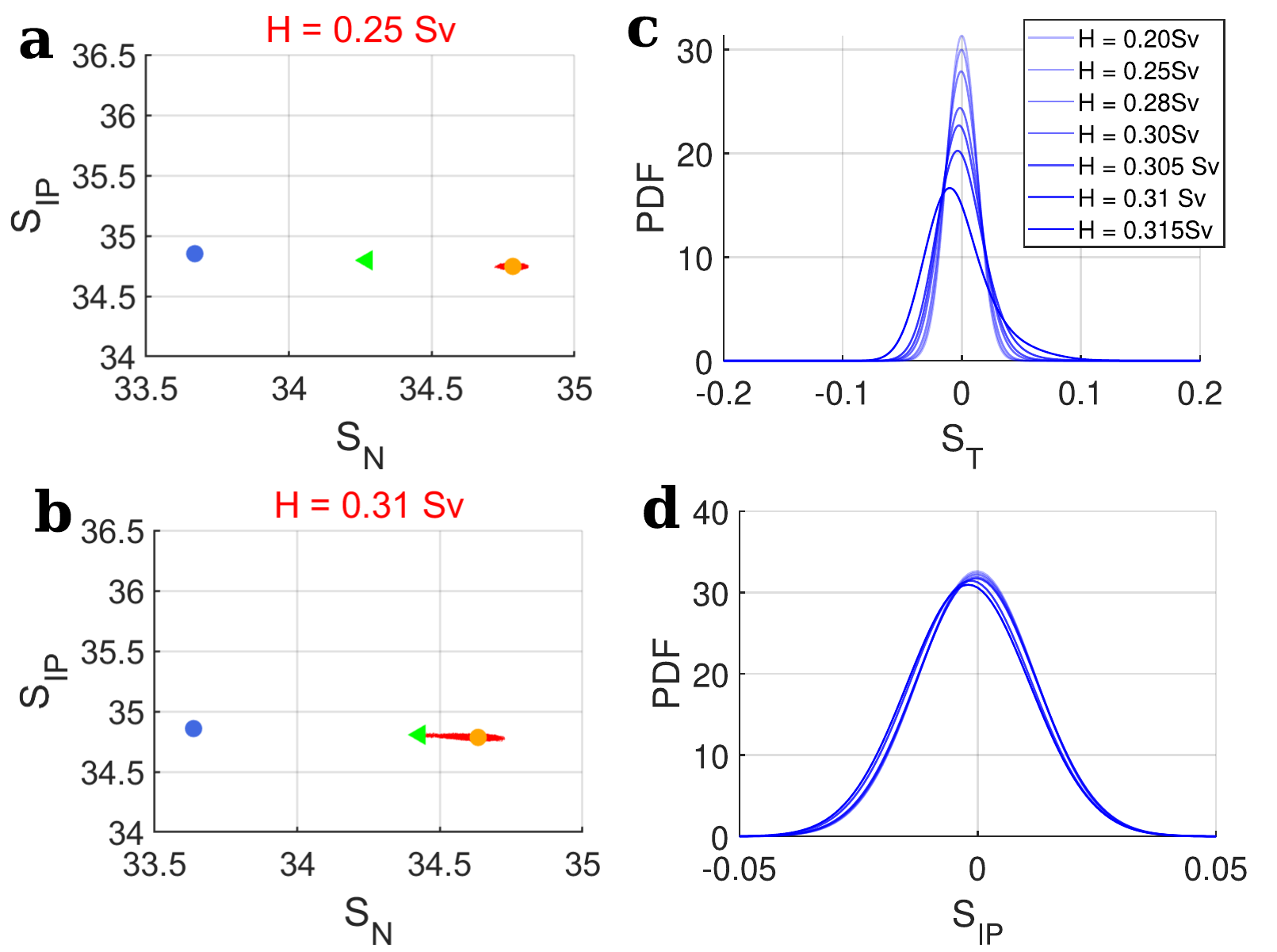}
\caption{\label{fig:5box_pdf} 
{\bf a,b}
Phase space projection of the conceptual 5-box AMOC model onto $S_N$ and $S_{IP}$ for the two control parameter values $H=0.25$~Sv ({\bf a}) and $H=0.305$~Sv ({\bf a}). The round dots are the stable fixed points, and the green triangle is the edge state. In red is an ensemble of noisy simulations (see main text). The values in the stable fixed points and the edge state for $S_{IP}$ differ only very slightly, and they do not change much as the bifurcation is approached (increasing $H$). Thus, a projection of the variability onto $S_{IP}$ does not lead to any early-warning signals via increases in variance or autocorrelation (see panel {\bf d}). 
{\bf c,d}) 
Distributions of the fluctuations around the ON fixed point in an ensemble of simulations of the 5-box conceptual AMOC model, projected onto the state variables $S_T$ ({\bf c}) and $S_{IP}$ ({\bf d}). Shown are distributions for seven values of the control parameter $H$, increasing towards the bifurcation. In panel {\bf d} the different distributions lie almost on top of each other, showing that in terms of this variable there is no change in variability as the bifurcation is approached, and thus no early-warning based on critical slowing down. 
}
\end{figure}

\begin{figure}[!htb]
\includegraphics[width=0.99\textwidth]{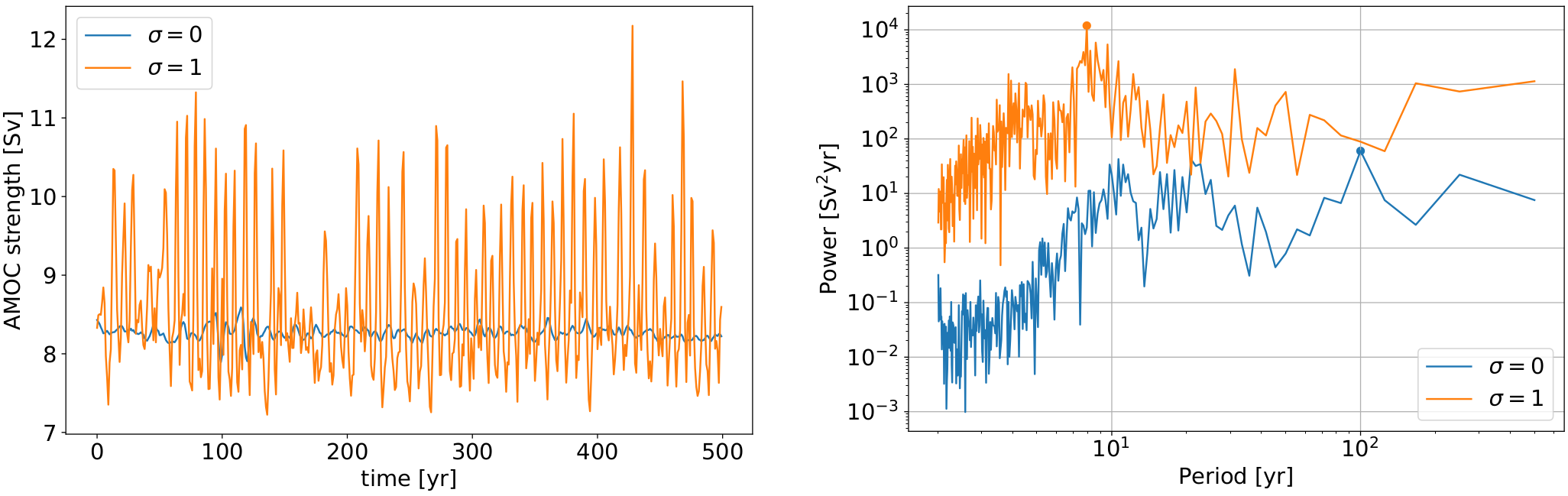}
\caption{\label{fig:rare_event_noise_control} 
Time series (left) and power spectra (right) of the yearly averaged AMOC strength for 500-year long control runs at $F=0.3515$ without ($\sigma_T=0$) and with ($\sigma_T=1$) Atlantic SST noise. While the noise-free run is dominated by a centennial timescale, with secondary peaks at 10 and 25 years, the run with noise exhibits a very strong 8-year oscillation, while showing relatively less pronounced centennial activity.
}
\end{figure}

\begin{figure}[!htb]
\includegraphics[width=0.8\textwidth]{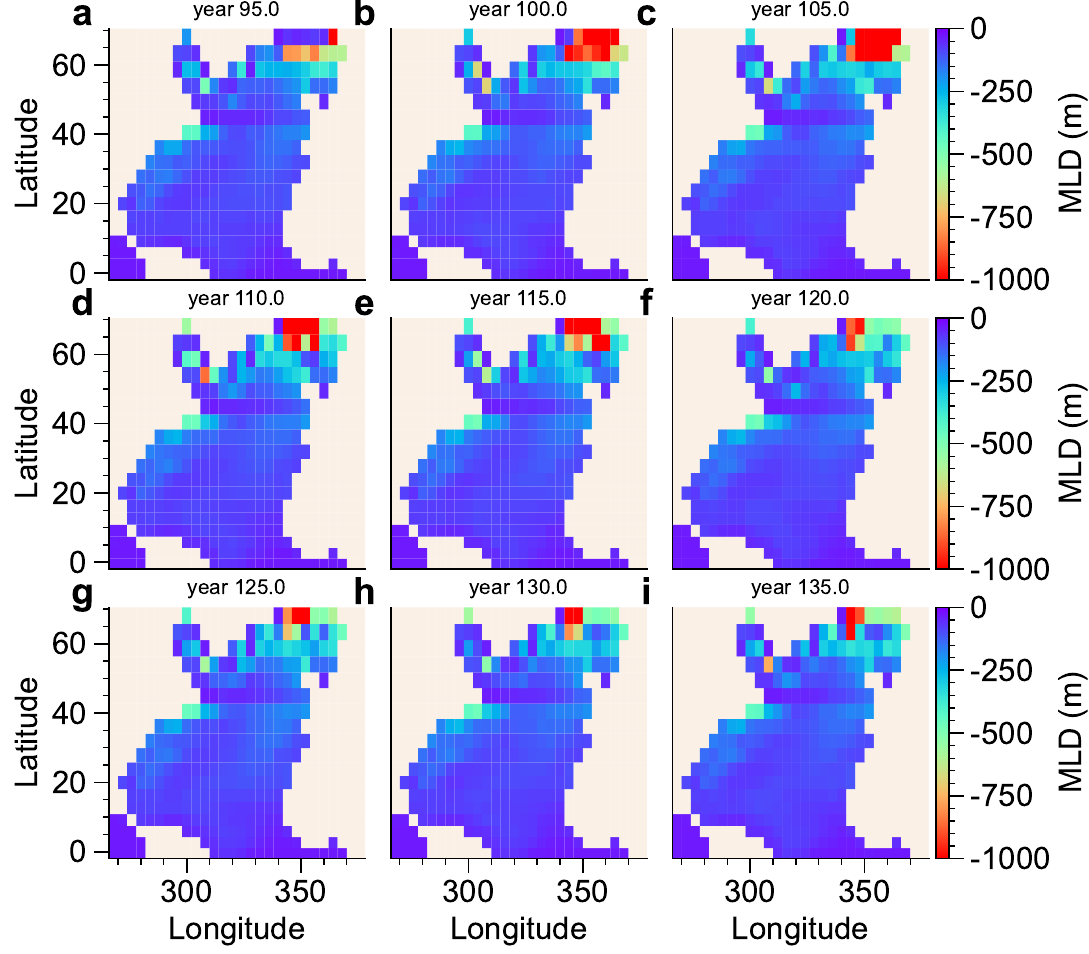}
\caption{\label{fig:convection_event} 
Abrupt convective event initiating the noise-induced transition from an AMOC OFF to an ON state in the Veros model, corresponding to the simulation shown in Fig.~10b of the main text. Shown are maps of the 5-year mean winter mixed layer depth in the North Atlantic. Time is shifted such that the convection event happens around the year 100. 
}
\end{figure}

\begin{figure}[!htb]
\includegraphics[width=0.99\textwidth]{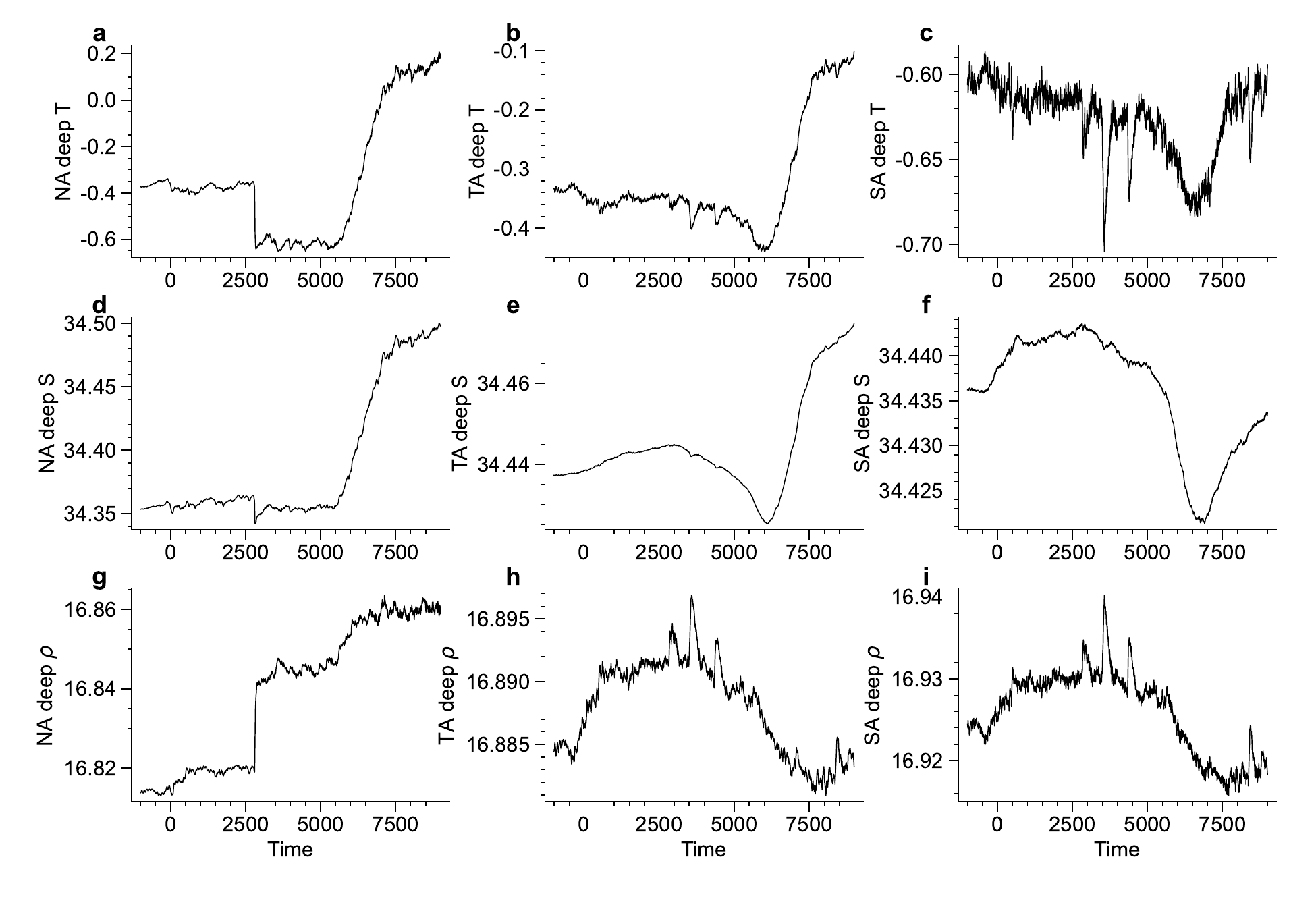}
\caption{\label{fig:timeseries_resurgence_deepocean_n2S3} 
Time series of different observables for a noise-induced transition from an AMOC OFF to an ON state in the Veros model, corresponding to the simulation shown in Fig.~10b of the main text. The observables shown are the mean deep ocean (below 1000 meter) temperature, salinity and density in three sectors of the Atlantic. 
}
\end{figure}

\begin{figure}[!htb]
\includegraphics[width=0.99\textwidth]{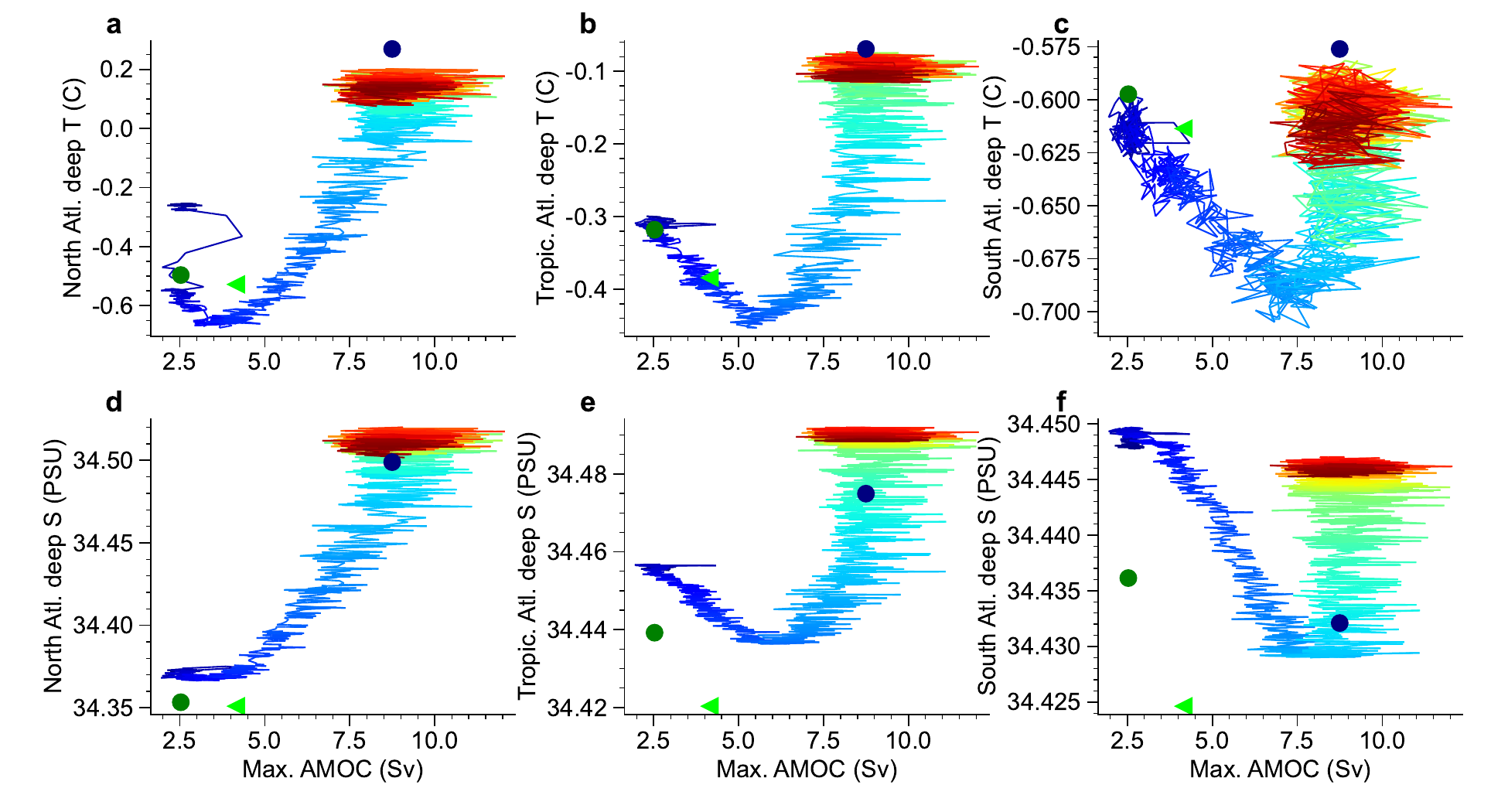}
\caption{\label{fig:resurgence_f81nrLy3} 
Noise-induced transition from an AMOC OFF (dark green dot) to an ON state (blue dot) in the Veros model, using a noise model with temperature-only forcing, and the noise strength $\sigma_T = 3$. The green triangle marks the location of the edge state. 
Shown are trajectories projected onto two observables, which are the AMOC strength on the horizontal axes and the deep ocean salinity and temperature in different sectors of the Atlantic. The color code indicates time.
}
\end{figure}

\begin{figure}[!htb]
\includegraphics[width=0.99\textwidth]{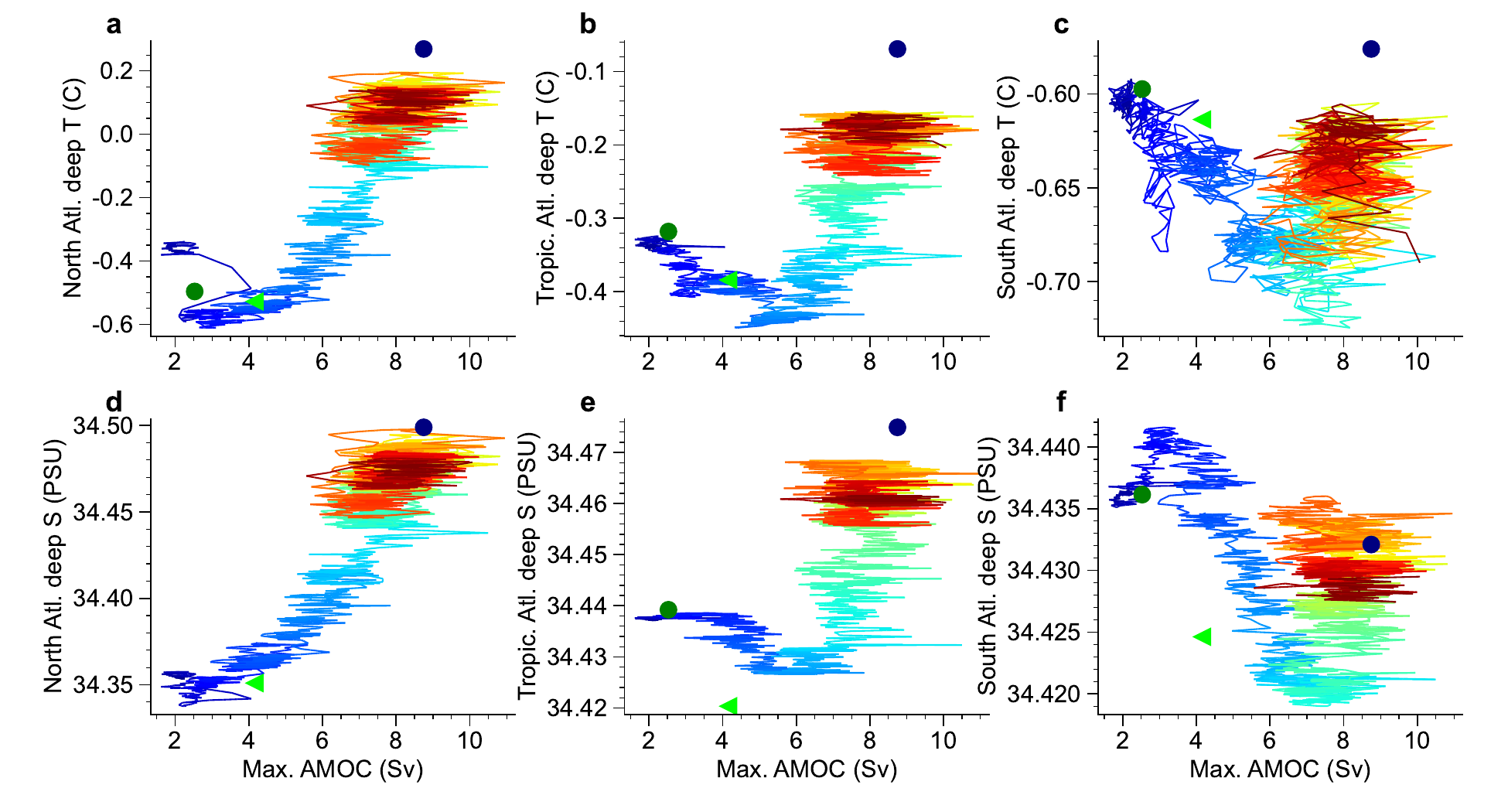}
\caption{\label{fig:edge_traj_n8} 
Same as in Fig.~S18, but for a noise-induced transition using a noise model with salinity-only forcing, and the noise strength $\sigma_S = 8$.
}
\end{figure}

\begin{figure}[!htb]
\includegraphics[width=0.99\textwidth]{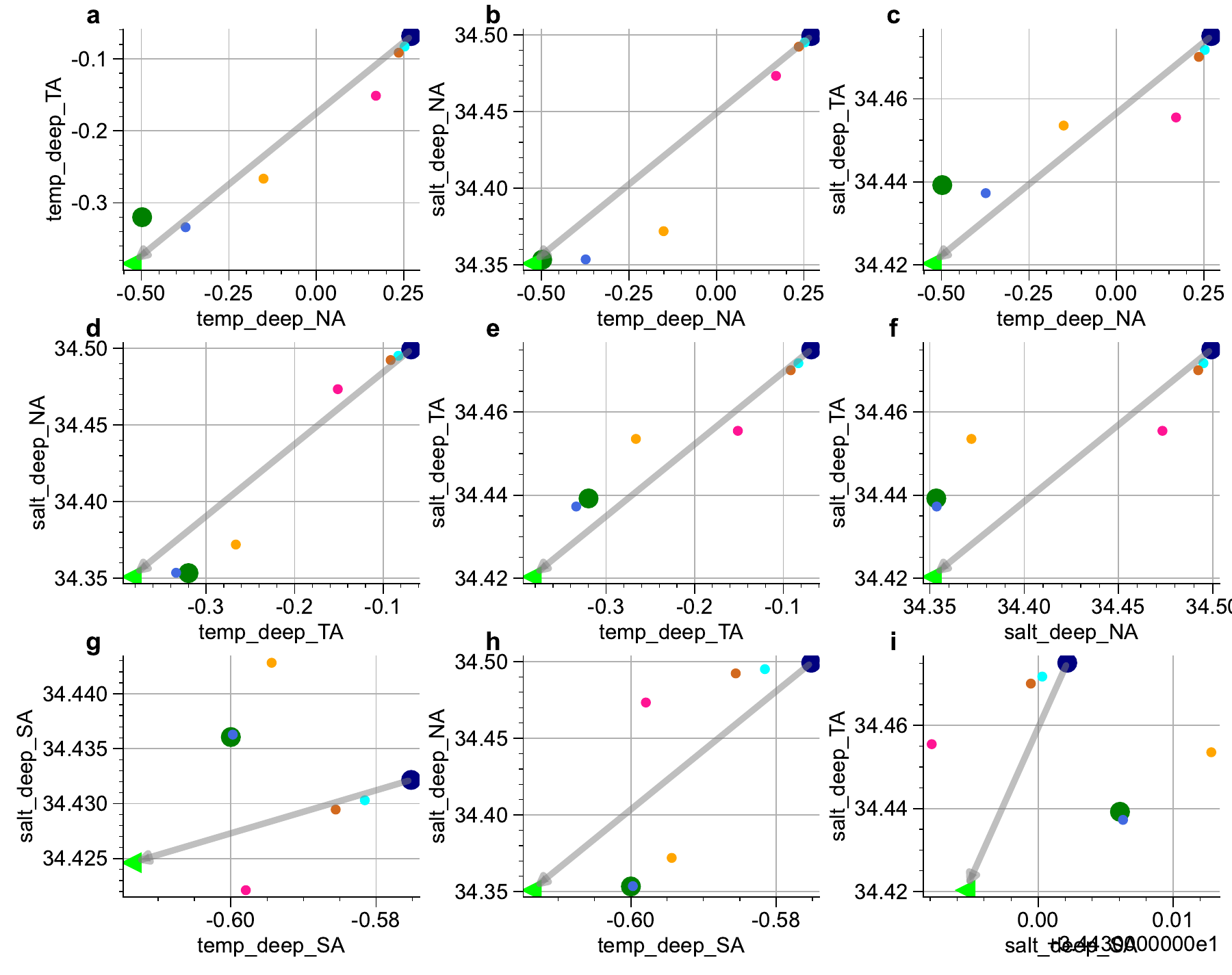}
\caption{\label{fig:projections_edge_vector} 
Vector (gray) pointing from the AMOC ON state to the edge state in the Veros model at the parameter value $F=0.3472$~Sv, shown in nine different projections comprised of observables concerning the Atlantic deep ocean temperature and salinity in different sectors. The symbols are the mean states in the different attractors and the edge state (green triangle). The largest symbols are the specific AMOC ON, OFF and edge states as defined in the main text, whereas the smaller symbols are the other co-existing attractors. 
}
\end{figure}

\begin{figure}[!htb]
\includegraphics[width=0.8\textwidth]{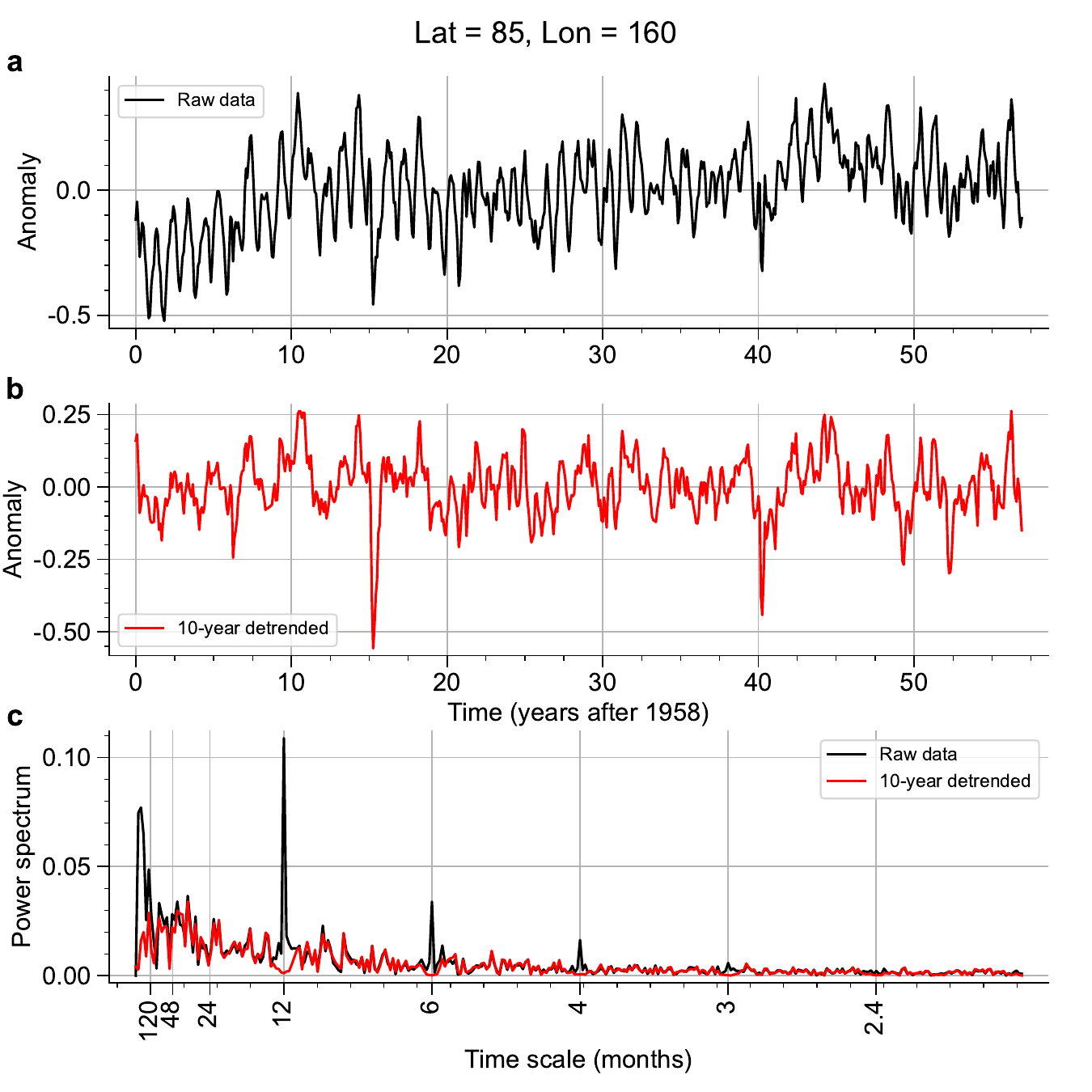}
\caption{\label{fig:detrending_salinity1} 
Data used for the construction of the salinity noise model (see Appendix~B), shown as an example time series at one particular grid point (see figure title). {\bf a} Raw anomaly data, and {\bf b} data detrended by the average climatology in a 10-year moving window. Panel {\bf c} shows the corresponding power spectra. 
}
\end{figure}

\begin{figure}[!htb]
\includegraphics[width=0.8\textwidth]{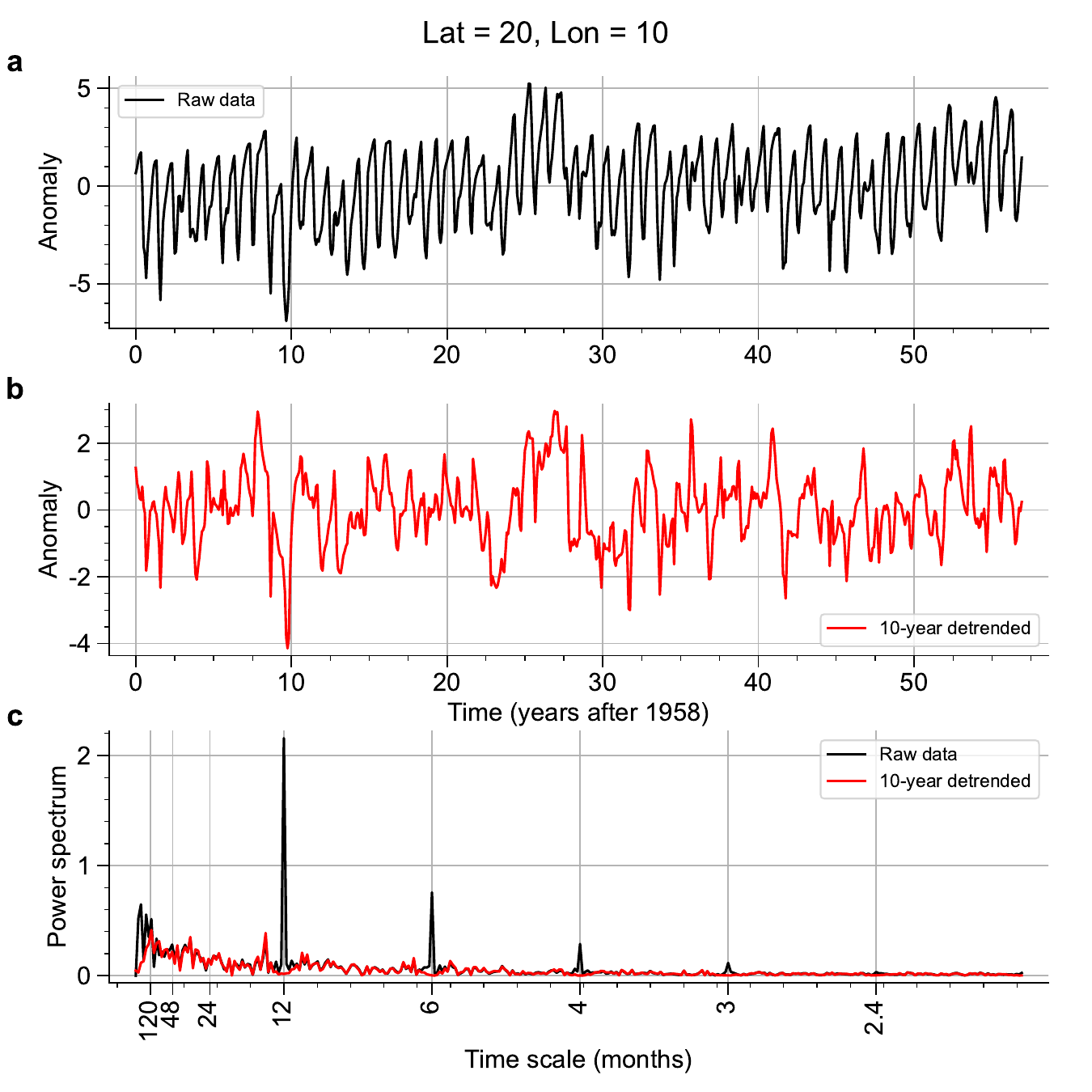}
\caption{\label{fig:detrending_salinity2} 
Same as Fig.~S21, but for a different grid point.
}
\end{figure}

\begin{figure}[!htb]
\includegraphics[width=0.99\textwidth]{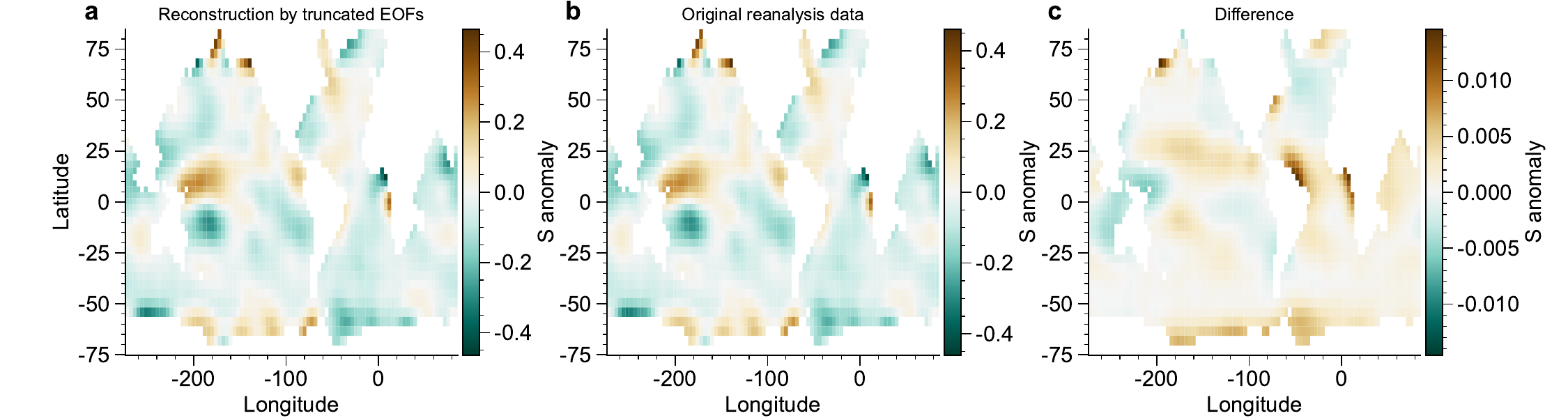}
\caption{\label{fig:eof_reconstruction_salinity} 
Demonstration of the error made by the chosen truncation of EOFs (see Appendix~B) to reconstruct the reanalysis data used for the salinity component of the noise model.
}
\end{figure}

\begin{figure}%[floatfix]%!htb
\includegraphics[width=0.99\textwidth]{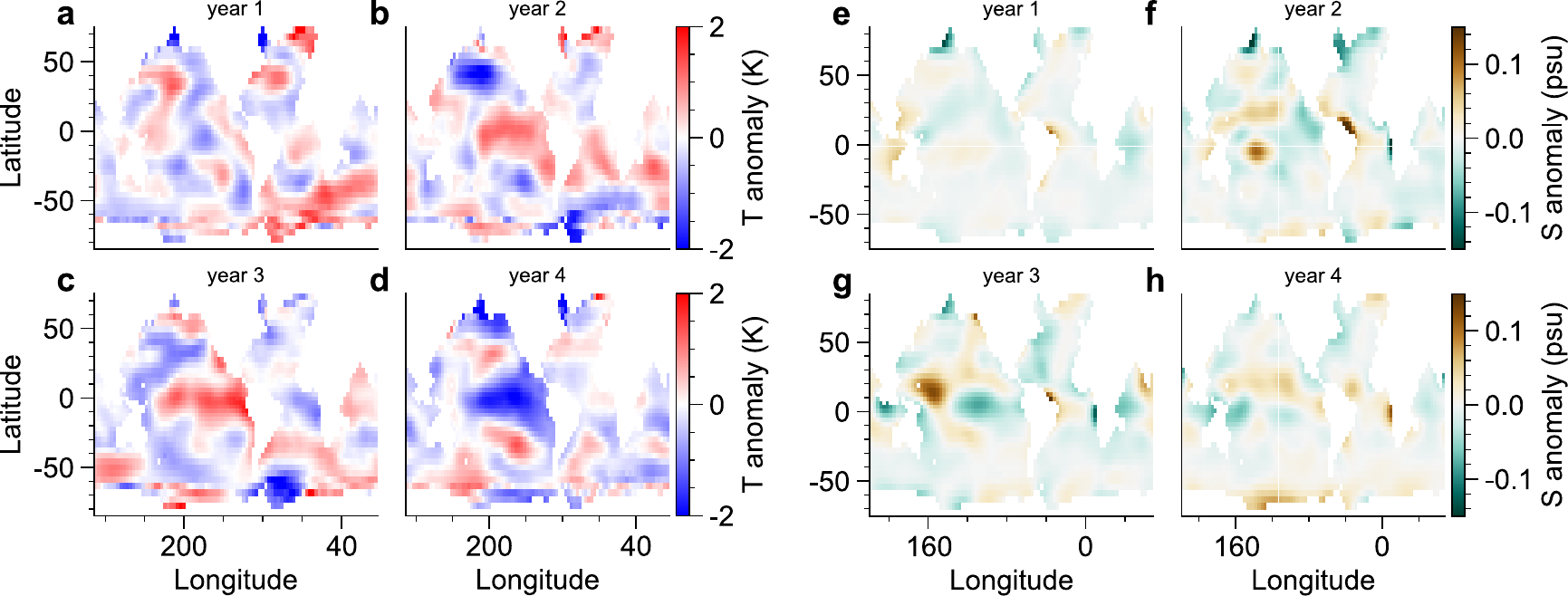}
\caption{\label{fig:noise_demo} 
Annual average output of one realization of the surface noise model for Veros in 4 consecutive years for the temperature ({\bf a-d}) and salinity components ({\bf a-d}).
}
\end{figure}

\begin{figure}[!htb]
\includegraphics[width=0.99\textwidth]{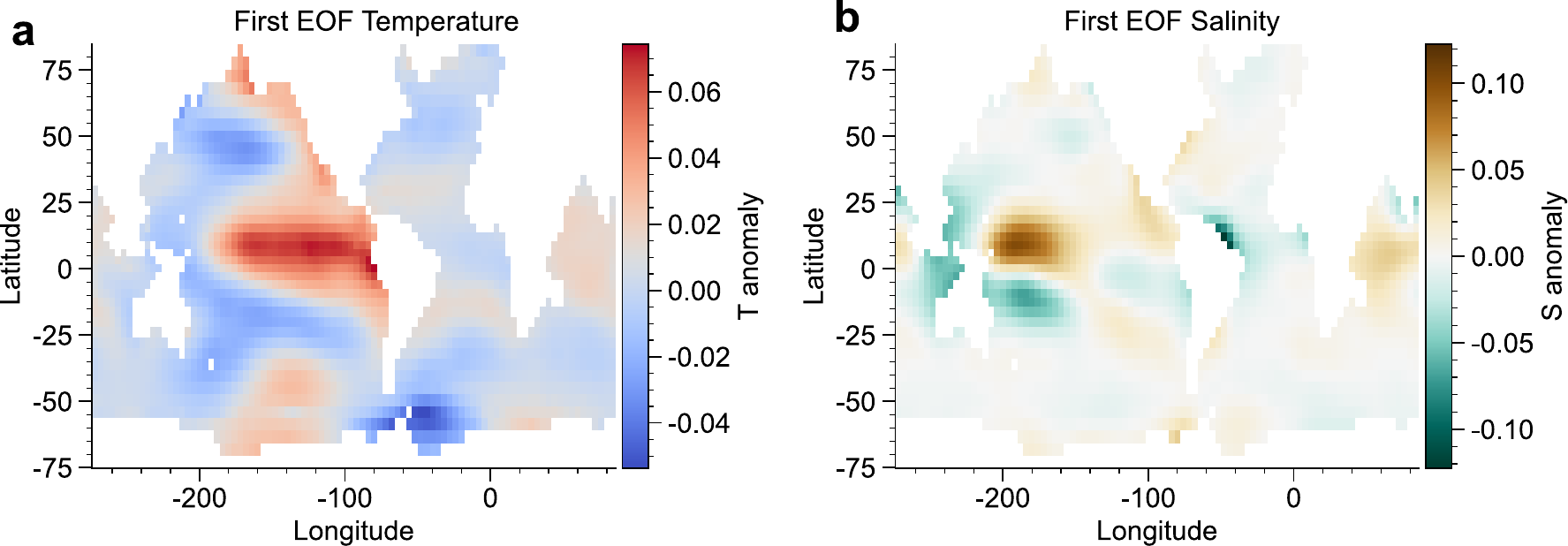}
\caption{\label{fig:eof1} 
First empirical orthogonal functions (EOF) of the re-analysis data after detrending and removal of seasonality, regridded on the Veros ocean model grid, for the {\bf a} temperature and {\bf b} salinity components of the Veros surface noise model (see main text and Appendix~B). 
}
\end{figure}